\documentclass[10pt,aps,prc,floatfix,twocolumn,nofootinbib]{revtex4-1}

\usepackage[subpreambles=true]{standalone}

\usepackage[labelfont={small},subrefformat=parens,caption=false]{subfig}
\usepackage{graphicx,amsmath,amssymb,bm}
\usepackage{amsfonts}
\usepackage[utf8]{inputenc}

\usepackage{verbatim}
\usepackage{float}
\usepackage{cancel}
\usepackage{multirow}
\usepackage{array}
\usepackage{xparse}

\usepackage{physics}
\usepackage{color}
\usepackage{dsfont}
\usepackage{tabularx}
\usepackage{longtable}

\usepackage[pdfencoding=auto, pdfpagelabels]{hyperref}

\usepackage[overload]{textcase}

\definecolor{linkcolor}{rgb}{0,0,0.40} 
\hypersetup{%
    pdfsubject=Paper,
    pdfkeywords={nuclear physics} {Bayesian} {chiral EFT},
    unicode = true,
    breaklinks = true,
    colorlinks = true,
    linkcolor = linkcolor,
    citecolor = linkcolor,
    menucolor = linkcolor,
    urlcolor = linkcolor
}

\usepackage[linesnumbered,algoruled,lined,commentsnumbered]{algorithm2e}

\usepackage{cellspace}
\graphicspath{{./figures2/}}

\usepackage{silence}
\usepackage[T1]{fontenc}

\usepackage[]{standalone}

\usepackage{makecell}

\newcommand{\phantomsublabel}[3]{%
\unitlength=1in%
\put(#1,#2){%
    \subfloat[]{%
        \label{#3}%
    }}%
}

\makeatletter
\newcommand\newsubcommand[3]{\newcommand#1{#2\sc@sub{#3}}}
\def\sc@sub#1{\def\sc@thesub{#1}\@ifnextchar_{\sc@mergesubs}{_{\sc@thesub}}}
\def\sc@mergesubs_#1{_{\sc@thesub#1}}

\newcommand\newsupcommand[3]{\newcommand#1{#2\sc@sup{#3}}}
\def\sc@sup#1{\def\sc@thesup{#1}\@ifnextchar^{\sc@mergesups}{^{\sc@thesup}}}
\def\sc@mergesups^#1{^{\sc@thesup#1}}
\makeatother

\DeclareMathAlphabet{\mathbcal}{OMS}{cmsy}{b}{n}

\newcommand{\cbar}{\bar c}

\newcommand{\diagnostic}{\textup{D}}
\DeclareMathOperator{\CI}{CI}
\newcommand{\DCI}{\diagnostic_{\CI}}
\newcommand{\MD}{\textup{MD}}
\newcommand{\DMD}{\diagnostic_{\MD}}
\newcommand{\DVAR}[1]{\inputvec{D}_{\textup{#1}}}

\newcommand{\iid}{\text{i.i.d.}}

\newcommand{\NLO}{\ensuremath{{\rm NLO}}\xspace}
\newcommand{\NNLO}{\ensuremath{{\rm N}{}^2{\rm LO}}\xspace}
\newcommand{\NNNLO}{\ensuremath{{\rm N}{}^3{\rm LO}}\xspace}
\newcommand{\NNNNLO}{\ensuremath{{\rm N}{}^4{\rm LO}}\xspace}
\newcommand{\NNNNLOp}{\ensuremath{{\rm N}{}^4{\rm LO}+}\xspace}

\DeclareMathOperator{\GP}{\mathcal{GP}}

\newcommand{\param}{\boldsymbol{\theta}}

\newcommand{\ordervec}{\vec}
\newcommand{\inputdimvec}{}
\newcommand{\inputvec}{\mathbf}

\newsubcommand{\ckvec}{\ordervec{c}}{k}

\newsubcommand{\bkvec}{\ordervec{b}}{k}
\newcommand{\kinparvec}{\inputdimvec{x}}
\newcommand{\kinparvecset}{\inputdimvec{\inputvec{x}}}

\newsubcommand{\ckvecset}{\ordervec{\inputvec{c}}}{k}

\newsubcommand{\ckvecapprox}{\mathbf{c}'}{k}
\newsubcommand{\ckvecapproxset}{\mathbf{C}'}{k}

\newsubcommand{\bkvecapprox}{\mathbf{b}'}{k}
\newsubcommand{\bkvecset}{\mathbf{B}}{k}
\newsubcommand{\bkvecapproxset}{\mathbf{B}'}{k}

\newcommand{\genobs}{y}

\newsubcommand{\genobsvec}{\ordervec{\genobs}}{k}
\newsubcommand{\genobsvecset}{\ordervec{\inputvec{\genobs}}}{k}
\newcommand{\genobsset}{\inputvec{\genobs}}

\newsubcommand{\akvec}{\mathbf{a}}{k}

\newsubcommand{\akvecapprox}{\mathbf{a}'}{k}
\newsubcommand{\akvecset}{\mathbf{A}}{k}
\newsubcommand{\akvecapproxset}{\mathbf{A}'}{k}

{}  

\DeclareMathOperator{\pr}{pr} 
\newcommand{\given}{\,|\,}  

\newcommand{\Elab}{E_{\rm lab}}

\newcommand{\prel}{p_{\rm rel}}

\newcommand{\chiEFT}{$\chi$EFT}

\newcommand{\genobsref}{\ensuremath{y_{\mathrm{ref}}}}

\newcommand{\sigmatot}{\sigma_{\text{tot}}}

\newcommand{\bigJ}{\mathcal{J}}
\newcommand{\littleJ}{j}

\def\diffd{\mathrm{d}}  

\DeclareDocumentCommand\differential{ o g d() }{ 
    \IfNoValueTF{#2}{
        \IfNoValueTF{#3}
            {\diffd\IfNoValueTF{#1}{}{^{#1}}}
            {\mathinner{\diffd\IfNoValueTF{#1}{}{^{#1}}\argopen(#3\argclose)}}
        }
        {\mathinner{\diffd\IfNoValueTF{#1}{}{^{#1}}#2} \IfNoValueTF{#3}{}{(#3)}}
    }
\DeclareDocumentCommand\dd{}{\differential} 

\newcommand{\pathd}{\mathcal{D}}  

\DeclareDocumentCommand\pathdifferential{ o g d() }{ 
    \IfNoValueTF{#2}{
        \IfNoValueTF{#3}
            {\pathd\IfNoValueTF{#1}{}{^{#1}}}
            {\mathinner{\pathd\IfNoValueTF{#1}{}{^{#1}}\argopen(#3\argclose)}}
        }
        {\mathinner{\pathd\IfNoValueTF{#1}{}{^{#1}}#2} \IfNoValueTF{#3}{}{(#3)}}
    }

\newcommand{\Lambdab}{\Lambda_b}
\newcommand{\mpi}{m_{\pi}}

\newcommand{\Qsum}{Q_{\mathrm{sum}}}
\newcommand{\qcm}{q_{\mathrm{CM}}}
\newcommand{\mpieff}{m_{\mathrm{eff}}} 

\newcommand{\negcos}{-\cos(\theta)}

\newcommand{\xE}{x_{E}}
\newcommand{\xtheta}{x_{\theta}}

\newcommand{\betag}{\beta_g}

\newcommand{\betaj}{\beta_j}

\begin{document}


\title{Assessing Convergence Patterns Across Modern Nucleon-Nucleon Potentials}

\author{P.~J. Millican}
\email{millican.7@osu.edu}
\affiliation{Department of Physics, The Ohio State University, Columbus, OH 43210, USA}

\author{R.~J. Furnstahl}
\email{furnstahl.1@osu.edu}
\affiliation{Department of Physics, The Ohio State University, Columbus, OH 43210, USA}

\author{J.~A. Melendez}
\email{melendez.27@osu.edu}
\affiliation{Department of Physics, The Ohio State University, Columbus, OH 43210, USA}

\author{D.~R.~Phillips}
\email{phillid1@ohio.edu}
\affiliation{Department of Physics and Astronomy and Institute of Nuclear and Particle Physics, Ohio University, Athens, OH 45701, USA}

\date{\today}

\begin{abstract}
The BUQEYE model for correlated effective field theory (EFT) truncation errors assumes a regular pattern of dimensionless coefficients extracted from order-by-order observable calculations. 
This enables results from lower orders to inform statistical predictions for error estimates of omitted higher orders.
We test the model for
multiple chiral EFT (\chiEFT) nucleon-nucleon (NN) potentials
using a suite of six common NN scattering observables represented as functions of relative momentum and scattering angle.
First, we flag irregularity in the convergence patterns of potentials with so-called ``soft'' regulator scales,
namely that the sizes of the coefficients in the observables' expansions are mismatched between the even and odd orders.
Second, we test the BUQEYE model's assumption of Gaussian process (GP) stationarity against the data
and find that the GP's correlation structure as encoded in its length scale is nonstationary: The GP length scale in the angular dimension $\ell_{\theta}$ is approximately inversely proportional to the relative momentum of the scattering process.
After we remediate this issue by allowing $\ell_{\theta}$ to vary with momentum,
diagnostics show significant improvement, validating the application of the BUQEYE model after this modification.
Third, we find that good performance of the BUQEYE model relies on properly choosing the \chiEFT\ breakdown scale $\Lambdab$.
With $\mpieff$ held fixed at the physical pion mass we find $\Lambdab=600$--$750$ MeV for various \NNNLO interactions.
Statistically consistent distributions for $\Lambdab$ across orders are only found for the SMS potential at a regulator scale of 450 or 500 MeV.
All our results can be reproduced using a publicly available Jupyter notebook, which can be straightforwardly modified to analyze other \chiEFT\ NN potentials.
\end{abstract}

\maketitle

\section{Introduction}
\label{sec:introduction}

In principle, chiral effective field theory (\chiEFT) provides a systematic expansion of nucleon-nucleon (NN) scattering observables in powers of a small expansion parameter $Q$, a ratio of the scattering momentum and the soft \chiEFT\ scale, $\mpieff$, to the hard scale, $\Lambdab$.
In practice, \chiEFT\ is predominantly implemented in nuclear physics calculations via NN potentials that are computed up to a fixed order in the EFT expansion and then inserted into the Schr\"odinger or Lippmann-Schwinger equation.
NN potentials constructed in this way are singular, so the standard implementation of \chiEFT\ in the NN system regularizes the \chiEFT\ potential by multiplying it by a regulator function that tames the high-momentum or short-distance behavior. 

Many \chiEFT\ NN potentials have been formulated and fitted to NN data.
These can be classified into NN potential ``families'': a given family is regulated using a particular functional form, but within that family the NN potential exists at different \chiEFT\ orders and has been fitted to data at different values of the regulator scale. 
Table~3.1 of
Ref.~\cite{machleidt2024recent} gives a comprehensive list of the \chiEFT\ NN potential families that have been developed in the last twenty years. 

Our previous work~\cite{Millican:2024yuz}   employed a set of statistical tools 
to assess the application of the BUQEYE model for correlated EFT truncation errors to the order-by-order convergence of NN observables computed with a single exemplary \chiEFT\ potential. 
In this paper we follow that progress by 
testing its application to
several families of potentials that encompass a variety of different regularization 
strategies.
We take all these potentials as they were
fit
in the original publications; that is, we do not do parameter estimation in this paper. 

The BUQEYE model uses a Bayesian statistical framework to describe a properly formulated EFT's truncation error.
The  model hypothesizes that the difference in predicted values of observables between successive orders can be parametrized using a series of dimensionless coefficient functions $c_n(\kinparvec)$, and that these functions 
can be described as draws from a single Gaussian process (GP).
The $c_n(\kinparvec)$'s are extracted by writing an observable $\genobs$ as
\begin{align} \label{eq:obs_k_expansion}
    \genobs_k(\kinparvec) = \genobsref(\kinparvec) \sum_{n=0}^k c_n(\kinparvec) Q^n(\kinparvec),
\end{align}
where $k$ is the highest computed order of the EFT, 
$x$ is the input space(s), and $\genobsref$ is a reference scale~\cite{Furnstahl:2015rha,Melendez:2019izc,Millican:2024yuz}.
%
In Sec.~\ref{sec:bayesian_stats} we
review the methods and process by which we learn from the $c_{n}$.
Statistical tests can address whether a set of coefficients satisfies the BUQEYE hypotheses~\cite{Melendez:2019izc,Millican:2024yuz}.

In Ref.~\cite{Millican:2024yuz} we considered the predictions of the $\Lambda=500$ MeV semi-local \chiEFT\ NN potential of Reinert, Krebs, and Epelbaum~\cite{Reinert:2017usi} for neutron-proton (\emph{np}) scattering observables in the 0--350 MeV lab energy range.
We first verified that---with suitable choices for $x$ and $Q$---the coefficient curves are consistent with random draws from a single GP.
We then used induction to produce error bars for the unknown orders' contribution---i.e., we estimated the truncation error.
Reference~\cite{Millican:2024yuz} both proved the applicability of the BUQEYE model to an exemplary modern NN potential and
recommended specific choices for $Q$ and $x$, which we adopt here.
In particular,
we take $Q$ to depend on $\prel$, the relative momentum, as:    
\begin{equation}
Q = \Qsum(\prel, \mpieff) = \frac{\prel + \mpieff}{\Lambdab + \mpieff} \;,
\label{eq:expansionparameter}
\end{equation}
where $\mpieff$ is the \chiEFT\ soft scale and $\Lambdab$ is the \chiEFT\ hard (or breakdown) scale.
Posterior probability distributions for $\mpieff$ and $\Lambdab$ using this interaction were also extracted in Ref.~\cite{Millican:2024yuz}.

In this work we apply and refine the insights of Ref.~\cite{Millican:2024yuz}.
We consider the following potentials:%
\footnote{As a matter of nomenclature, we will refer to each of the following potentials for short by its abbreviation followed by its cutoff value. For instance, the potential from the SCS family with cutoff 0.9 fm will be called ``SCS 0.9 fm.''}
\begin{itemize}
    \item The 
    semi-local
    momentum-space (``SMS'') potential of Reinert, Krebs, and Epelbaum, which has been fitted with regulator scales of 400, 450, 500,%
    \footnote{The SMS potential with cutoff 500 MeV was the potential tested in Ref.~\cite{Millican:2024yuz}.}
    and 550 MeV~\cite{Reinert:2017usi};
    \item The 
    semi-local
    coordinate space potential (``SCS'') of Epelbaum, Krebs, and Mei{\ss}ner, which has regulator scales of 0.8, 0.9, 1.0, 1.1, and 1.2 fm~\cite{Epelbaum:2014efa};
    \item The 
    nonlocal
    momentum-space (``EMN'') potential of Entem, Machleidt, and Nosyk, which has regulator scales of 450, 500, and 550 MeV~\cite{Entem:2017gor}; and
    \item The 
    local
    coordinate-space (``GT+'') potential of Gezerlis, Tews, et al., which has regulator scales of 0.9, 1.0, 1.1, and 1.2 fm~\cite{Gezerlis:2014zia}.
\end{itemize}
We rely on \chiEFT\ being a systematic expansion of the NN interaction, so we examine only those potentials for which order-by-order fits are available; this excludes from consideration the sim~\cite{Carlsson:2015vda} and NNLO$_{\mathrm{sat}}$~\cite{Ekstrom:2015rta} potentials, for example.
Other recently developed \chiEFT\ potentials, such as those proposed in Refs.~\cite{Piarulli:2014bda} and~\cite{Ekstrom:2017koy}, include $\Delta$ isobars as explicit low-energy degrees of freedom.
BUQEYE-framework analysis of the convergence of these $\Delta$-ful potentials
is left for future work.

In Sec.~\ref{subsec:convergence_diagnosis} we 
apply the BUQEYE model to a subset of this list to gauge
which regulator scales lead to the most regularity, in particular focusing on the implications of low-momentum cutoffs for EFT convergence.
Previous work~\cite{Melendez:2017phj} has observed that these ``soft'' potentials  
exhibit irregular convergence patterns.
We revisit these observations in this paper using the correlated truncation model,
outline diagnostic criteria for irregular order-by-order convergence in \chiEFT\ potentials, and apply them to potentials on our list. 

The results of Ref.~\cite{Millican:2024yuz} also require refinement, since there we observed that the correlation-length $\ell_\theta$ associated with EFT coefficient variability in angle is very large at low momentum and becomes smaller almost monotonically as the momentum rises.
This is unsurprising: 
Lower-momentum scattering states access only low partial waves, which have less angular structure and therefore are best characterized by a GP with a long length scale, whereas higher-momentum states can access higher partial waves with structure at smaller angular scales.
This physics violates the assumption that the coefficient data at all energies share a common angular length scale and variance: the ``stationarity'' of the GP. Section~\ref{subsec:nonstationarity_evidence} examines the nature of the non-stationarity in the GP, presenting evidence that it is largely confined to the momentum dependence of $\ell_\theta$. 
In Sec.~\ref{sec:nonstationarity} we show how to remediate this issue by transforming the angular input space and demonstrate that diagnostics~\cite{Millican:2024yuz} show significant improvement, validating
the application of the BUQEYE model after it is modified to incorporate this physics. 

In Sec.~\ref{sec:Lambdab_meff_posteriors} we execute our modified analysis for a number of non-soft potentials, and
examine what can be learned 
about
$\Lambdab$ and $\mpieff$.
We find that properly choosing values of $\Lambdab$ is important for good performance of the BUQEYE model, but statistically consistent distributions for $\Lambdab$ across orders are only found for a subset of interactions when  $\mpieff$ is held fixed.
Our summary and outlook are given in Sec.~\ref{sec:outlook}.
Additional figures are provided in the Supplemental Material. 
All figures can be reproduced or modified using freely available software.

\section{Statistical methods}
\label{sec:bayesian_stats}

\subsection{Formulas for the BUQEYE model}
\label{subsec:buqeye_gp_review}

First, we review the key expressions of the BUQEYE model from Refs.~\cite{Millican:2024yuz,Melendez:2019izc}.
The coefficient functions in Eq.~\eqref{eq:obs_k_expansion} are independent and identically distributed (i.i.d.) draws from a GP,
\begin{align}
    c_n(\kinparvec) \given \param \overset{\text{\tiny \iid}}{\sim} \GP[m(x), \kappa(x, x'; \param)] ,
    \label{eq:gp_generic}
\end{align}
where $\param$ are any parameters required to define the GP kernel (in this case, $\{\cbar^2, \ell, \dots\}$)
and $G \sim H$ should be read as ``$G$ is distributed as $H$.''
As in previous work~\cite{Melendez:2017phj, Melendez:2019izc, Millican:2024yuz}, we assume that the mean function $m(x)$ is identically zero and the kernel $\kappa$ factorizes into a marginal variance $\cbar^2$ and a correlation kernel $r(\kinparvec, \kinparvec')$:
\begin{align}
    \kappa(\kinparvec, \kinparvec') = \cbar^2 r(\kinparvec, \kinparvec') .
\end{align}
Further, we assume that $\kappa\left(x, x^{\prime}; \boldsymbol{\theta} = \left\{\ell, \cbar^{2}\right\} \right)$ is a squared-exponential radial basis function (SERBF):
\begin{align}
    \kappa\left(x, x^{\prime}\right) &= \cbar^{2} r(\kinparvec, \kinparvec';\ell) 
    \notag \\
    &= \cbar^{2} \exp\left\{-\frac{1}{2}\left(x - x^{\prime}\right)^{\intercal}L^{-1}\left(x - x^{\prime}\right)\right\} \;.
    \label{eq:rbf_kernel}
\end{align}
Here, $x$ and $x^{\prime}$ are arbitrary points in the input space, $\cbar^{2}$ is the variance of the distribution encoding the rough size of the $c_{n}$, and $L$ is a square matrix that collects the (squares of the) length scales in the different dimensions of the input space.

In particular, for NN scattering observables we take $x = \left(\xE, \xtheta\right)$ and use a  diagonal $2 \times 2$ matrix for $L$.
Under these assumptions,
\begin{equation}
    r(x, x'; \ell_{E}, \ell_{\theta}) = \exp{-\frac{(\xE - \xE^{\prime})^2}{2\ell_{E}^2} - \frac{(\xtheta - \xtheta^{\prime})^2}{2\ell_{\theta}^2}} \;,
    \label{eq:rbf_2d}
\end{equation}
where $\xE$ is the input variable that encodes the variation across the input space with energy and is taken to be $\prel$ throughout this paper, $\xtheta$ varies across the scattering-angle input space and is taken to be $\xtheta=\negcos$, and $\ell_{E}$ and $\ell_{\theta}$ are the corresponding length scales, respectively. Equations~\eqref{eq:gp_generic}--\eqref{eq:rbf_2d} imply that the coefficients share a common size ($\cbar^{2}$) and characteristic scale of variability ($\ell$) and are Gaussian-distributed at any given point in $x$.

The assumption that $\ell_{E}$ and $\ell_{\theta}$ are constant across $x$ (i.e., that the GP is {\it stationary}) is assessed in Sec.~\ref{sec:buqeye_irregular} and found wanting.
We revise it in Sec.~\ref{sec:nonstationarity}.

\subsection{Inferring parameters and hyperparameters}
\label{subsec:parameter_estimates}

A common theme throughout the following sections will be the inference of quantities, either physical or statistical, from the data.
It is worth describing this process in generality before
applying it in specific cases later on.

To begin, we observe that there are at least three data spaces to keep track of here:
\begin{itemize}
    \item the set of $N$ training points $\kinparvecset$, denoted in bold font, which could be one-dimensional or multi-dimensional;
    
    \item the set of $k$ orders, denoted with an arrow, with individual orders indexed by $n$, e.g., $\vec{y}_k$ is the set and $y_n$ is the $n^{\text{th}}$ order for a specific observable. $\genobsvecset = \{\genobsset_1, \cdots, \genobsset_k\}$ is then the set of $k$  curves for that observable, where each is evaluated at the $N$ training locations $\kinparvecset$ (these are assumed to be the same for each order).
    \item the set of different observables, denoted with braces $\{\cdots\}_\bigJ$ for the full set and $\{\cdots\}_j$ for individual observables.
\end{itemize}
The set of observables used for training is then $\{\genobsvecset\}_\bigJ$. This is the most general set of data available for training in this context. An individual member of this set is then denoted $\{\genobsvecset\}_\littleJ$, although we will sometimes drop the set notation and subscript $j$ when referring to results for an individual observable. 

There are two categories of parameters: those that are global and shared across observables, and those that are observable-specific.
Suppose we want to learn about the complete set of global parameters $\betag$ from the data.
In this case, $\betag=\{\Lambdab, \mpieff\}$, which determine the dimensionless expansion parameter $Q$ according to Eq.~\eqref{eq:expansionparameter}.
To determine $\betag$ we use
\begin{align}
    \pr(\betag \given \{\genobsvecset\}_\bigJ) \propto \pr(\{\genobsvecset\}_\bigJ \given \betag) \pr(\betag)\;,
    \label{eq:betag_posterior}
\end{align}
which follows from Bayes' theorem.
All the information shared across observables is through the complete set $\betag$, therefore given $\betag$ we can factorize the likelihood in Eq.~\eqref{eq:betag_posterior}:
\begin{align}
    \pr(\{\genobsvecset\}_\bigJ \given \betag) = \prod_{j\in \bigJ} \pr(\{\genobsvecset\}_j \given \betag) \;.
\end{align}

The observable-specific parameters for observable $j$ are denoted $\betaj$. 
Here these are the parameters of the SERBF kernel in Eq.~\eqref{eq:rbf_kernel}, namely $\cbar$, $\ell_{E}$, and $\ell_{\theta}$.
Numerical and graphical consideration of these quantities in Sec.~\ref{subsec:nonstationarity_evidence} reveals that $\cbar$ and $\ell_{E}$ can be treated as local and stationary whereas $\ell_{\theta}$ ought to be treated as nonstationary and described by a power law in terms of $\xE$.
The exponent of that power law derives from physics that is common to all observables, and so could be treated as part of the set $\beta_g$; however, the pre-factor in $\ell_\theta$ is observable specific and so remains a ``local'' parameter.

To estimate $\betaj$, we start by integrating them in (for readability dropping the $j$ subscripts when not needed):
\begin{align}
    \pr(\genobsvecset \given \betag) = \int \dd{\betaj} \pr(\genobsvecset \given \betaj, \betag) \pr(\betaj) .
\end{align}
At this stage, we can change variables from the data $\genobsvecset$ for observable $j$ to the corresponding coefficients using Eq.~\eqref{eq:obs_k_expansion} (see \cite{Melendez:2019izc} for full details),
\begin{align} \label{eq:genobsvec_given_ell_Q} 
    \pr(\genobsvecset\given\betaj, \betag) = 
\frac{\pr(\ordervec{\inputvec{c}}_k\given\betaj, \betag)}{\prod_{n,i} \abs{\genobsref(\kinparvec_i)Q^n(\kinparvec_i)}}.
\end{align}%
This change of variables is critical if one is sweeping through values of $Q$, but otherwise the factor it introduces is just part of the normalization of the pdf.
Note that $\betag$ has two different roles here: enabling factorization through conditional independence Eq.~\eqref{eq:betag_posterior} and dictating the change of variables from data to coefficients (given $\genobsref$).
Finally, we have:
\begin{align}
    \pr(\beta_j \given \ordervec{\inputvec{c}}_k, \betag) & \propto \pr(\ordervec{\inputvec{c}}_k \given \beta_j, \betag) \pr(\beta_j) \notag \\
    & = \pr(\beta_j) \prod_n \pr(\{c_{n}\} \given \beta_j, \betag) . 
    \label{eq:betajposterior}
\end{align}

In the application we are discussing here the observable-specific  
parameters $\beta_j$ are, as already mentioned, the GP hyperparameters appearing in Eqs.~\eqref{eq:gp_generic}--\eqref{eq:rbf_2d}. Equation~\eqref{eq:betajposterior} tells us how their pdf is obtained from  the extracted $c_{n}$. To implement that equation we must declare the prior for those parameters, $\pr(\beta_j)$.
The $\ell$'s are assigned the Heaviside prior $H(x)$, while $\cbar^{2}$'s prior is taken to be the scaled inverse-chi-squared (scaled $\chi^{-2}$):
\begin{equation}
    \cbar^{2} \sim \chi^{-2}\left(\nu_{0}, \tau_{0}^2\right) \;,
    \label{eq:scaled_inv_chi_sq}
\end{equation}
where $\nu_{0}$ is the prior number of degrees of freedom and $\tau_{0}$ is the scale parameter. Since this is the conjugate prior for the variance of a Gaussian likelihood the $\cbar^2$ distribution can be updated from prior to posterior by 
replacing the prior parameters $\nu_0$ and $\tau_0^2$ by updated values $\nu$ and $\tau^2$. The updating formulae are given in Appendix A, Section 2, of Ref.~\cite{Melendez:2019izc}. 

Although the $\chi^{-2}$ distribution becomes more and more sharply peaked around $\tau^2$ as $\nu \rightarrow \infty$, the analysis of any finite set of coefficient curves, $\{c_n(x)\}_{n=0}^k$, will still produce a finite-width posterior for $\cbar^2$.
The use of a point estimate (e.g., the maximum \emph{a posteriori} or MAP value) for $\cbar^2$ neglects this aspect of uncertainty in the truncation error model.
In order to include it in our predictions for the higher-order EFT curves $\{c_n(x)\}_{n=k+1}^\infty$ that are unavailable to us,  we must marginalize over the posterior for $\cbar^2$. 
This integration over the variance of our GP converts it into a Student \emph{t}-process (TP):
\begin{equation}
    c_n(\kinparvec) \given \ell,\nu,\tau^2 \overset{\text{\tiny \iid}}{\sim} \int d\cbar^2 \chi^{-2}(\nu,\tau^2) \GP[m(x), \kappa\left(x, x^{\prime}; \boldsymbol{\theta}\right)]\; .
    \label{eq:tprocess}
\end{equation}
If $\cbar^2$ is not tightly constrained by the lower-order curves then the higher-order curves are described by correlated Student \emph{t}-distributions, rather than by correlated Gaussians. 

\section{You can see a lot by looking}
\label{sec:buqeye_irregular}

\subsection{Looking at the size of the curves: diagnosis of irregular convergence patterns}
\label{subsec:convergence_diagnosis}

\begin{figure*}[ptbh!]
    \centering
    \includegraphics{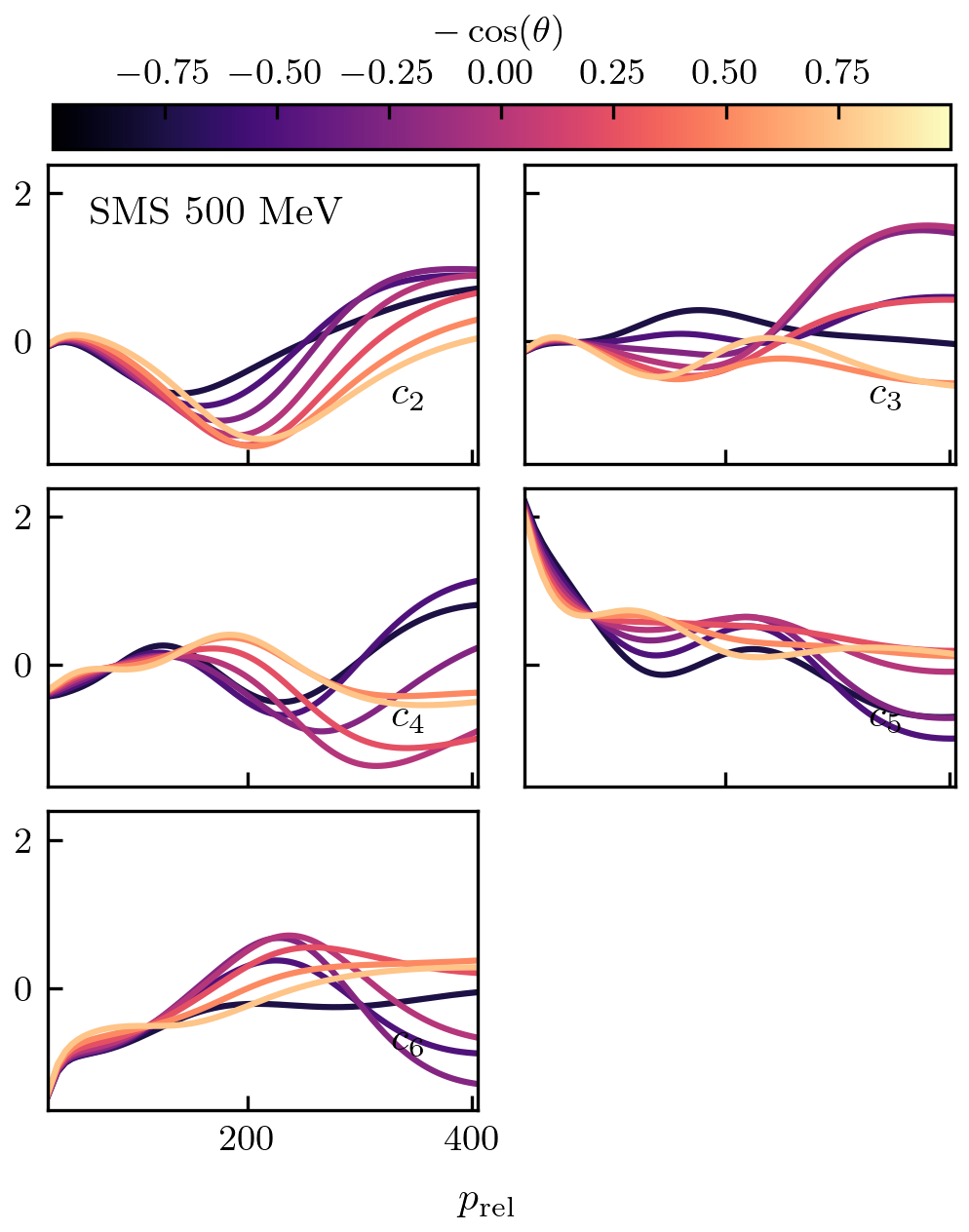}
    \phantomsublabel{-3.0}{0.0}{fig:coeffs_SMS500MeV_DSG} \includegraphics{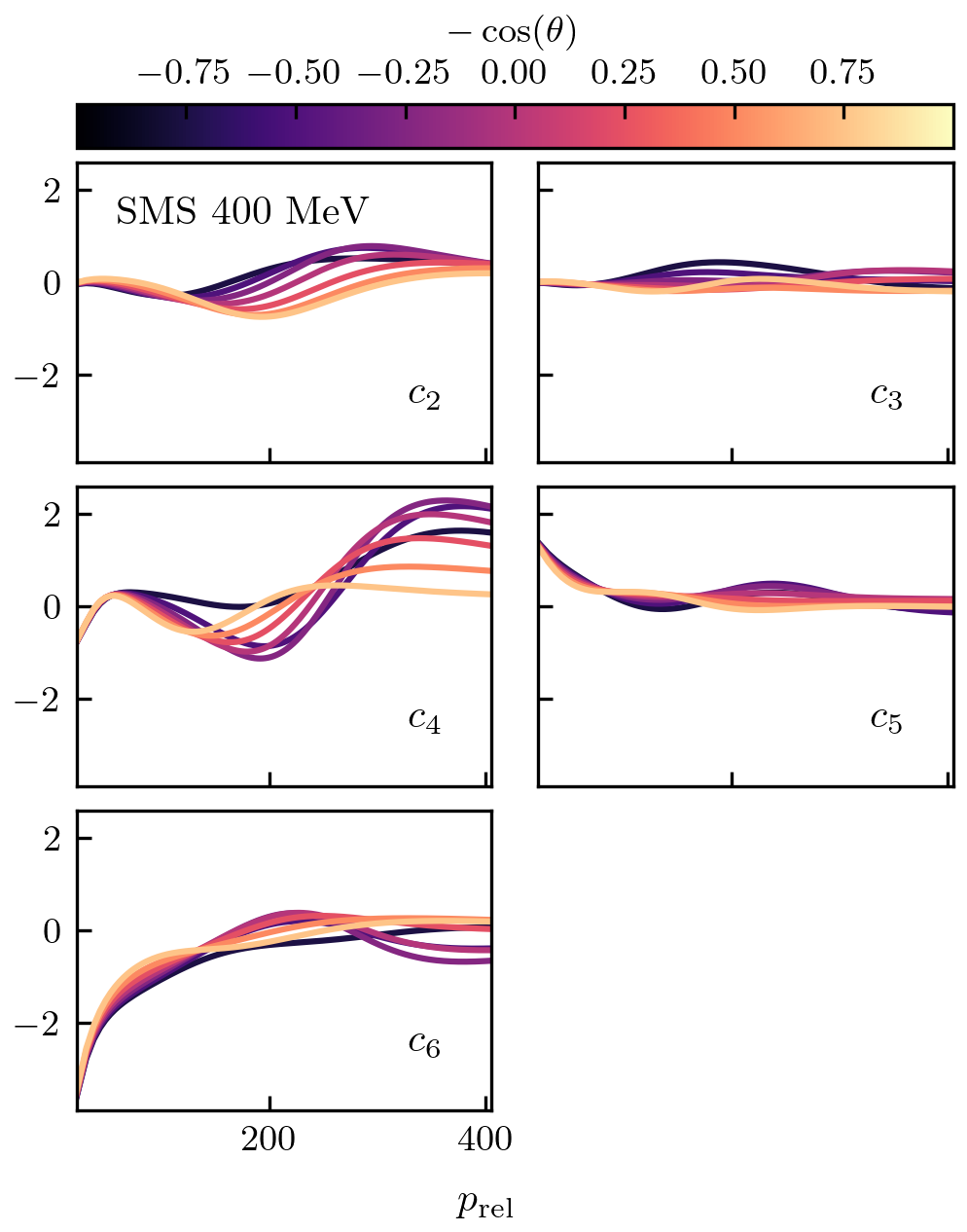}
    \phantomsublabel{-2.8}{0.0}{fig:coeffs_SMS400MeV_DSG}

    \includegraphics{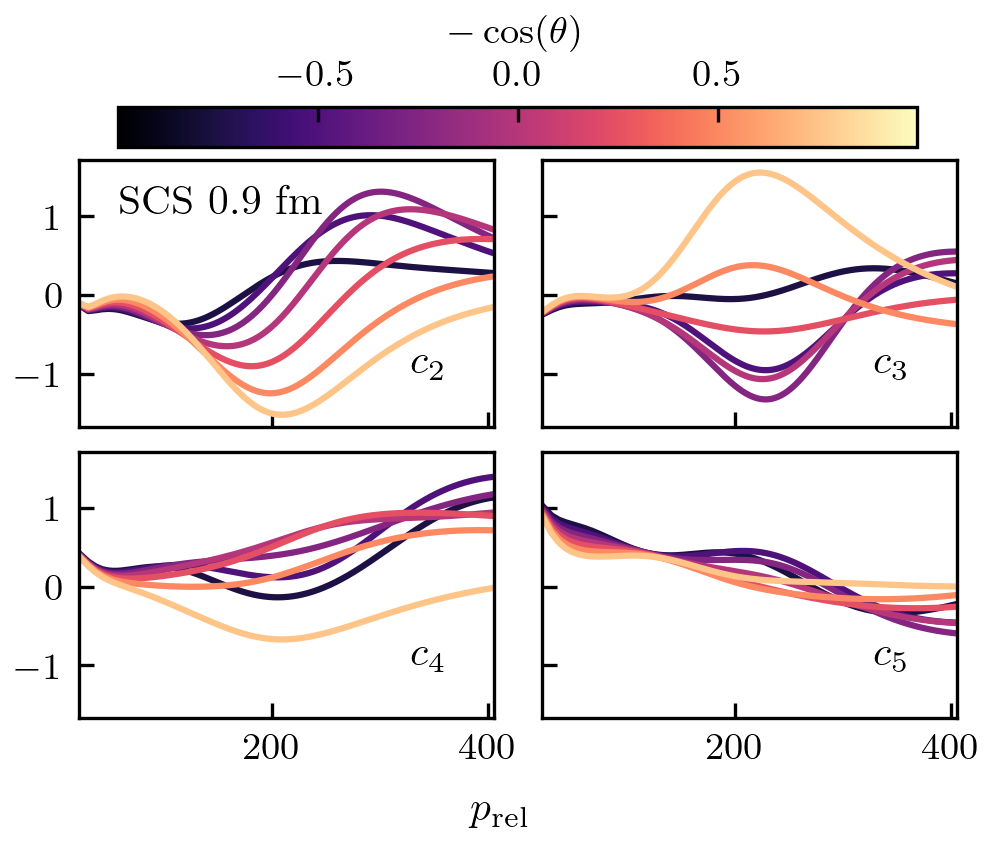}
    \phantomsublabel{-3.0}{0.0}{fig:coeffs_SCS0p9fm_DSG}\includegraphics{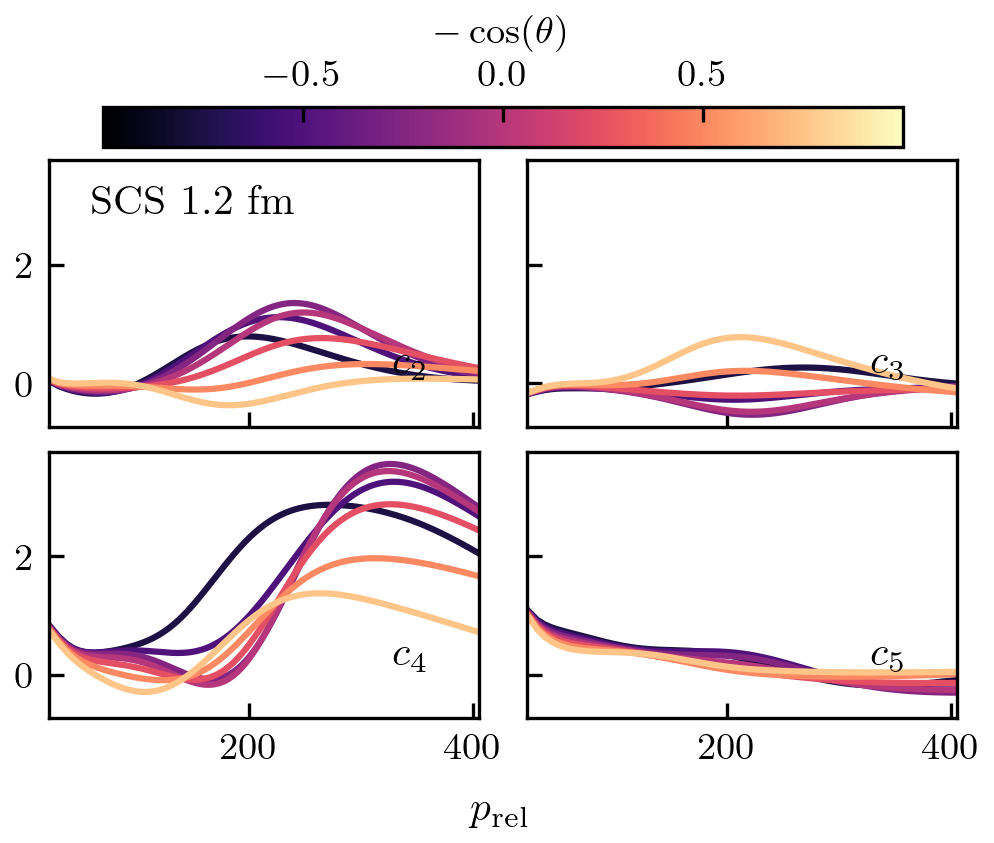}
    \phantomsublabel{-3.0}{0.0}{fig:coeffs_SCS1p2fm_DSG}
   \caption{Order-by-order coefficient functions for $d\sigma / d\Omega$ as functions of $\prel$ at fixed $\negcos$ for 
   \protect\subref{fig:coeffs_SMS500MeV_DSG} SMS 500 MeV and 
   \protect\subref{fig:coeffs_SMS400MeV_DSG} 400 MeV and 
   \protect\subref{fig:coeffs_SCS0p9fm_DSG} SCS 0.9 fm and 
   \protect\subref{fig:coeffs_SCS1p2fm_DSG} 1.2 fm extracted at fixed $\negcos$ with the choice $\left(\Lambdab = 570\,\mathrm{MeV}, \mpieff = 138\,\mathrm{MeV}\right)$ from Ref.~\cite{Millican:2024yuz}. 
    Note that the coefficients of the soft potentials (SMS 400 MeV and SCS 1.2 fm) exhibit a marked even-odd variation in their rough sizes, which is absent for the coefficients of the harder potentials (SMS 500 MeV and SCS 0.9 fm).
    Quantitative measures of these patterns are given in Table~\ref{tab:rms_deg_slices}. 
}
    \label{fig:coeffs_eachorder_smsscs_vsprel}
\end{figure*}

The BUQEYE model---and hence the entire inference structure of the previous section---assumes that the dimensionless coefficients $c_{n}$ can be treated as draws from the same GP regardless of order $n$.
An initial question to ask, then, is: ``Is such behavior in evidence when we adopt the MAP value of $\Lambda_b$?''

We base our analysis on plots of coefficient functions for a given potential.
We first plot the coefficient functions $c_n(p_{\rm rel},- \cos(\theta))$ as functions of $p_{\rm rel}$ for a representative sample of angles. Then we plot them as a function of $-\cos(\theta)$ for different $p_{\rm rel}$.
We expect that if the potential respects the BUQEYE model assumption that the GP is indifferent to order, we will see that the different $c_n$ functions vary at about the same rate and have roughly the same size, irrespective of order.

In Figs.~\ref{fig:coeffs_SMS500MeV_DSG} and~\ref{fig:coeffs_SMS400MeV_DSG}, we show the coefficients extracted for the SMS 500 MeV and 400 MeV predictions of the differential cross section $d\sigma / d\Omega$.
The coefficient curves are extracted at different fixed values of $\xtheta = \negcos$ and plotted against $\xE = \prel$.
In the plots of the SMS 500 MeV coefficients, the $c_{n}$ do not behave very differently, regardless of the value of $n$.
In contrast, the SMS 400 MeV 
curves \emph{do} care, so to speak, about order.
Specifically, the $c_{n}$ of even $n = \{2, 4, 6\}$ exhibit much greater variance across $x$ than do those of odd $n = \{3, 5\}$. 
This dependence on order is not accommodated by the current implementation of our truncation model.

A similar contrast is visible in Figs.~\ref{fig:coeffs_SCS0p9fm_DSG} and~\ref{fig:coeffs_SCS1p2fm_DSG} in the juxtaposition between the $c_{n}$ extracted from SCS 0.9 fm's prediction of $d\sigma / d\Omega$, and those extracted from SCS 1.2 fm's prediction of the same.
Once again, the curves for the former potential look the similar across orders, while those for the latter have a notably larger variance at even orders than they do at odd orders.

The fact that some potentials exhibit this even-odd splitting in coefficient size and some do not, can be quantified  by computing the root-mean-square (rms) values for the coefficient functions across order. These values are given in Table~\ref{tab:rms_deg_slices} for the four potentials discussed thus far. 

\begin{table}[tb]
    \centering
    \renewcommand{\arraystretch}{1.2}
    \begin{ruledtabular}
    \begin{tabular}{ c c c }
         & SMS 500 MeV & SMS 400 MeV \\
        \hline
        $c_{2}$ & $0.7 \pm 0.1$ & $0.40 \pm 0.04$ \\
        $c_{3}$ & $0.4 \pm 0.2$ & $0.14 \pm 0.04$ \\
        $c_{4}$ & $0.5 \pm 0.2$ & $0.9 \pm 0.3$ \\
        $c_{5}$ & $0.6 \pm 0.1$ & $0.35 \pm 0.04$ \\
        $c_{6}$ & $0.5 \pm 0.1$ & $0.87 \pm 0.07$ \\
        \hline
         & SCS 0.9 fm & SCS 1.2 fm \\
        \hline
        $c_{2}$ & $0.6 \pm 0.2$ & $0.5 \pm 0.2$ \\
        $c_{3}$ & $0.5 \pm 0.2$ & $0.2 \pm 0.1$ \\
        $c_{4}$ & $0.6 \pm 0.1$ & $1.7 \pm 0.4$ \\
        $c_{5}$ & $0.38 \pm 0.07$ & $0.36 \pm 0.05$ \\
    \end{tabular}
    \end{ruledtabular}
    \caption{
    Root-mean-square (rms) values for the coefficients shown in Fig.~\ref{fig:coeffs_eachorder_smsscs_vsprel}.
    In each case, the rms value is calculated for each coefficient, and then all the coefficients for a given potential and order are averaged to obtain these values.
    Note the similar values across for SMS 500 MeV and and across SCS 0.9 fm, two relatively hard potentials, and the significantly larger values at even order than odd order for SMS 400 MeV and SCS 1.2 fm, two relatively soft potentials.
    }
    \label{tab:rms_deg_slices}
\end{table}

Previous work with an uncorrelated error model~\cite{Melendez:2017phj} flagged such
irregular convergence patterns in the softest two SCS potentials, those with regulator cutoffs of 1.1 and 1.2 fm.
It was observed that the coefficients at the even orders \NLO ($\sim Q^{2}$) and \NNNLO ($\sim Q^{4}$) were significantly larger than those at the odd orders \NNLO ($\sim Q^{3}$) and \NNNNLO ($\sim Q^{5}$).
By comparison, coefficients derived from SCS 0.9 fm predictions did not show an appreciable discrepancy in size between even and odd orders.

The emergence of a staggering in the size of the observable coefficients as the momentum cutoff is lowered is
explained by the shuffling of pion-exchange physics,
which occur in pion-exchange mechanisms at all orders in the \chiEFT\ expansion, to short-range even-order mechanisms.
Softer potentials suppress the contribution of mid-range pion-exchange and the contact terms at even orders must adjust to produce those effects in observables.
Lower-momentum regulators thus tend to generate observable coefficients that approach those for a potential with only even orders in the EFT expansion, as in pionless EFT, which includes only contact terms~\cite{Bub:2024gyz,Chen:1999tn,Weinberg:1991um}.

The results of this section show that this feature is not peculiar to the semi-local regulator scheme or coordinate-space potentials; indeed, it is a feature of softer potentials generally, since softer regulators  generate larger regulator artifacts.
Plots of coefficients for other potentials are shown in  Figs.~\ref{fig:coeffs_eachorder_sms}--\ref{fig:coeffs_eachorder_gt} as part of Appendix~\ref{app:additional_potentials}.

Table~\ref{tab:rms_deg_slices_full} in 
Appendix~\ref{app:additional_potentials} shows an expanded version of Table~\ref{tab:rms_deg_slices} for all 16 potentials tested in this paper.
Based on the diagnostic criteria outlined in this section and our intuition about softness gained from examining exemplary hard and soft SMS and SCS potentials, we can exclude from further consideration SMS 400 MeV, SCS 1.1 fm, and SCS 1.2 fm on the grounds that there is no modification of $\Lambdab$ or $\mpieff$ that can rescue them from an irregular convergence pattern.
For the purposes of analysis in the remainder of Sec.~\ref{sec:buqeye_irregular} and in Sec.~\ref{sec:nonstationarity} and~\ref{sec:Lambdab_meff_posteriors}, we will focus on the following potentials:
\begin{itemize}
    \item SMS 450, 500, and 550 MeV;
    \item SCS 0.9 and 1.0 fm; and
    \item EMN 500 MeV.
\end{itemize}
We exclude SCS 0.8 fm from consideration not because it has an irregular convergence pattern, but because it is not commonly used in \emph{ab initio} calculations of nuclear observables.

Per Table~\ref{tab:rms_deg_slices_full} in the Appendix, potentials in the EMN family exhibit somewhat irregular convergence when $c_{2}$ and $c_{5}$ are compared, but we include one potential---namely EMN 500 MeV---as a representative because potentials in this family, too, are commonly used in nuclear physics.
In this case the irregularity in coefficient sizes across orders cannot be easily attributed to a low cutoff, and we will see below that it is not so severe as to violate the BUQEYE GP assumption.

Discussion of the GT+ family of potentials is postponed until Sec.~\ref{subsec:fixed_meff} because those potentials are calculated only to \NNLO.

\subsection{Looking at the curves' length scale: evidence for nonstationarity}
\label{subsec:nonstationarity_evidence}

\begin{figure*}[!h]
    \centering
    \includegraphics{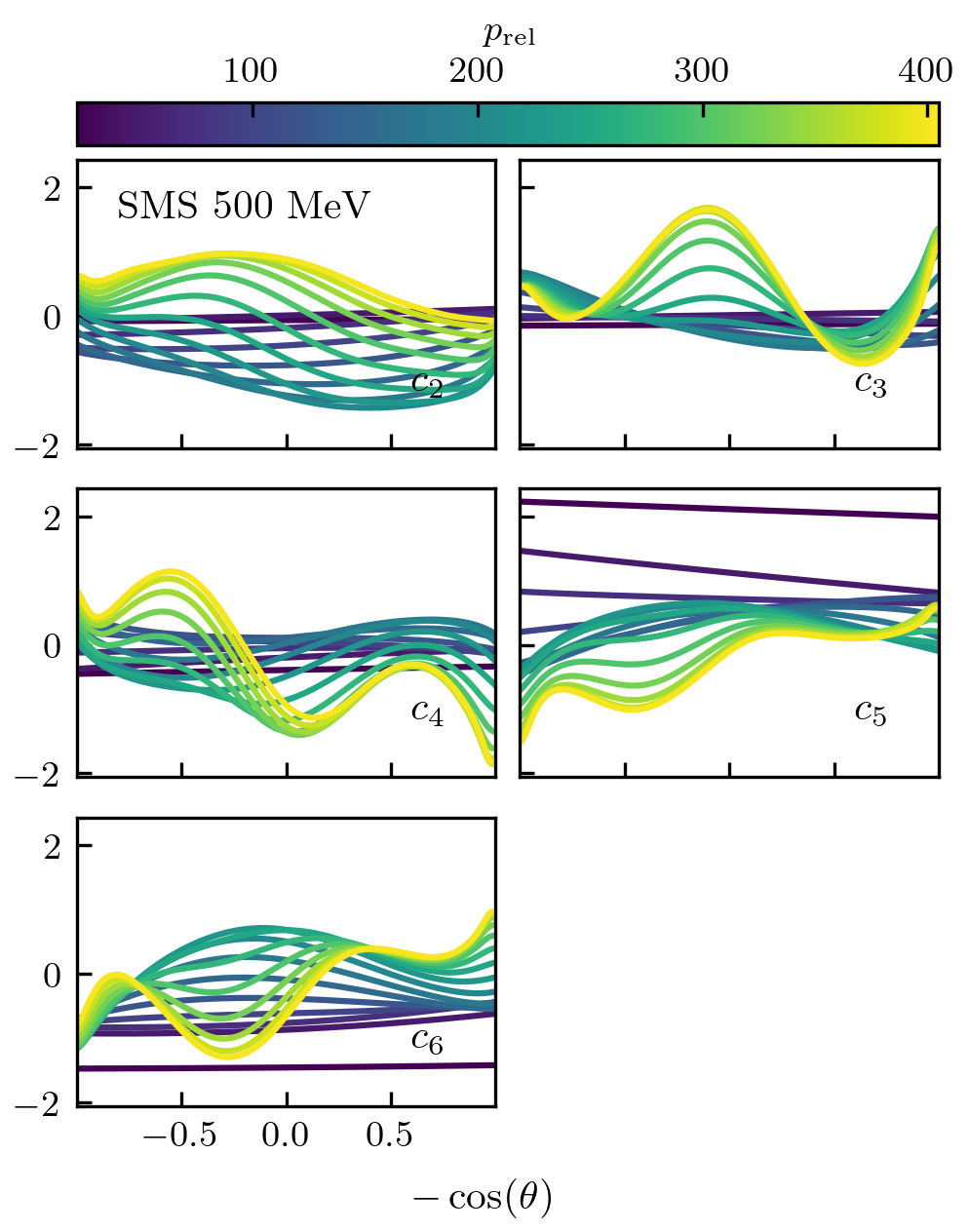}
    \phantomsublabel{-3.0}{0.0}{fig:coeffs_tlab_SMS500MeV_DSG} \includegraphics{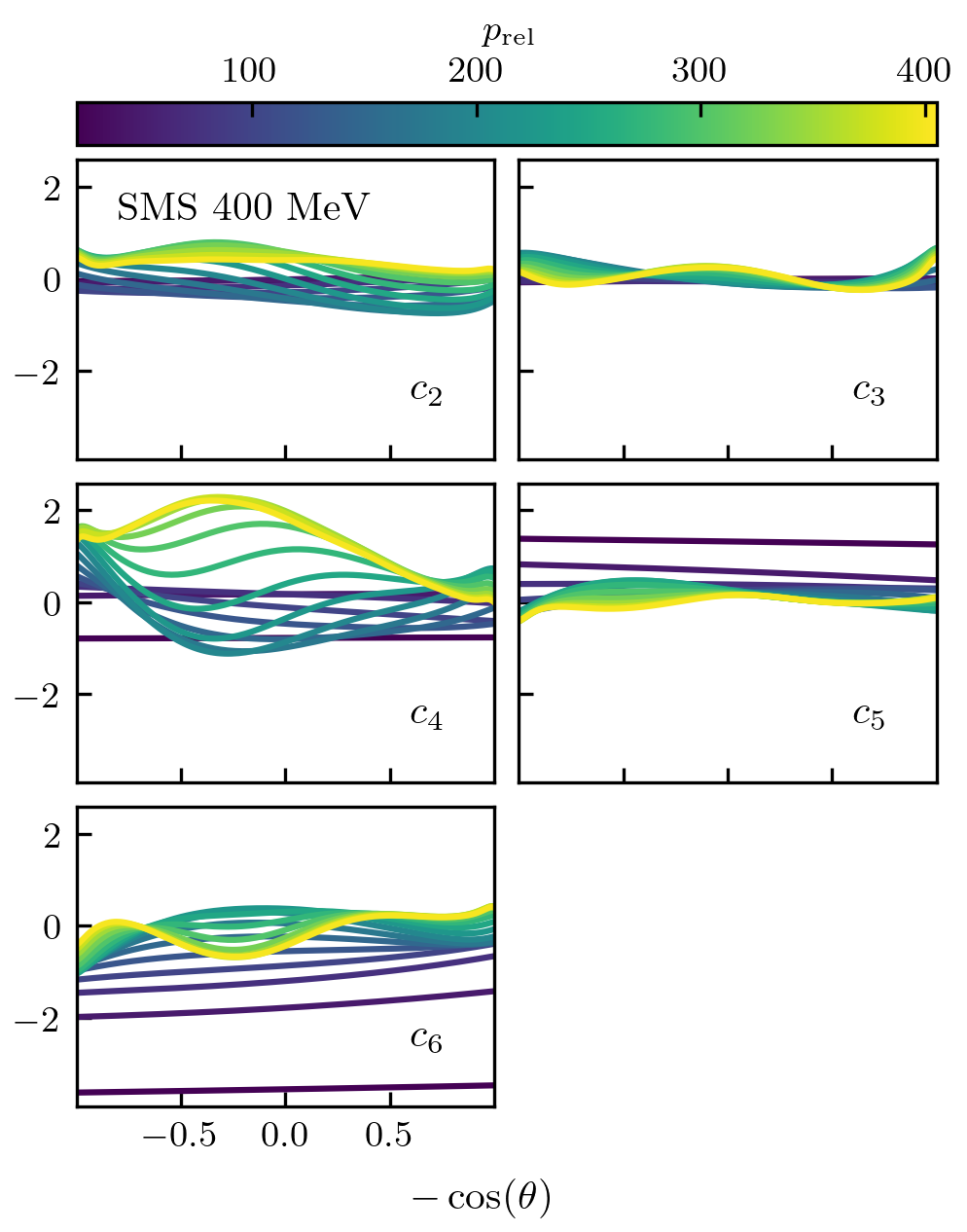}
    \phantomsublabel{-2.9}{0.0}{fig:coeffs_tlab_SMS400MeV_DSG} 

    \includegraphics{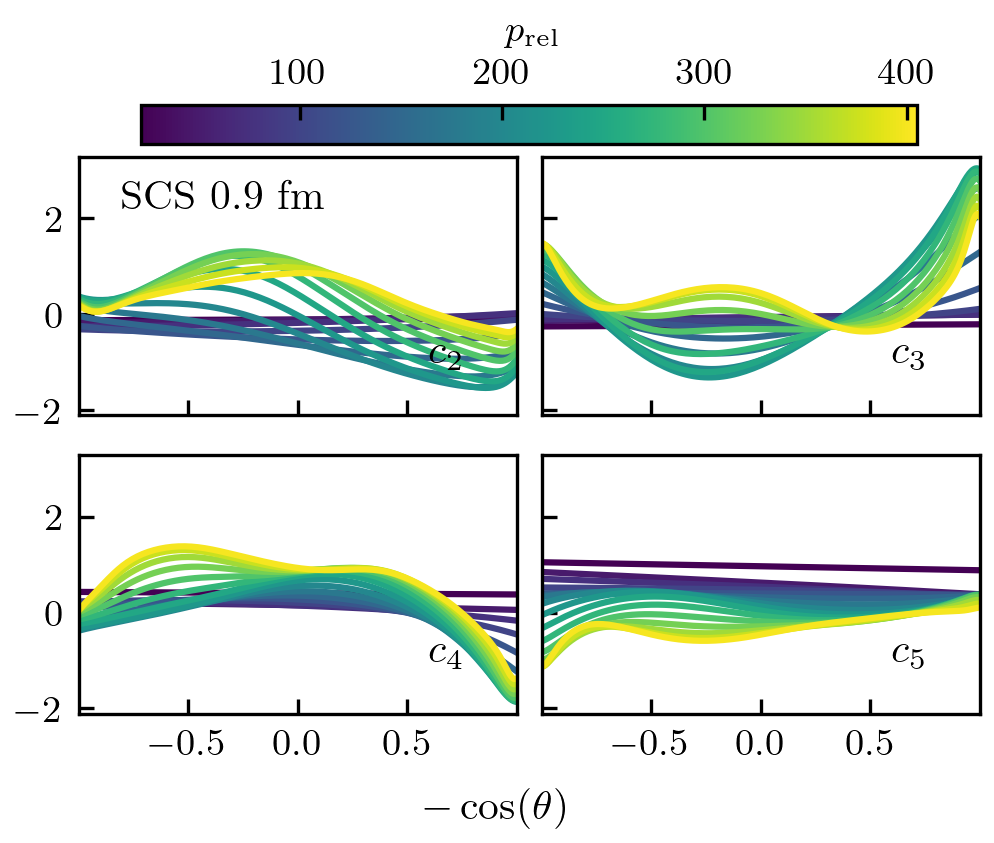}
      \phantomsublabel{-3.0}{0.0}{fig:coeffs_tlab_SCS0p9fm_DSG} \includegraphics{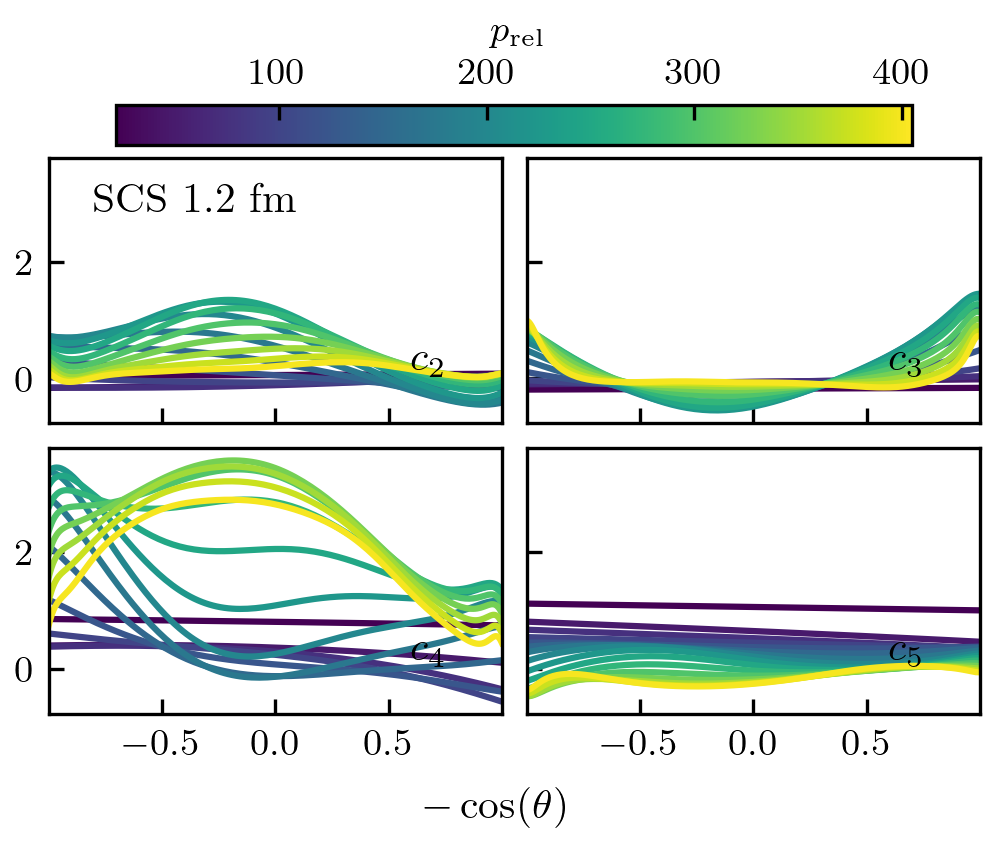}
        \phantomsublabel{-3.0}{0.0}{fig:coeffs_tlab_SCS1p2fm_DSG} 
    \caption{Order-by-order coefficients for $d\sigma / d\Omega$ for 
    \protect\subref{fig:coeffs_tlab_SMS500MeV_DSG} SMS 500 MeV and 
    \protect\subref{fig:coeffs_tlab_SMS400MeV_DSG} 400 MeV and 
    \protect\subref{fig:coeffs_tlab_SCS0p9fm_DSG} SCS 0.9 fm and 
    \protect\subref{fig:coeffs_tlab_SCS1p2fm_DSG} 1.2 fm extracted at fixed $\prel$ with the choice $\left(\Lambdab = 570\,\mathrm{MeV}, \mpieff = 138\,\mathrm{MeV}\right)$ from Ref.~\cite{Millican:2024yuz}.
    Coefficients extracted at low $\prel$ are shown in purple and those at high $\prel$ in yellow.
    The low-$\prel$ coefficients exhibit distinctly long-length-scale behavior (even resembling straight lines, though they are not actually linear), which transitions to a length scale whose value is consistently near 0.4 for $\prel \geq \mpi$.}
    \label{fig:coeffs_eachorder_smsscs_vscos}
\end{figure*}

In all previous work concerning NN scattering,
the BUQEYE model has assumed that the GP to which the coefficient functions would be compared is stationary with respect to the input space---that is, that $\ell$ and $\cbar^{2}$ are constant across the input space $x$ over which they are fit (e.g., energy or scattering angle).%
\footnote{Previous works involving the BUQEYE model (such as Ref.~\cite{Melendez:2020ikd}) have tested nonstationary approaches to Gaussian processes in contexts other than NN scattering.}
There are, of course, typically nonstationarities inherent in 
observables; for instance, the \emph{$np$} total cross section $\sigmatot$ spans orders of magnitude over the \chiEFT\ energy regime as its value decreases with energy.
To factor this behavior of $\sigmatot$ out, when formulating the error model for this observable we set the reference scale $\genobsref$ [see Eq.~\eqref{eq:obs_k_expansion}] to the highest available order of $\sigmatot$ for a given potential.
However, even with careful and informed choices of $\genobsref$ and the parametrization of $Q$, the investigations in our prior paper~\cite{Millican:2024yuz}, which were premised on the assumption of a fully stationary GP kernel, identified marked nonstationarity in the residual coefficient functions.
This section examines whether the assumption of stationarity comports with the available evidence from the potentials under test.

\subsubsection{Length scale}
\label{subsubsec:ls_nonstat}

We begin with the length scale.
Figure~\ref{fig:coeffs_eachorder_smsscs_vscos} shows the coefficients extracted from both hard and soft potentials' calculations of $d\sigma / d\Omega$ at fixed values of $\xE = \prel$ plotted against $\xtheta = \negcos$.
Low-momentum coefficients are shown in purple and high-momentum in bright yellow.
The flat, almost linear shape of the low-momentum curves gives way to a more consistently wiggly shape as momentum rises.

\begin{figure*}[hbt]
    \centering
    \includegraphics[width=0.98\textwidth]{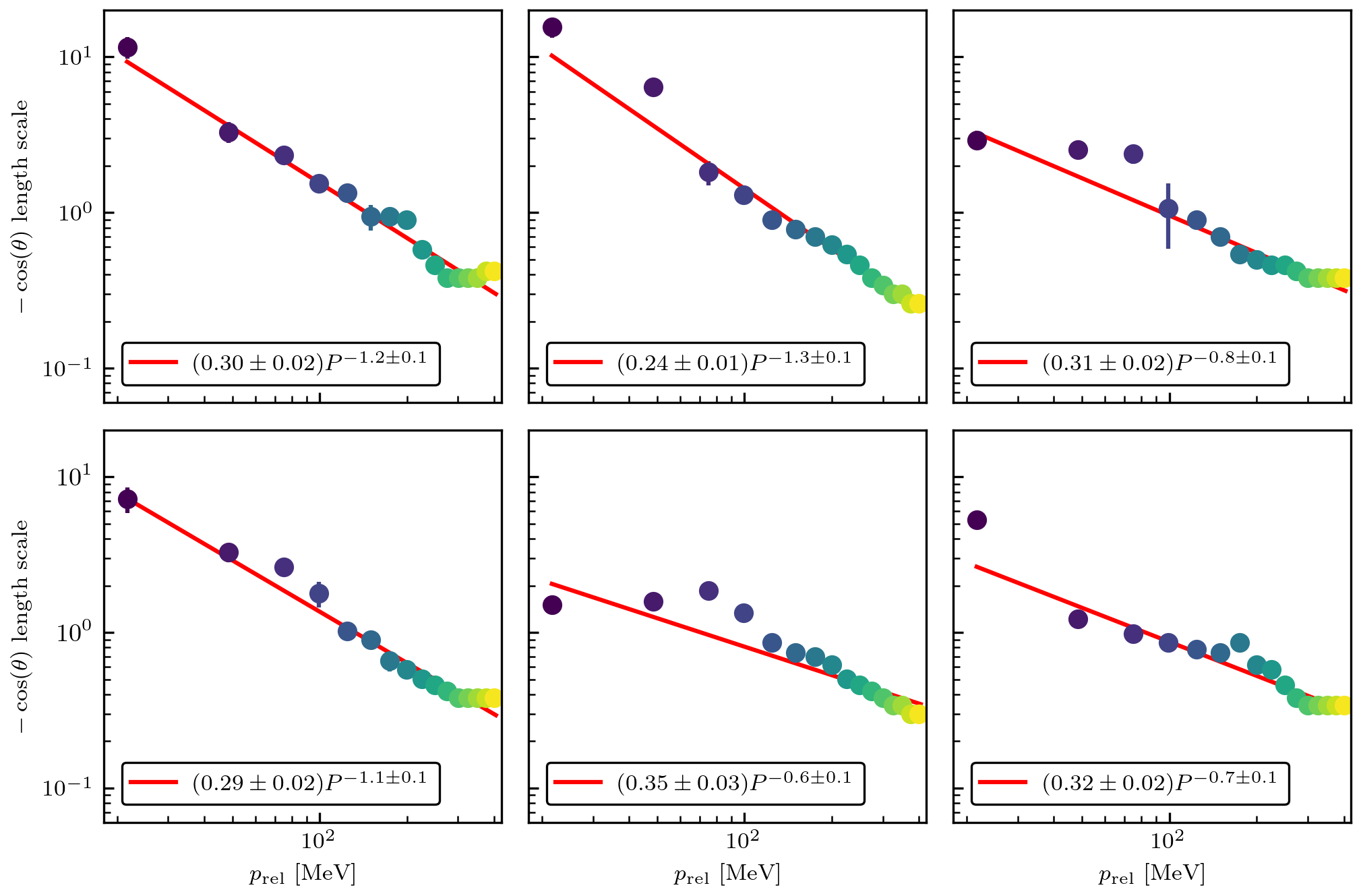}
    \phantomsublabel{-4.75}{3.80}{fig:ls_tlab_slices_eachobs_dsg}
    \phantomsublabel{-2.65}{3.80}{fig:ls_tlab_slices_eachobs_d}
    \phantomsublabel{-0.55}{3.80}{fig:ls_tlab_slices_eachobs_axx}
    \phantomsublabel{-4.75}{1.70}{fig:ls_tlab_slices_eachobs_ayy}
    \phantomsublabel{-2.65}{1.70}{fig:ls_tlab_slices_eachobs_a}
    \phantomsublabel{-0.55}{1.70}{fig:ls_tlab_slices_eachobs_ay}
    \caption{Plots of $\ell_{\theta}$ in $\negcos$ at fixed $\prel$ for 
    \protect\subref{fig:ls_tlab_slices_eachobs_dsg} $d\sigma / d\Omega$, 
    \protect\subref{fig:ls_tlab_slices_eachobs_d} $D$, 
    \protect\subref{fig:ls_tlab_slices_eachobs_axx} $A_{xx}$, 
    \protect\subref{fig:ls_tlab_slices_eachobs_ayy} $A_{yy}$, 
    \protect\subref{fig:ls_tlab_slices_eachobs_a} $A$, and 
    \protect\subref{fig:ls_tlab_slices_eachobs_ay} $A_{y}$ for SMS 500 MeV extracted with the choice $\left(\Lambdab = 570\,\mathrm{MeV}, \mpieff = 138\,\mathrm{MeV}\right)$ from Ref.~\cite{Millican:2024yuz}.
    Here, $P$ is the dimensionless quantity $\prel / 405\,\mathrm{MeV}$.
    Note the consistent power-law behavior across observables, as seen in the red line of best fit to Eq.~\eqref{eq:ls_powerlaw}.}
    \label{fig:ls_tlab_slices_eachobs}
\end{figure*}

For more direct evidence of momentum-dependent nonstationarity in the angular length scale, it behooves us to look more closely observable by observable.
We will do so for the same potential as we considered in Ref.~\cite{Millican:2024yuz}, SMS 500 MeV.
Figure~\ref{fig:ls_tlab_slices_eachobs} shows plots for all six 2D observables---$d\sigma / d\Omega$, $D$, $A_{yy}$, $A_{xx}$, $A$, and $A_{y}$---for the length scale in the $\negcos$ input space, $\ell_{\theta}$ determined independently at fixed $\prel$ in intervals of 25 MeV from 25 to 400 MeV.
$1\sigma$ error bars on $\ell_\theta$ are  included in the figure, although in most cases they are smaller than the point size. 
Here $\Lambdab$ and $\mpieff$ 
take the optimal values from Ref.~\cite{Millican:2024yuz}'s Eq.~(17) (570 and 138 MeV, respectively), $Q$ is parametrized by $\Qsum$, $p$ is parametrized by $\prel$, $\genobsref$ is given by the usual forms in Ref.~\cite{Millican:2024yuz}'s Table~I, and all orders are included. As elsewhere in the paper, we marginalize out the variance $\cbar^2$, so effectively rendering our description of the coefficient curves into a Student-t process rather than a Gaussian process, see Eq.~\eqref{eq:tprocess}.

In each case, the pattern is stark: Very long length scales, longer in every case at lowest momentum than the extent of the $\negcos$ input space itself, give way to shorter length scales at higher momentum. 
We fit this $\prel$ dependence of $\ell_\theta$ to a power law:
\begin{equation}
    \ell_{\theta} = a \times \left(\prel / 450\,\mathrm{MeV}\right)^{-b} \;.
\label{eq:ls_powerlaw}
\end{equation}
The maximum $\prel$ in our input space is $405\,\mathrm{MeV}$; it acts as a reference value to ensure that the bracketed quantity remains dimensionless.
Then $a$ 
corresponds to 
the length scale in $-\cos(\theta)$ at that momentum. The parameters $a$ and $b$ are fit observable by observable, i.e., they are part of the set $\beta_j$ in terms of the presentation of Sec.~\ref{subsec:parameter_estimates}.

The fit values in Fig.~\ref{fig:ls_tlab_slices_eachobs} show some striking trends: 
The values for $a$ are quite stable and consistent across all six observables and the values of $b$ are not far from $1$.
The roughly $1 / \prel$ dependence of $\ell_{\theta}$ can be explained as in Sec.~IV of Ref.~\cite{Millican:2024yuz}; namely, that wiggliness of the dimensionless coefficient functions varies with the number of partial-wave states that are accessible.
Semi-classically, we know that the maximum $l$ probed at a particular momentum $p$ is of order $p R$ where $R$ is the range of the potential.
Since the minimum angular scale decreases roughly as $1/l_{\rm max}$ this semi-classical argument suggests that the GP length scale will drop as $1/p$, just as we see in these results.
This behavior of the GP length scale emerges also for the other potentials we examine in this paper.

We could treat $b$ as a global parameter as in Sec.~\ref{subsec:parameter_estimates}, which would convert this multistep pointwise procedure to a fully Bayesian approach that includes correlation information.
However, the multistep approach yields illuminating intermediate diagnostics, as in Fig~\ref{fig:ls_tlab_slices_eachobs}.
Following the global approach also indicates that a range of values of $b$ around $b=1$ is consistent with the posterior probability distribution function. 
Therefore, given the results in Fig.~\ref{fig:ls_tlab_slices_eachobs}, we take a fixed value of $b$, denoted by $B$, and set $B=1$, in accord with the semi-classical argument.

Now, in order to probe length-scale nonstationarity that depends upon $x_{\theta}$ instead of $x_{E}$, we will switch from the angular length at fixed $\prel$, to the momentum length scale at fixed $\negcos$, plotted as a function of the fixed $\negcos$ and examine its behavior.
The results of this analysis for SMS 500 MeV are shown in Fig.~\ref{fig:ls_deg_slices_eachobs} in Appendix~\ref{app:additional_potentials}.
The momentum length scale for all observables stays in the range 40--80 MeV, and the variation for any one observable is not more than 10\%. There is no gross behavior across all observables that could be approximated by a function.
We also examined length scales aggregated across observables for the other five potentials on which we focus and saw no trend there, either.
Therefore, $\ell_{E}$ is henceforth treated as fully stationary under all circumstances.

\subsubsection{Variance}
\label{subsubsec:var_nonstat}

\begin{figure*}[bht!]
    \centering
    \includegraphics{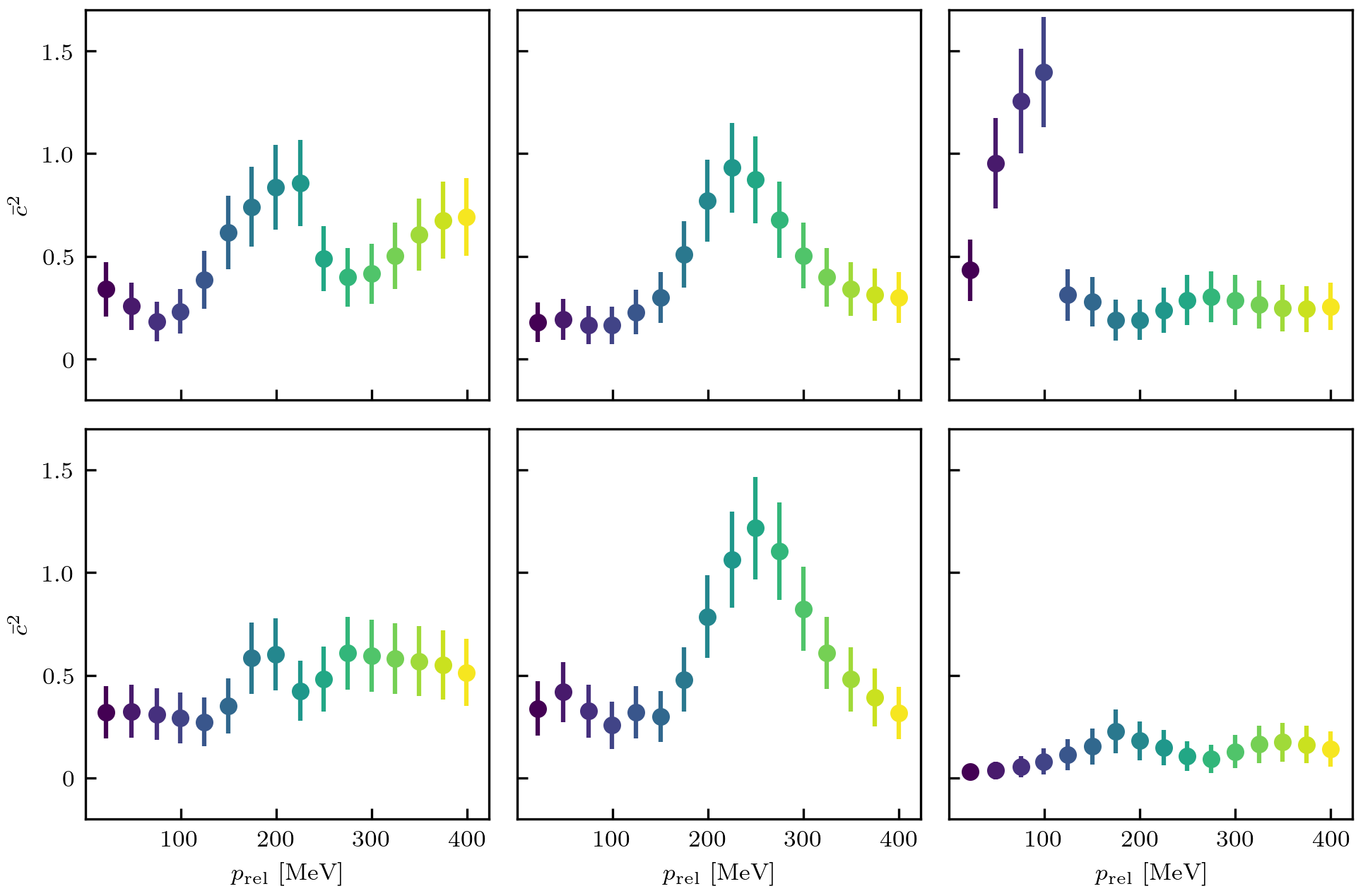}
    \phantomsublabel{-4.65}{3.70}{fig:var_tlab_slices_eachobs_dsg}
    \phantomsublabel{-2.55}{3.70}{fig:var_tlab_slices_eachobs_d}
    \phantomsublabel{-0.45}{3.70}{fig:var_tlab_slices_eachobs_axx}
    \phantomsublabel{-4.65}{1.60}{fig:var_tlab_slices_eachobs_ayy}
    \phantomsublabel{-2.55}{1.60}{fig:var_tlab_slices_eachobs_a}
    \phantomsublabel{-0.45}{1.60}{fig:var_tlab_slices_eachobs_ay}
    \caption{Plots of $\cbar^{2}$ at fixed $\prel$ for \protect\subref{fig:var_tlab_slices_eachobs_dsg} $d\sigma / d\Omega$, 
    \protect\subref{fig:var_tlab_slices_eachobs_d} $D$, 
    \protect\subref{fig:var_tlab_slices_eachobs_axx} $A_{xx}$, 
    \protect\subref{fig:var_tlab_slices_eachobs_ayy} $A_{yy}$, 
    \protect\subref{fig:var_tlab_slices_eachobs_a} $A$, and 
    \protect\subref{fig:var_tlab_slices_eachobs_ay} $A_{y}$ for SMS 500 MeV extracted with the choice $\left(\Lambdab = 570\,\mathrm{MeV}, \mpieff = 138\,\mathrm{MeV}\right)$ from Ref.~\cite{Millican:2024yuz}.}
    \label{fig:var_tlab_slices_eachobs}
\end{figure*}

Now we will move on to consideration of 
the variance $\cbar^{2}$.
Section~IV of Ref.~\cite{Millican:2024yuz} pointed out that $\cbar^2$ 
shows some signs of nonstationarity, and that the spread of the variance is significantly lower when training points below roughly $\prel = 125\,\mathrm{MeV}$ are omitted.

Here, we proceed as for the GP length scales in the previous section.
We examine results for $\cbar^2$ for each observable with coefficients extracted from the predictions of SMS 500 MeV;  $\Lambdab$ and $\mpieff$ were fixed to the same values as in Fig.~\ref{fig:ls_tlab_slices_eachobs}.
The distribution for $\cbar^2$ is a scaled inverse chi-squared, so to carry out this exercise we compute the mean and 68\% interval of that distribution, evaluated for the number of extracted degrees of freedom 
used in the training data at each value of $\prel$.

Results for the variance of the GPs that describe the $\theta$ dependence at equally spaced values of $\prel$ from 25--400 MeV are shown in  Fig.~\ref{fig:var_tlab_slices_eachobs}. 

Although $d\sigma / d\Omega$, $D$, and $A$ show a ``low, high, low'' pattern of behavior from left to right, the other three observables do not, so attempts to parametrize all six observables by the same function (such as a Lorentzian) fail.
Further efforts to treat the parameters of a fitting function as Bayesian random variables and combine the resulting probability distributions across observables for a given potential likewise fail.
A ready functional form approximating $\cbar^{2}$, such as Eq.~\eqref{eq:ls_powerlaw} for $\ell_{\theta}$, is thus not available, and the variability of $\cbar^2$, while significant, is not an order of magnitude as was the case for the length scale.
We therefore take $\cbar^{2}$ to be stationary with respect to  (i.e. independent of) $\xE$.

We did an analogous analysis of the $\cbar^{2}$ for the momentum dependence of the coefficients at different values of $\negcos$ across observables. The results are presented in Fig.~\ref{fig:var_deg_slices_eachobs} in Appendix~\ref{app:additional_potentials}.
Little nonstationarity was observed in the variance in this case, also.

\subsubsection{Conclusion}
\label{subsubsec:conclusion_nonstat}

We showed in Sec.~\ref{subsubsec:ls_nonstat} that the RBF kernel length scale $\ell_{\theta}$ has a $1/\prel$ dependence on $\xE = \prel$, while in Sec.~\ref{subsubsec:var_nonstat} we found no systematic trend of variance nonstationarity.
We can implement a two-dimensional GP for the coefficients $c_n$ that includes this behavior of the GP length scale.
The extraction of probability distributions for GP hyperparameters like $\Lambdab$ and $\mpieff$ will then not be distorted by the mismatch between a stationary GP kernel and nonstationary coefficients.
Two different, but equivalent, ways of doing this are explained in the next section.

\section{Addressing nonstationarity in the BUQEYE model}
\label{sec:nonstationarity}

In this section, we will consider two ways to implement the nonstationarity we uncovered in the $c_n$ data in the previous section: warping, and use of a nonstationary kernel.
We also show that ultimately these two methods, while they appear different, give the same results as long as they are assessed using equivalent training and testing data.

\subsection{Warping}
\label{subsec:warping}

One option for remediating the nonstationarity in the length scale is a technique called \emph{warping}.
This technique involves applying a function to the input and/or output space of the data modeled by a GP.
It has an extensive pedigree in the statistical literature and has been applied fruitfully in 
multiple contexts~\cite{5539975, LALLY2018124, e24030321, KOU2013410, NIPS2003_6b5754d7}.

In practice we have warped the input space in previous work: When we tested the difference between the input spaces in Ref.~\cite{Millican:2024yuz}, we ``deformed,'' so to speak, the input space by means of simple functions and thereby changed our GP's assessment of the data.%
\footnote{The two $x_{E}$ input spaces tested in Ref.~\cite{Millican:2024yuz} were $\Elab$ and $\prel$. Because they are related by the equation $\Elab = 2 \prel^{2} / M_{N}$ (with the nucleon mass $M_{N} \approx 940\,\mathrm{MeV}$), the transformation from one input space to the other is a 1D warping of the input space.
Additionally, diagnostics were identical when the input space $x_{\theta}$ was cast as $\negcos$ or as $\qcm^{2}$ because $\qcm^{2} \sim 1 - \cos(\theta)$, which implies that switching $x_{\theta}$ from $\negcos$ to $\qcm = \sqrt{\qcm^{2}}$ (or vice-versa) is \emph{also} just a power-law warping.}

In this section we will do analogous warping on the full 2D space in order to obtain an input space in which a stationary length scale is a reasonable description of the data. 
This would take the form of shrinking $x_{\theta}$ at small values of $x_{E}$ and leaving it  undeformed at larger values of $x_{E}$.
In the 2D space in which the six observables under test live, warping takes the form
\begin{equation}
    c_{n}\left(\xE, \xtheta\right) \longrightarrow c_{n}\left(w_{E}\left(\xE, \xtheta\right), w_{\theta}\left(\xE, \xtheta\right)\right) \;,
    \label{eq:warping_transform}
\end{equation}
where $c_{n}$ is a dimensionless coefficient, $\xE = \prel$, $\xtheta = \negcos$, $w_{E}\left(\xE, \xtheta\right) = \xE = \prel$ and 
\begin{equation}
    w_{\theta}\left(\xE, \xtheta\right) = \xtheta \left(\prel / 405\; \mathrm{MeV} \right)^{B} \;.
    \label{eq:warping_fn}
\end{equation}
We choose this functional form for the warping function because Eq.~\eqref{eq:ls_powerlaw} so heavily suggests it; in general, however, a warping function need not operate as a multiplicative scaling of one or more dimensions of the input space in terms of the other and may take an arbitrarily complicated functional form.
Here, $B=1$ is consistent with the distribution for $b$ in Eq.~\eqref{eq:ls_powerlaw}, and the factor of $405\; \mathrm{MeV}$ keeps $\xtheta$'s range (namely, $\left[-1, 1\right]$) equal between the unwarped and warped plots at high momentum.

\begin{figure*}[p]
    \centering
    \includegraphics[scale = 0.97]{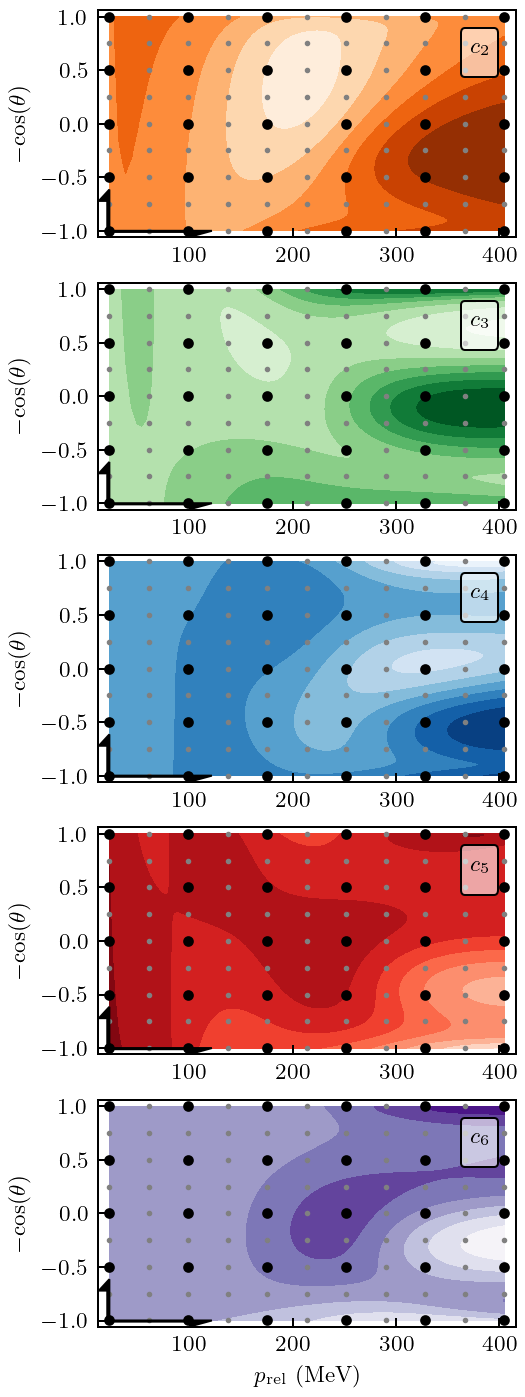}
    \includegraphics[scale = 0.97]{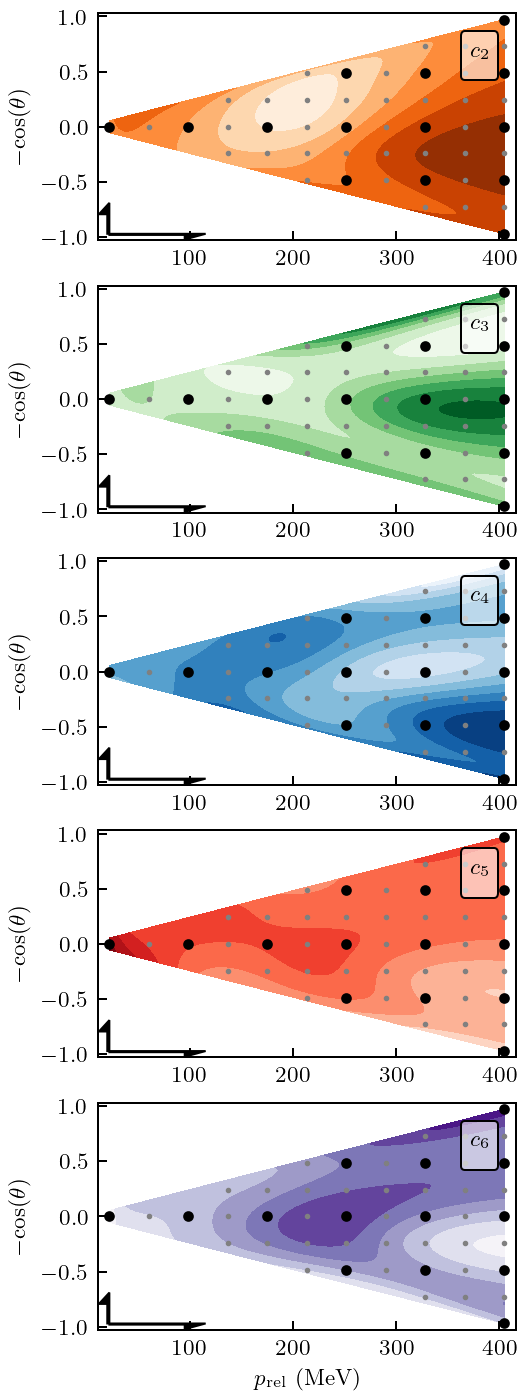}
    \caption{Unwarped (left) and warped (right) coefficients for $d\sigma / d\Omega$ for the SMS 500 MeV potential.
    On the right, the space is warped according to Eq.~\eqref{eq:warping_fn}.}
    \label{fig:warped_coeffs}
\end{figure*}

The result of this warping is shown in Fig.~\ref{fig:warped_coeffs}; here, the input space on the right has been shrunk at low momentum by the warping, which means the length scale in the warped space at low momentum has also shrunk and no longer changes as dramatically from low to high momentum.
The 2D coefficients for different orders from SMS 500 MeV are presented with no warping (left panel) and with warping (right panel).

\subsection{Non-stationary kernel implementation}
\label{subsec:nsk}

The second approach is to create a nonstationary kernel (NSK).
This was accomplished by using \texttt{scikit-learn}'s \texttt{Kernel} class (and associated functionality) as a template~\cite{scikit-learn}.
In their stationary incarnations, these classes expect an array of scalar values for the length scale and variance, but they have been modified so that they expect functions [such as Eq.~\eqref{eq:ls_powerlaw}] that depend on new random variables, which are the parameters that describe the non-stationarity in the input space.

The form of the most general NSK in the context of our treatment of observables in $\left(\xE, \xtheta\right)$ is a modification of Eqs.~\eqref{eq:gp_generic}--\eqref{eq:rbf_2d} 
\begin{align}
    \kappa\left(x, x^{\prime}\right) &\sim \cbar^{2}\left(\xE, \xtheta\right) 
    \notag \\
    & \ \times \null
    \exp{-\frac{(\xE - \xE^{\prime})^2}{2\ell_{E}^2 \left(\xE, \xtheta\right)} - \frac{(\xtheta - \xtheta^{\prime})^2}{2\ell_{\theta}^2 \left(\xE, \xtheta\right)}} \;.
    \label{eq:gp_specific_nsk}
\end{align}
This form encodes nonstationarity in all three of $\ell_{\theta}$, $\ell_{E}$, and $\cbar^{2}$, whereas our findings in Sec.~\ref{subsec:nonstationarity_evidence} lead us to implement nonstationarity in $\ell_{\theta}$ alone.
Thus, our NSK approach will evaluate posteriors using a kernel of the form
\begin{equation}
    \kappa\left(x, x^{\prime}\right) \sim \cbar^{2} \exp{-\frac{(\xE - \xE^{\prime})^2}{2\ell_{E}^2} - \frac{(\xtheta - \xtheta^{\prime})^2}{2\ell_{\theta}^2 \left(\xE\right)}} \;,
    \label{eq:gp_nsk_final}
\end{equation}
where $\ell_{\theta}\left(\xE\right)$ takes the form of Eq.~\eqref{eq:ls_powerlaw}.

Note that for each new Bayesian random variable that factors into the NSK and is marginalized over, it is necessary to add another dimension to the random-variable mesh.
The mesh's size scales as the power of the number of random variables, which can impose significant computational costs in the NSK compared to the warping approach. 

\begin{figure*}[tbh!]
\centering
    \includegraphics[width=0.4\textwidth]{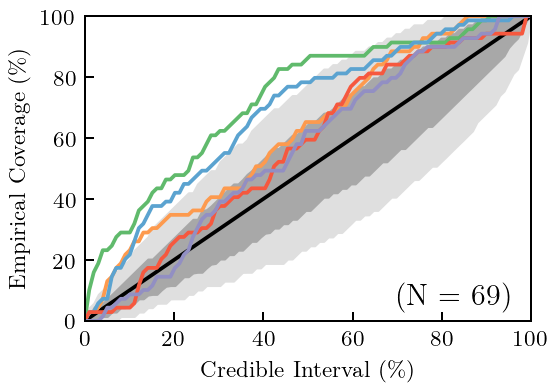}
    \includegraphics[width=0.14\textwidth]{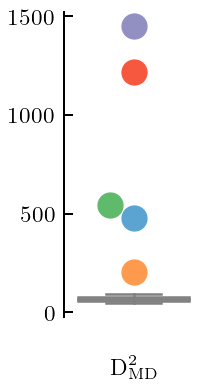}
    \includegraphics[width=0.4\textwidth]{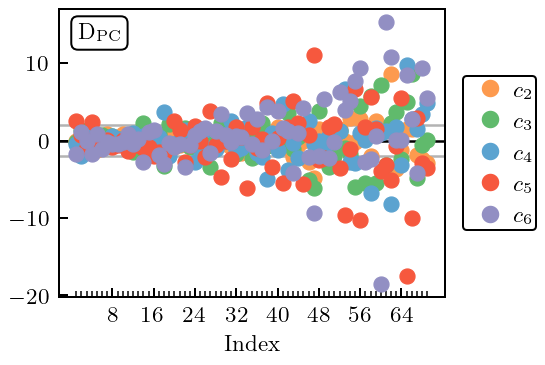}
    
    \caption{$\DCI$, $\DMD^{2}$, and $\DVAR{PC}$ diagnostics for the differential cross section $d\sigma / d\Omega$ from SMS 500 MeV.
    They are generated with ($\xE$, $\xtheta$) = ($\prel$, $\negcos$) and 
    $Q = \Qsum(p = \prel, \mpieff = 138\, \mathrm{MeV}, \Lambdab = 600\, \mathrm{MeV})$ and calculated with 30 training points and 69 testing points, no warping or downsampling, and a fully stationary RBF kernel.
}
    \label{fig:dsg_sms500_unwarped_default}
\end{figure*}

\begin{figure*}[tbh!]
\centering
    \includegraphics[width=0.4\textwidth]{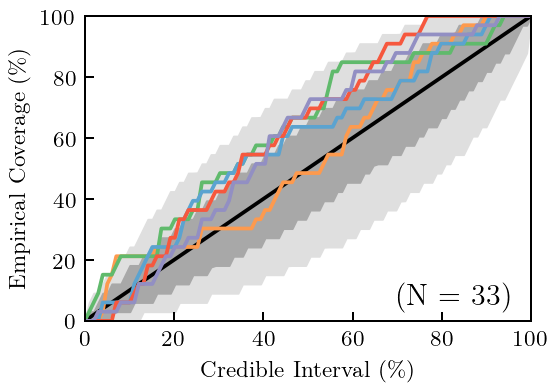}
    \includegraphics[width=0.14\textwidth]{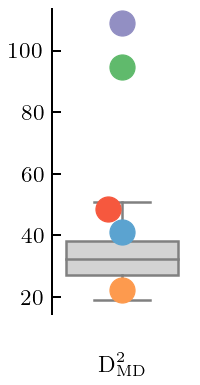}
    \includegraphics[width=0.4\textwidth]{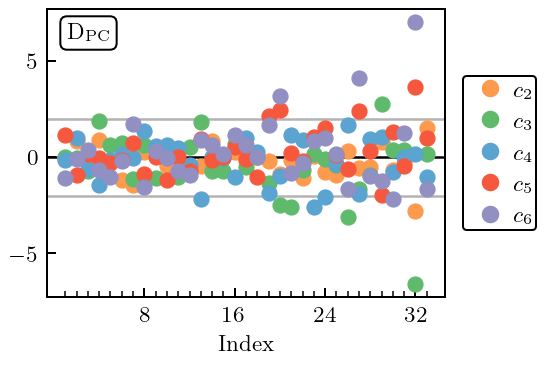}
    \caption{$\DCI$, $\DMD^{2}$, and $\DVAR{PC}$ diagnostics for the differential cross section $d\sigma / d\Omega$ from SMS 500 MeV.
    They are generated with ($\xE$, $\xtheta$) = ($\prel$, $\negcos$) and $Q = \Qsum(p = \prel, \mpieff = 138\, \mathrm{MeV}, \Lambdab = 600\, \mathrm{MeV})$ and calculated with 14 training points and 33 testing points, warping and downsampling, and a fully stationary RBF kernel.
    }
    \label{fig:dsg_sms500_warped_default}
\end{figure*}

\begin{figure*}[tbh!]
\centering
    \includegraphics[width=0.4\textwidth]{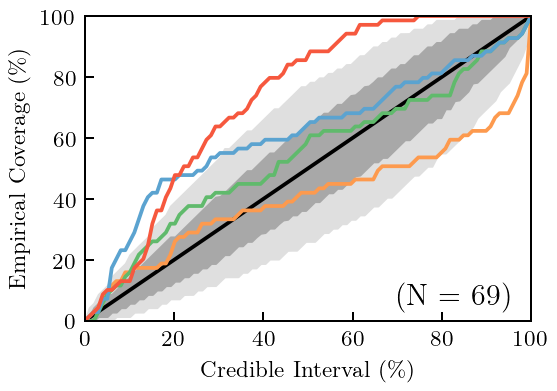}
    \includegraphics[width=0.14\textwidth]{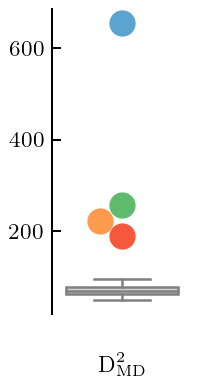}
    \includegraphics[width=0.4\textwidth]{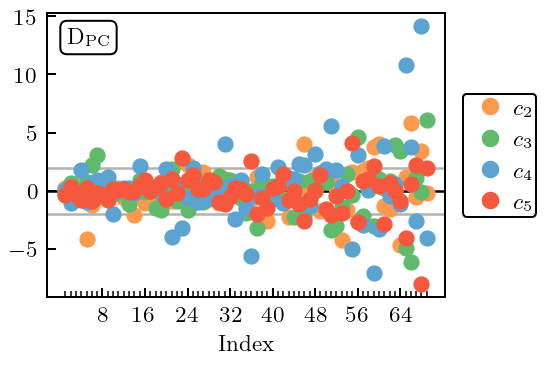}
    
    \caption{$\DCI$, $\DMD^{2}$, and $\DVAR{PC}$ diagnostics for the differential cross section $d\sigma / d\Omega$ from EMN 500 MeV.
    They are generated with ($\xE$, $\xtheta$) = ($\prel$, $\negcos$) and 
    $Q = \Qsum(p = \prel, \mpieff = 138\, \mathrm{MeV}, \Lambdab = 600\, \mathrm{MeV})$ and calculated with 30 training points and 69 testing points, no warping or downsampling, and a fully stationary RBF kernel.
}
    \label{fig:dsg_emn500_base}
\end{figure*}

\begin{figure*}[tbh!]
\centering
    \includegraphics[width=0.4\textwidth]{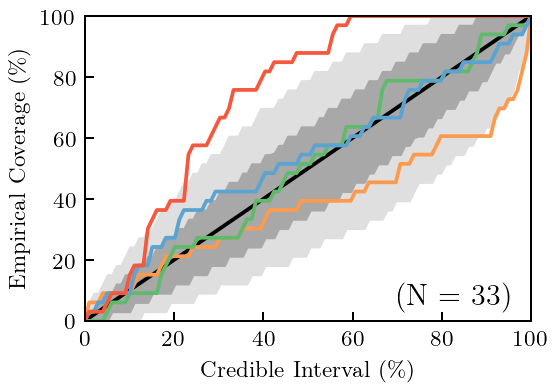}
    \includegraphics[width=0.14\textwidth]{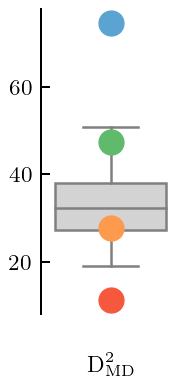}
    \includegraphics[width=0.4\textwidth]{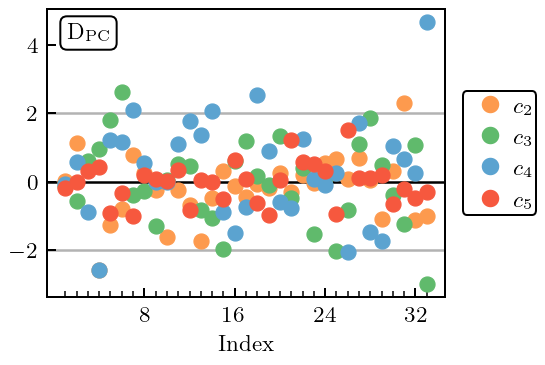}
    \caption{$\DCI$, $\DMD^{2}$, and $\DVAR{PC}$ diagnostics for the differential cross section $d\sigma / d\Omega$ from EMN 500 MeV.
    They are generated with ($\xE$, $\xtheta$) = ($\prel$, $\negcos$) and $Q = \Qsum(p = \prel, \mpieff = 138\, \mathrm{MeV}, \Lambdab = 600\, \mathrm{MeV})$ and calculated with 14 training points and 33 testing points, warping and downsampling, and a fully stationary RBF kernel.
    }
    \label{fig:dsg_emn500_warped}
\end{figure*}

\subsection{Downsampling and the equivalence of the non-stationary kernel and warping approaches}
\label{subsec:downsampling_equivalence}

The equivalence between warping and a non-stationary kernel is clear when we consider Eq.~\eqref{eq:rbf_kernel}, the equation for the stationary SERBF kernel.
In this context, it becomes manifest that warping is a manipulation of the numerator of that exponent (since it warps the input space but treats the length scale as stationary), while the NSK approach manipulates the denominator of that fraction by casting $\ell$  as a function that depends on $x$ and $x^{\prime}$.
Thus, the two methods can be seen as equivalent---as long as they are provided with the same (or more accurately, equivalent given the mapping of the space) choices of training and testing data.

The choice of train-test split, then, is a more important decision than the specific method of remediating nonstationarity. 
If one starts with a given train-test split in the warped input space, applies the inverse transformation to map the points at which that data resides to the original input space, and then uses those training and testing points in the analysis with the NSK,  the two approaches produce the same result. 
Our practice heretofore has been to array training and testing points at regular intervals in one dimension or in a grid pattern in two dimensions.
But a grid that is uniform in the warped input space will not be uniform once mapped to the original space and employed for the NSK: It will have fewer points at lower momentum and be ``downsampled'' in that regime.
We have good statistical reasons to do this: Because $\ell_{\theta}$ is longer for smaller $\prel$, each training point in the low-momentum region contributes less information, and we should be able to make do with fewer of them. 

What this looks like in practice can be seen in Fig.~\ref{fig:warped_coeffs}, where the unwarped coefficients and the grid of training and testing points at left contrast with the warped coefficients and corresponding training and testing points, which are roughly triangular, at right; here, the coefficients were warped and those training and testing points outside the bounds of the new coefficients were dropped, which achieved the desired low-momentum downsampling.

Our previous conclusion that the warping and NSK approaches are equivalent remains true, but, with the need for low-momentum downsampling in mind, we can say that warping offers a very geometrically intuitive way of guiding the process of deciding where points should be placed, as Fig.~\ref{fig:warped_coeffs} demonstrates.
However, 
since the warping method only transforms the input space in angle and therefore only remediates nonstationarity in $\ell_\theta$, 
there may be instances where the NSK is more fully able to capture sources of nonstationarity outside of $\ell_\theta$.

\subsection{Results after remediation of non-stationarity}
\label{subsec:nonstat_remediation}

We tested whether warping and downsampling produce (approximate) stationarity of the two-dimensional GP in the warped input space. 
After warping and accompanying downsampling with $B=1$, we re-fitted the momentum-dependence of the angular length scale using Eq.~(\ref{eq:ls_powerlaw}), thus effectively turning the power of $p$ with which the space is warped into one of the set of global parameters, $\beta_g$.
Now, when $b$ is extracted across all observables for a given potential, the resulting posterior probability distribution function (pdf) is consistent with $b=0$.
In fact $b=0$ is within the 68\% interval for any warping/down-sampling exponent $B \gtrsim 0.5$.

This improvement in the covariance structure of the GP after warping and down-sampling should also be visible in the three statistical diagnostics that characterize how well a GP fit to a set of training data describes a (different) set of testing data. 

We first recapitulate what these diagnostics are; more detail on each of them is given in Sec.~III of Ref.~\cite{Millican:2024yuz}.
The first is the credible interval or ``weather plot'' ($\DCI$).
This compares the credibility intervals predicted by the GP uncertainty model at order N$^k$LO with the empirical coverage that those intervals have for the N$^{k+1}$LO \chiEFT\ result.
This diagnostic thus allows us to visualize the quality of the \chiEFT\ error bars across orders.
The second and third diagnostics assess a slightly different aspect of the statistical model of \chiEFT\ convergence.
They test the extent to which the EFT expansion coefficients $c_n$ are described by the GP chosen as the statistical model.
The Mahalanobis distance squared ($\DMD^{2}$) is the analog of the $\chi^2$ in the (correlated) GP context, and so should not fall very far away from the number of degrees of freedom.
The pivoted Cholesky decomposition ($\DVAR{PC}$) breaks down the contributions to the Mahalanobis distance across the input space, and so can help diagnose why $\DMD^{2}$ is too large or too small. 

What happens to our statistical diagnostics for $d\sigma / d\Omega$ from the SMS 500 MeV potential when we warp and downsample is seen in the contrast between Fig.~\ref{fig:dsg_sms500_unwarped_default}, where no remediation of nonstationarity takes place, and Fig.~\ref{fig:dsg_sms500_warped_default}, where the warping approach is paired with low-momentum downsampling.
The order-by-order empirical coverage in the $\DCI$ plot in Fig.~\ref{fig:dsg_sms500_unwarped_default} is systematically higher than the corresponding credible interval percentage; this is remedied in Fig.~\ref{fig:dsg_sms500_warped_default}, where all curves lie within the one-sigma (dark gray) or two-sigma (light gray) bands~\cite{Millican:2024yuz}.
The $\DMD^{2}$ and $\DVAR{PC}$ diagnostics are also greatly improved, although not perfect.
A similar story is told in Figs.~\ref{fig:dsg_emn500_base} and \ref{fig:dsg_emn500_warped} for EMN 500 MeV, although the initial $\DCI$ plot is already not bad. 
Overall, the diagnostics after warping and downsampling demonstrate strong consistency between our GP model and the $c_n$ data for these two potentials---as long as warping and downsampling is applied.

\section{Learning about  \texorpdfstring{$\Lambdab$}{Lambdab} and \texorpdfstring{$\mpieff$}{mpieff}}
\label{sec:Lambdab_meff_posteriors}

\begin{figure*}[tbh!]
\centering
    \includegraphics[width=0.4\textwidth]{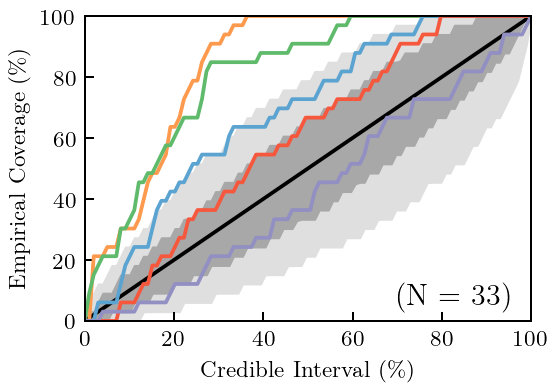}
    \includegraphics[height=0.27\textwidth]{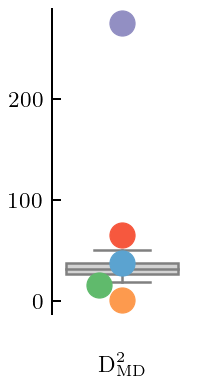}
    \includegraphics[width=0.4\textwidth]{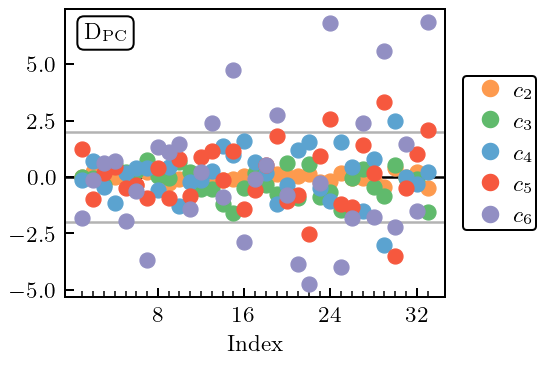}
    \caption{Diagnostics for the differential cross section $d\sigma / d\Omega$ from SMS 500 MeV.
    They are generated with ($\xE$, $\xtheta$) = ($\prel$, $\negcos$) and $Q = \Qsum(p = \prel, \mpieff = 138\, \mathrm{MeV}, \Lambdab = 1\, \mathrm{GeV})$ and calculated with 14 training points and 33 testing points, warping and downsampling, and a fully stationary RBF kernel.
    }
    \label{fig:dsg_sms500_warped_bad}
\end{figure*}

In the previous section we demonstrated by the comparison of statistical diagnostics that our efforts to remediate nonstationarity in the GP length scale have borne fruit.
In this section, we analyze what we can learn about the optimal hyperparameters $\mpieff$ and $\Lambdab$ and their impact on the diagnostics after remediation.
Before we attempt to
extract distributions for $\Lambdab$ at fixed $\mpieff$ in Sec.~\ref{subsec:fixed_meff}, we first demonstrate that there actually {\it is} sensitivity to $\Lambdab$ revealed by the diagnostics. 
If we take $\mpieff = 138\, \mathrm{MeV}$ and $\Lambdab = 1\, \mathrm{GeV}$, we obtain the diagnostics for the differential cross section shown in Fig.~\ref{fig:dsg_sms500_warped_bad}.
The statistical diagnostics here compare very unfavorably to those in Fig.~\ref{fig:dsg_sms500_warped_default}, and in particular the $\DCI$ in Fig.~\ref{fig:dsg_sms500_warped_bad} shows a clear order-by-order trend; specifically, the sysetmatic separation of the curves from \NLO in the upper left to \NNNNLOp in the lower right, indicates that $\Lambdab$ has not been well-chosen for this value of $\mpieff$. Ultimately, the posterior pdf of $\Lambdab$ (and $\mpieff$) peaks at the value of the breakdown scale that makes the coefficients as similar as possible across different orders. The diagnostics will then show a good match between the curves $c_n(x)$ and our GP model for them. 
Having given this preliminary example to stress the importance of properly choosing values of $\Lambdab$ and $\mpieff$,
we now extract posteriors for $\Lambdab$ at fixed $\mpieff$ for several different families of potentials. We do this using different maximum orders $k$ within a particular family, and then 
discuss the extent to which the $\Lambdab$ posteriors within a given potential are consistent for different values of $k$.

\subsection{Fixed \texorpdfstring{$\mpieff$}{mpieff}}
\label{subsec:fixed_meff}

For each potential, evaluations are performed up to \NNLO, \dots, N$^{K}$LO (where $K$ is the highest order for that potential).
The reason that \NNLO is chosen as the lowest maximum order is that it generates two coefficients ($c_{2}$ and $c_{3}$), and this is the minimum necessary for $\Lambdab$ inference in which the $\Lambdab$ and $\cbar^2$ priors do not play a significant role. 

The space is warped, with the value of $B$ in Eq.~\eqref{eq:warping_fn} held fixed at $1$ for each scheme/scale combination.
The new configuration of the warped space is used to downsample low-momentum testing and training points.
We treat $\Lambdab$, the length scale in the warped space $\ell_\theta$, and $\ell_{E}$, as random variables on meshes. (Note that the length scales in the warped space are taken to be fully stationary.)
Then, we extract the posterior $\mathrm{pr}\left(\Lambdab, \{\ell_{E}\}_\bigJ, \{\ell_{\theta}\}_\bigJ \given \{\genobsvecset\}_\bigJ,\mpieff\right)$ with 
$\mpieff$  fixed first at $138\,\mathrm{MeV}$ and then at $200\,\mathrm{MeV}$. This posterior is then marginalized over the set of length scales $\{\ell_{E}\}_\bigJ, \{\ell_{\theta}\}_\bigJ$. This is the specific implementation of the general procedure for extracting global parameters from the order-by-order results for observables that is laid out in Sec.~\ref{subsec:parameter_estimates}.

\begin{figure*}[tbh!]
\centering
    \includegraphics[width=0.4\textwidth]{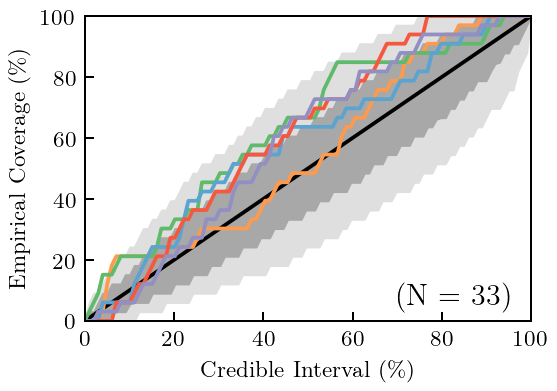}
    \includegraphics[height=0.27\textwidth]{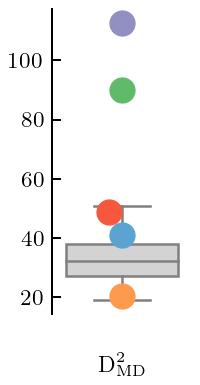}
    \includegraphics[width=0.4\textwidth]{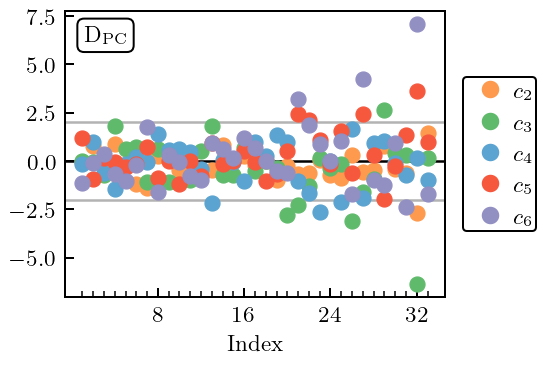}
    \caption{Diagnostics for the differential cross section $d\sigma / d\Omega$ from SMS 500 MeV.
    They are generated with ($\xE$, $\xtheta$) = ($\prel$, $\negcos$) and $Q = \Qsum(p = \prel, \mpieff = 138\, \mathrm{MeV}, \Lambdab = 610\, \mathrm{MeV})$ (the optimal value of $\Lambdab$ from the second column in Table~\ref{tab:nonstat_6potentials}) and  calculated with 14 training points and 33 testing points, warping and downsampling, and a fully stationary RBF kernel.
    }
    \label{fig:dsg_sms500_warped_mpi138_Lambdabopt}
\end{figure*}

\begin{figure*}[tbh!]
\centering
    \includegraphics[width=0.4\textwidth]{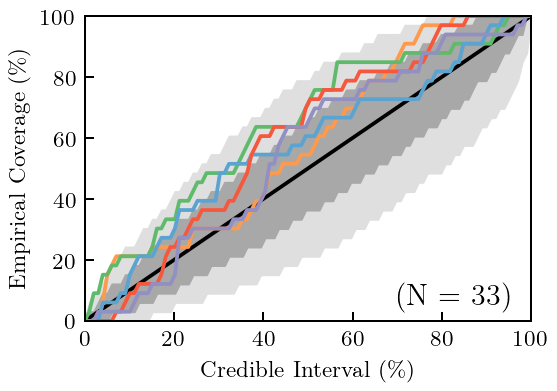}
    \includegraphics[height=0.27\textwidth]{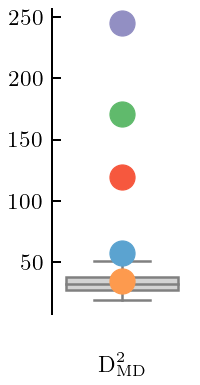}
    \includegraphics[width=0.4\textwidth]{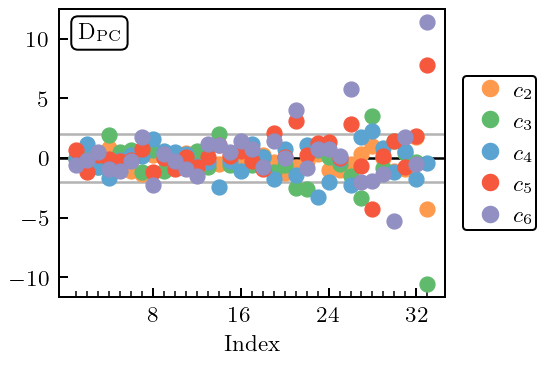}
    \caption{Diagnostics for the differential cross section $d\sigma / d\Omega$ from SMS 500 MeV.
    They are generated with ($\xE$, $\xtheta$) = ($\prel$, $\negcos$) and $Q = \Qsum(p = \prel, \mpieff = 200\, \mathrm{MeV}, \Lambdab = 710\, \mathrm{MeV})$ (the optimal value of $\Lambdab$ from the third column in Table~\ref{tab:nonstat_6potentials}) and  calculated with 14 training points and 33 testing points, warping and downsampling, and a fully stationary RBF kernel.
    }
    \label{fig:dsg_sms500_warped_mpi200_Lambdabopt}
\end{figure*}

\begin{table*}[tbh!]
\renewcommand{\arraystretch}{1.2}
\centering
\setlength{\tabcolsep}{3pt}
\begin{ruledtabular}
\begin{tabular}{ccccc}
    & \multicolumn{1}{c}{$\mpieff = 138\,\mathrm{MeV}$}
    & \multicolumn{1}{c}{$\mpieff = 200\,\mathrm{MeV}$} 
    & \multicolumn{2}{c}{Variable $\mpieff$} \\
    Potential
    & $\Lambdab$
    & $\Lambdab$
    & $\Lambdab$
    & $\mpieff$
    \\
    \hline
    \textbf{SMS 550 MeV} & & & & \\
    \NNLO ($c_{3}$) & $430 \pm 30$ & $450 \pm 30$ & $400 \pm 20$ & $63 \pm 8$ \\
    \NNNLO ($c_{4}$) & $460 \pm 20$ & $480 \pm 20$ & $420 \pm 20$ & $61 \pm 5$ \\
    \NNNNLO ($c_{5}$) & $550 \pm 20$ & $620 \pm 20$ & $550 \pm 20$ & $139\pm 6$ \\
    \NNNNLOp ($c_{6}$) & $630 \pm 20$ & $730 \pm 20$ & $670 \pm 20$ & $155 \pm 5$ \\
    \textbf{SMS 500 MeV} & & & & \\
    \NNLO ($c_{3}$) & $490 \pm 30$ & $520 \pm 40$ & $450 \pm 30$ & $59 \pm 8$ \\
    \NNNLO ($c_{4}$) & $580 \pm 30$ & $610 \pm 30$ & $530 \pm 20$ & $85 \pm 6$ \\
    \NNNNLO ($c_{5}$) & $610 \pm 20$ & $710 \pm 20$ & $640 \pm 20$ & $154 \pm 7$ \\
    \NNNNLOp ($c_{6}$) & $610 \pm 10$ & $710 \pm 20$ & $670 \pm 20$ & $168 \pm 6$ \\
    \textbf{SMS 450 MeV} & & & & \\
    \NNLO ($c_{3}$) & $580 \pm 40$ & $620 \pm 40$ & $530 \pm 30$ & $61 \pm 9$ \\
    \NNNLO ($c_{4}$) & $550 \pm 20$ & $600 \pm 30$ & $550 \pm 30$ & $140 \pm 10$ \\
    \NNNNLO ($c_{5}$) & $620 \pm 20$ & $730 \pm 20$ & $700 \pm 30$ & $180 \pm 10$ \\
    \NNNNLOp ($c_{6}$) & $580 \pm 10$ & $690 \pm 10$ & $680 \pm 20$ & $193 \pm 8$ \\
    \hline
    \textbf{SCS 0.9 fm} & & & & \\
    \NNLO ($c_{3}$) & $520 \pm 40$ & $550 \pm 40$ & $490 \pm 30$ & $70 \pm 10$ \\
    \NNNLO ($c_{4}$) & $730 \pm 40$ & $790 \pm 40$ & $670 \pm 30$ & $94 \pm 7$ \\
    \NNNNLO ($c_{5}$) & $840 \pm 30$ & $990 \pm 30$ & $880 \pm 30$ & $151 \pm 7$ \\
    \textbf{SCS 1.0 fm} & & & & \\
    \NNLO ($c_{3}$) & $620 \pm 40$ & $650 \pm 50$ & $590 \pm 40$ & $80 \pm 10$ \\
    \NNNLO ($c_{4}$) & $590 \pm 30$ & $640 \pm 30$ & $570 \pm 20$ & $111 \pm 9$ \\
    \NNNNLO ($c_{5}$) & $760 \pm 20$ & $880 \pm 30$ & $790 \pm 30$ & $149 \pm 7$ \\
    \hline
    \textbf{EMN 500 MeV} & & & & \\
    \NNLO ($c_{3}$) & $530 \pm 40$ & $560 \pm 40$ & $480 \pm 30$ & $44 \pm 5$ \\
    \NNNLO ($c_{4}$) & $590 \pm 30$ & $630 \pm 30$ & $500 \pm 20$ & $55 \pm 4$ \\
    \NNNNLO ($c_{5}$) & $810 \pm 30$ & $890 \pm 30$ & $700 \pm 20$ & $84 \pm 4$ \\
\end{tabular}
\end{ruledtabular}
\caption{%
Posterior-extracted values (with errors) for $\Lambdab$ and $\mpieff$ for the six potentials on which we focus in this paper.
All values here are extracted under the assumption of a stationary squared-exponential kernel in a space warped according to Eq.~\eqref{eq:warping_transform} with accompanying downsampling.
In the second (third) column, $\mpieff$ is fixed at $138\,\mathrm{MeV}$ ($200\,\mathrm{MeV}$).
In the fourth and fifth columns, both $\Lambdab$ and $\mpieff$ are treated as Bayesian random variables and a 2D mesh of their values is considered.
}
\label{tab:nonstat_6potentials}
\end{table*}

With $\mpieff= 138\,(200)\,\mathrm{MeV}$ we obtain the intervals for $\Lambdab$ in the second (third) column of Table~\ref{tab:nonstat_6potentials}.
For the SMS 500 MeV potential,
when $\mpieff$ is fixed at the pion rest mass $138\,\mathrm{MeV}$ and $\Lambdab$ is set to $610\,\mathrm{MeV}$ from the table, warping and low-$\prel$ downsampling have all three of the diagnostics showing a successful match between our model and the data (see Fig.~\ref{fig:dsg_sms500_warped_mpi138_Lambdabopt}).
If we fix $\mpieff = 200\,\mathrm{MeV}$ and readjust $\Lambdab$, the diagnostics do not really change (see Fig.~\ref{fig:dsg_sms500_warped_mpi200_Lambdabopt}) although the MAP value of $\Lambdab$ is now 710 MeV, 15\% higher than if we take the physical pion mass as $\mpieff$.

\begin{figure*}
    \centering
    \includegraphics{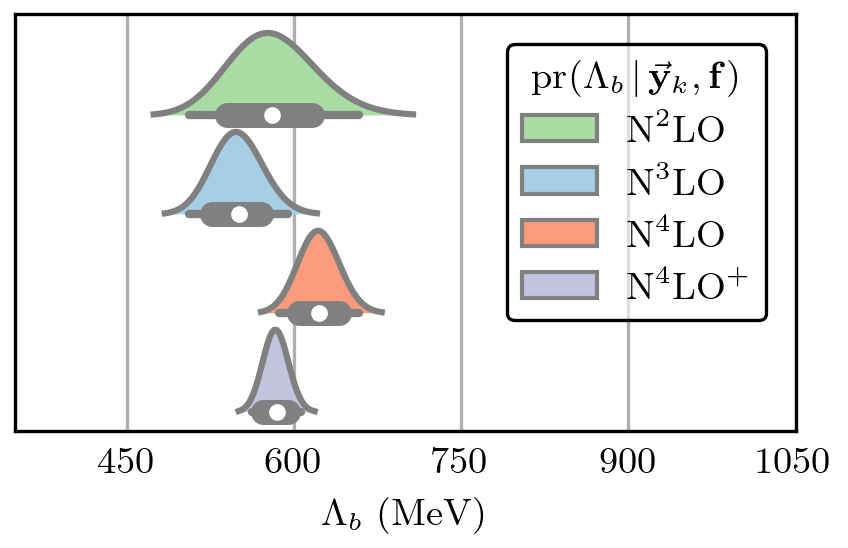}
    \includegraphics{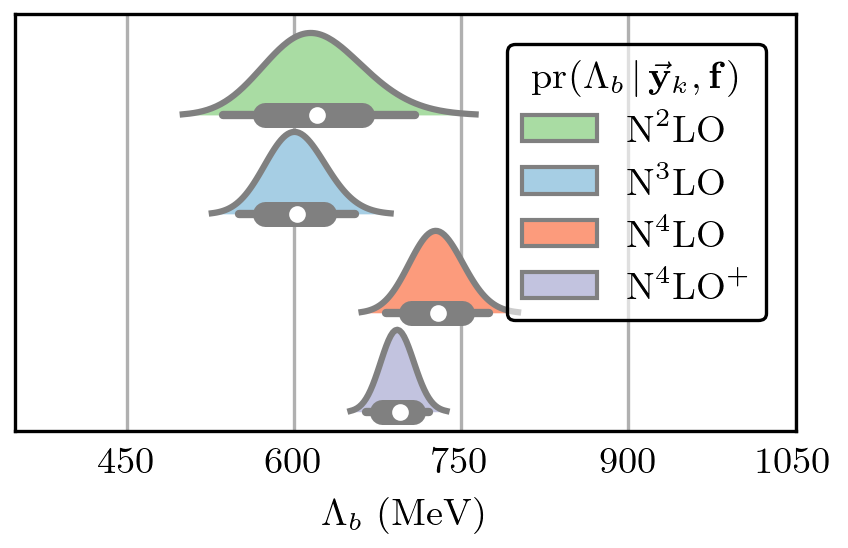}
    \includegraphics{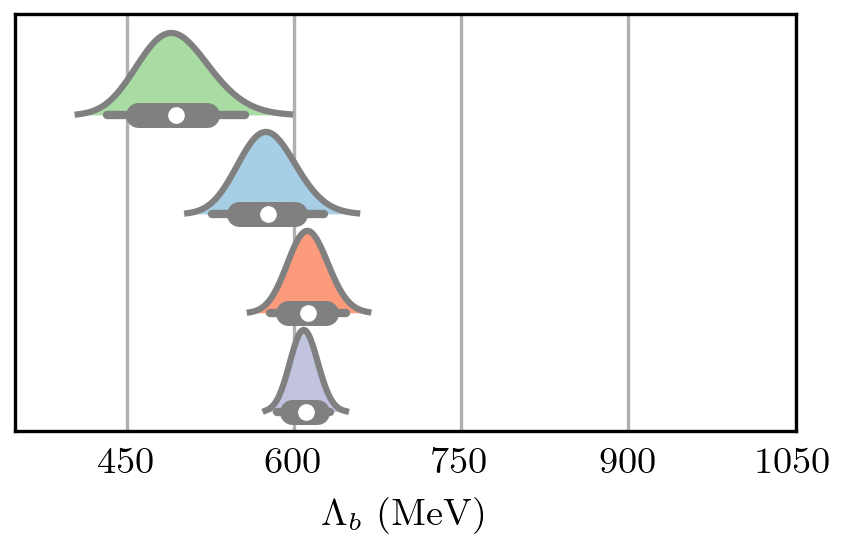}
    \includegraphics{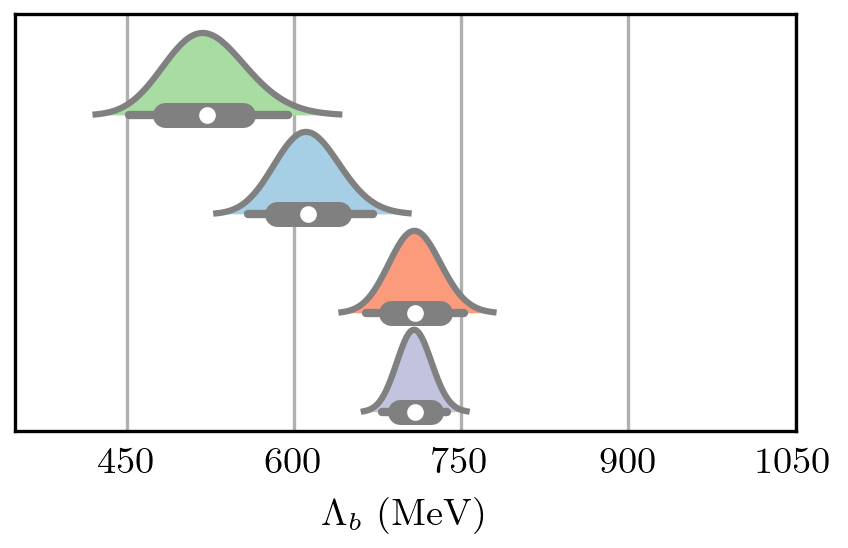}   
    \includegraphics{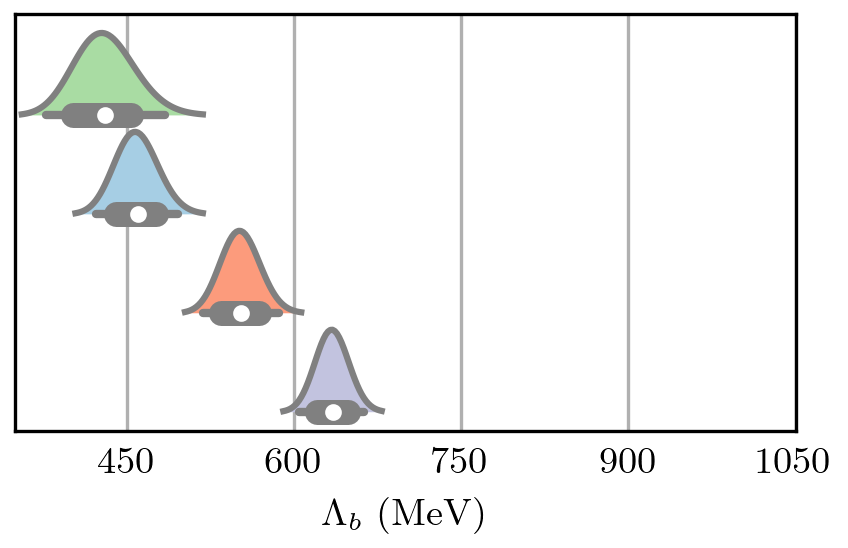}
    \includegraphics{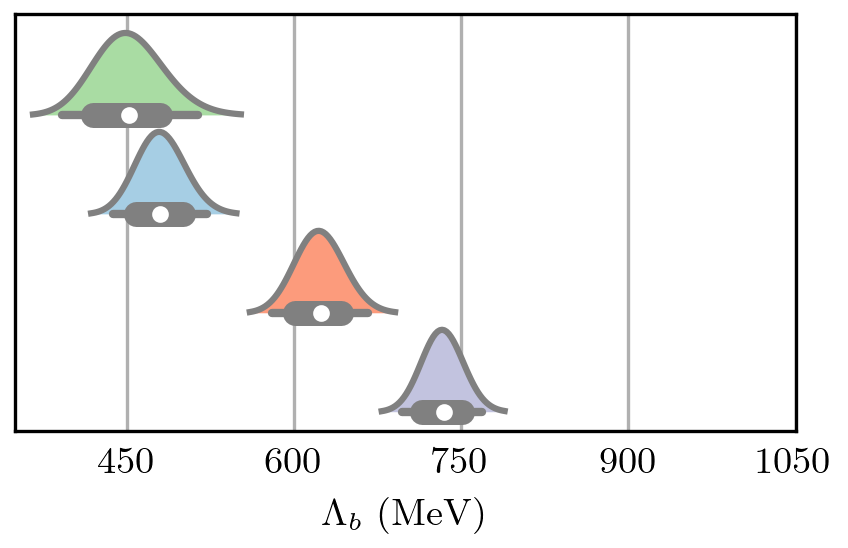}
    \caption{$\Lambdab$ posterior pdfs for SMS 450, 500, and 550 MeV with $\mpieff$ fixed at 138 MeV (left) and 200 MeV (right) generated using all six 2D observables.
    Note the discrepancies between the \NNNLO distribution and higher-order distributions.
    }
    \label{fig:Lambdab_mpifixed_sms}
\end{figure*}

\begin{figure*}
    \centering
    \includegraphics{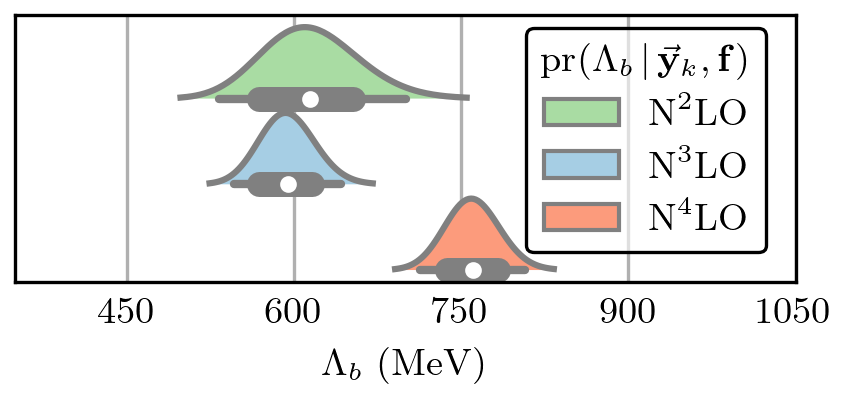}
    \includegraphics{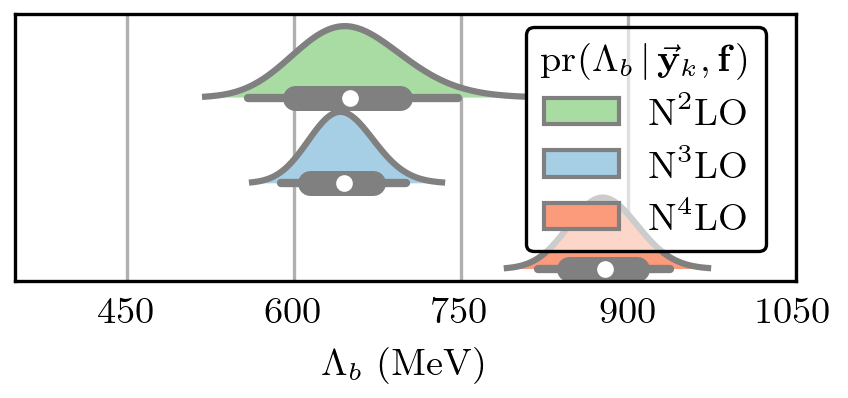}   
    \includegraphics{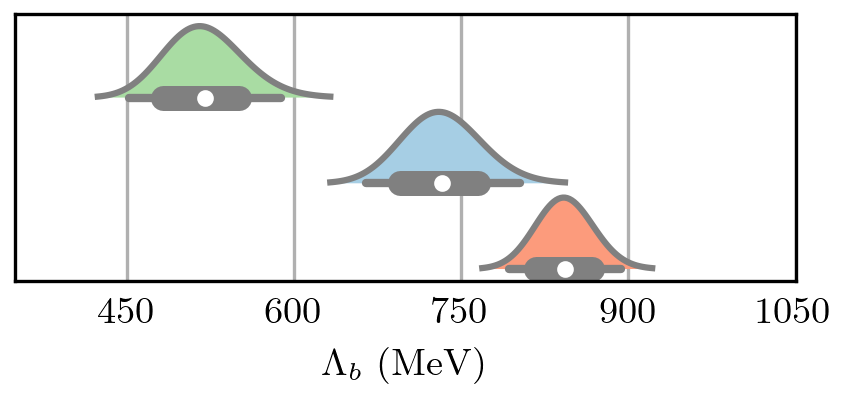}
    \includegraphics{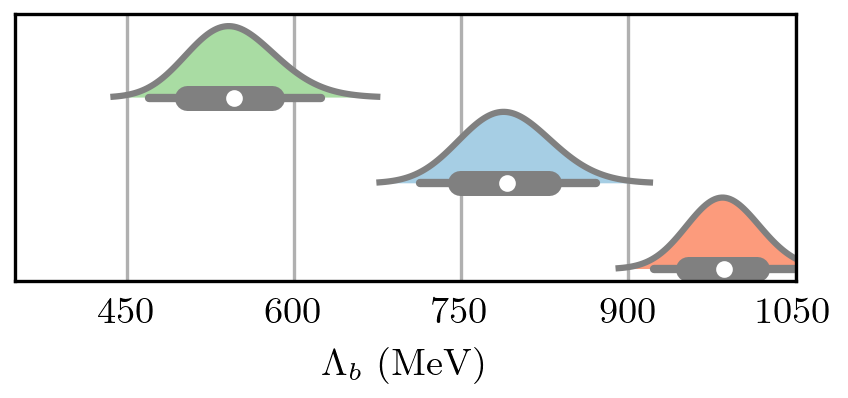}
    \includegraphics{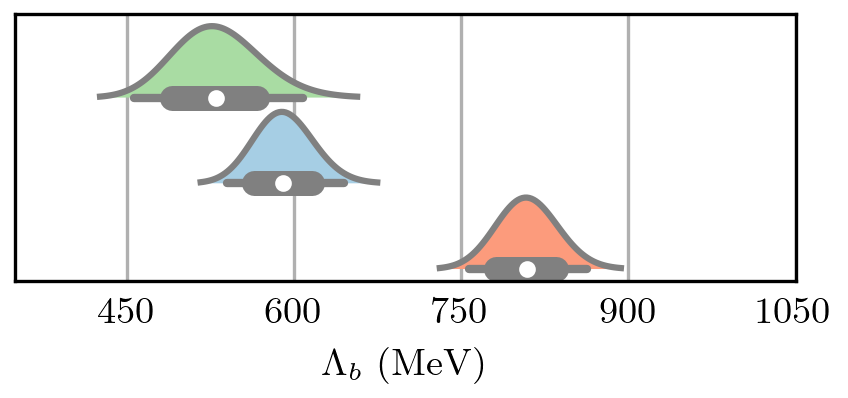}
    \includegraphics{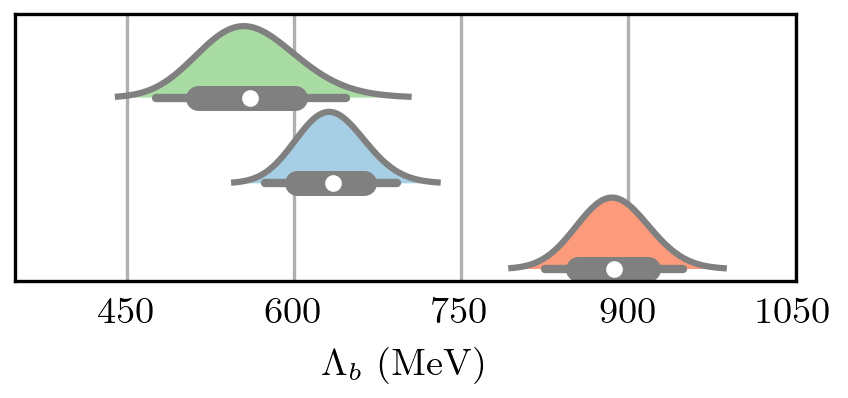}
    \caption{$\Lambdab$ posterior pdfs for SCS 1.0 fm, SCS 0.9 fm, and EMN 500 MeV with $\mpieff$ fixed at 138 MeV (left) and 200 MeV (right) generated using all six 2D observables.
    Note the discrepancies between the \NNNLO distribution and \NNNNLO distribution.
    }
    \label{fig:Lambdab_mpifixed_scs_emn}
\end{figure*}

We move now to assess the robustness of inference of $\Lambdab$ within a family of potentials as the order of the potential is increased.

Figure~\ref{fig:Lambdab_mpifixed_sms} plots the $\Lambdab$ posterior pdfs for orders up to \NNLO, \dots, \NNNNLOp for SMS 450, 500, and 550 MeV, and Fig.~\ref{fig:Lambdab_mpifixed_scs_emn} for \NNLO, \dots, \NNNNLO for SCS 0.9 fm, SCS 1.0 fm, and EMN 500 MeV.
In both figures $\mpieff = 138\,\mathrm{MeV}$ (the pion rest mass) in the left-hand panel and $\mpieff = 200\,\mathrm{MeV}$ (an estimate from Ref.~\cite{Epelbaum:2019wvf}) in the right-hand panel.
These figures are thus graphical representations of the second and third columns of Table~\ref{tab:nonstat_6potentials}.

In all cases we see the steady narrowing of the $\Lambdab$ distributions with rising order that we know should be taking place if the coefficients are drawn from the same GP.
SMS 450 MeV and 500 MeV appear most consistent across order.
There are are notable jumps between orders for different potentials (namely, \NNNLO to \NNNNLO for SCS 1.0 fm and EMN 500 MeV, and \NNLO to \NNNLO for SCS 0.9 fm).
It is particularly notable that the \NNNNLO extraction of $\Lambdab$ for the SCS and EMN potentials yields a markedly higher result than the \NNLO value and even the \NNNLO one. 
Consistent $\Lambdab$ inference across orders occurs at best only within a certain range of regulator values and perhaps only in particular regulator schemes.

It may also be the case that how potentials are fitted matters here: We note that the LECs in the SCS and EMN potentials were not obtained in a way that took into account the theoretical uncertainty at each order.
If potentials at lower orders are fine-tuned to data, then the pattern of order-by-order changes in observables may not be in accord with the EFT expansion.  

The distribution of $\Lambdab$ extracted from the results through \NNLO differs well outside its error bars from the distributions extracted at \NNNNLO for five of the six potentials considered.
Only for the SMS 450 MeV potential (and the choice $\mpieff=138$ MeV) does the \NNLO 68\% interval for $\Lambdab$ cover the results obtained in the \NNNLO 
 and \NNNNLO analyses. 
This suggests that the values for the breakdown scale extracted at \NNLO are not robust with respect to increase of order within a family of  potentials.
This is pertinent because some families of \chiEFT\ NN potentials, e.g. the GT+ potentials, are only calculated up to third order (\NNLO) in the expansion.
It is difficult to draw strong conclusions about $\Lambdab$ for those potentials, since we see here that having only two coefficients in the expansion does not pin down $\Lambdab$ very well. 

And indeed, the corresponding analysis of the GT+ potentials yields (for $\mpieff = 138\,\mathrm{MeV}$)
the $\Lambdab$ values listed in the second column of Table~\ref{tab:nonstat_6potentials}, which are implausibly large.
The preferred mean values for $\Lambdab$ (with error bands of about 7\%) were anywhere between $700\,\mathrm{MeV}$ for GT+ 0.9 fm and $1300\,\mathrm{MeV}$ for GT+ 1.2 fm.
It is particularly surprising that the breakdown scale increases as the EFT momentum cutoff decreases.

\subsection{Variable \texorpdfstring{$\mpieff$}{mpieff}}
\label{subsec:variable_meff}

In the previous subsection we showed that  fixing $\mpieff$ did not yield reliably consistent pdfs for $\Lambdab$.
It is conceivable that treating both parameters as Bayesian random variables might yield the consistency that we expect.
In this subsection we generate pdfs using the approach employed in the previous subsection but with $\mpieff$ as a random variable on its own mesh.
The resulting 68\% intervals for $\mpieff$ and $\Lambdab$ are displayed in the fourth and fifth columns in Table~\ref{tab:nonstat_6potentials}.

\begin{figure*}[tbh!]
\centering
    \includegraphics[width=0.4\textwidth]{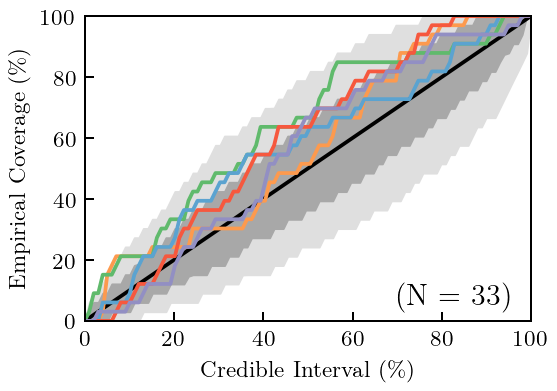}
    \includegraphics[height=0.27\textwidth]{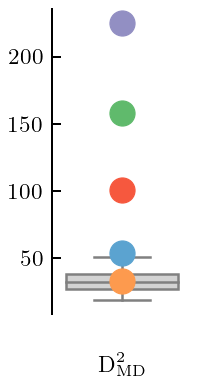}
    \includegraphics[width=0.4\textwidth]{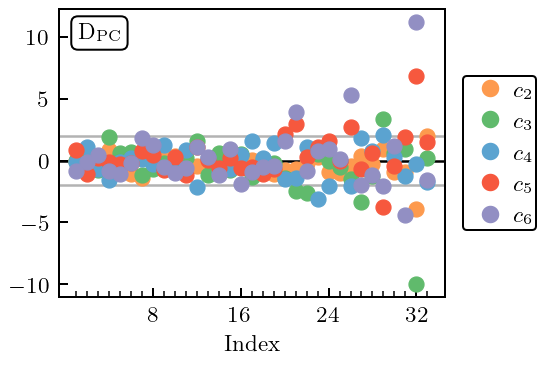}
    \caption{Diagnostics for the differential cross section $d\sigma / d\Omega$ from SMS 500 MeV.
    They are generated with ($\xE$, $\xtheta$) = ($\prel$, $\negcos$) and $Q = \Qsum(p = \prel, \mpieff = 168\, \mathrm{MeV}, \Lambdab = 670\, \mathrm{MeV})$ (optimal values for $\Lambdab$ and $\mpieff$ from the fourth and fifth columns in Table~\ref{tab:nonstat_6potentials}) and calculated with 14 training points and 33 testing points with warping and downsampling and a fully stationary RBF kernel.
    }
    \label{fig:dsg_sms500_warped_jointopt}
\end{figure*}

Perhaps unsurprisingly, given that there are only small differences between the diagnostics in Figs.~\ref{fig:dsg_sms500_warped_mpi138_Lambdabopt} and \ref{fig:dsg_sms500_warped_mpi200_Lambdabopt}, turning $\mpieff$ and $\Lambdab$ to the MAP values of the two-dimensional posterior does not yield a noticeable improvement over those obtained in the fixed $\mpieff$ studies of the previous subsection (see Fig.~\ref{fig:dsg_sms500_warped_jointopt}).

\begin{figure}[htb!]
\centering
    \includegraphics{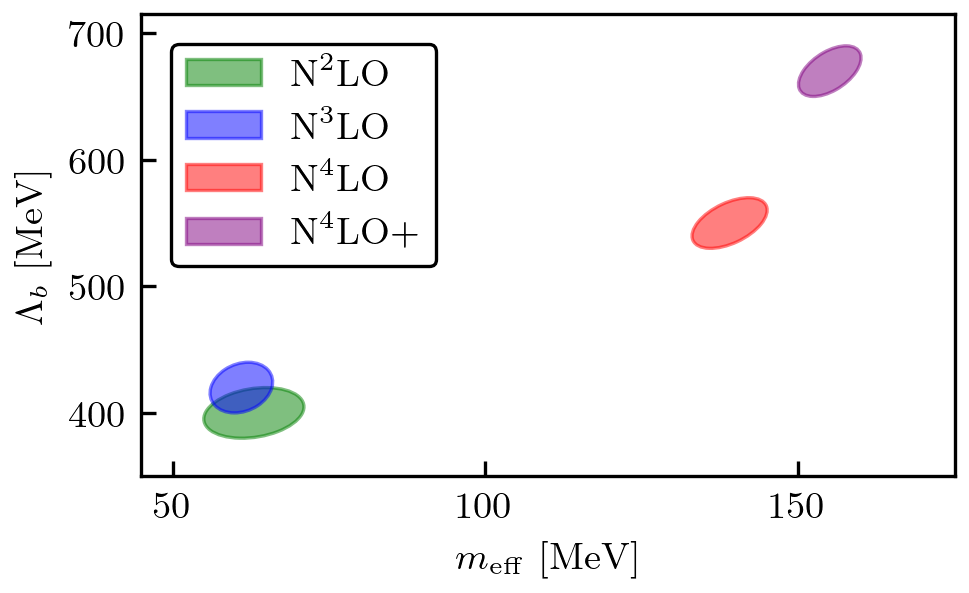}
    \caption{Plot of the joint $\Lambdab$-$\mpieff$ probability distributions for all six 2D observables combined from SMS 500 MeV, order by order.
    These are the ellipses (plotted with 95\% credibility intervals) from which the values in the fourth and fifth columns of Table~\ref{tab:nonstat_6potentials} are extracted.
    Note the inconsistency between different orders' distributions, bearing in mind that the \NNLO and \NNNLO ellipses do not overlap at the 68\% confidence interval level.
    }
    \label{fig:ellipses_sms500mev}
\end{figure}

Figure~\ref{fig:ellipses_sms500mev} shows the joint $\Lambdab$-$\mpieff$ 95\% credibility regions for SMS 500 MeV.
The consistency across orders is not improved by choosing to treat $\mpieff$ as a Bayesian random variable.
In fact, it is worse, since the posteriors become narrower upon the addition of this extra parameter.
Na{\"i}vely, one might interpret these results as evidence of an underlying identifiability problem between $\Lambdab$ and $\mpieff$.
If that were true, however, the ellipses would be very narrow in one direction and very broad in the orthogonal one; that is clearly not the case here.
The shape and size of the distributions, along with their incompatibility, point to an overfitting issue, specifically with parametrizing the dimensionless expansion parameter $Q$ in terms of $p$, $\mpieff$, and $\Lambdab$ and treating the last two as Bayesian random variables.
Under that assumption, the extracted distributions are relatively small and ordinary-looking ellipses (a sign that our algorithm is quite sure that it has located a bounded area of high probability) that do not get much narrower as information is added with each successive order (a sign that our data is not converging with order).
These aspects of the results point to a mismatch between the data and our assumptions about how to parametrize $Q$.

In spite of these issues, even when we let $\mpieff$ vary,
different potentials tend to agree on a preferred value of $\Lambdab$ between 650 and 800 MeV.
This result is consistent with the commonly accepted range $500 \lesssim \Lambdab \lesssim 800\,\mathrm{MeV}$.
The results for $\mpieff$, however, are somewhat harder to reconcile with the identification of $\mpieff$ with a physical parameter. 
At highest order (\NNNNLOp), the three SMS potentials favor values of $\mpieff$ between 150 and 200~MeV, which is fairly consistent with the discussion by Epelbaum in Ref.~\cite{Epelbaum:2019wvf}.
(His results drew on data for only $\sigmatot$ and assumed $\Lambdab \approx 600\,\mathrm{MeV}$.)
SCS 0.9 and 1.0~fm favor values of $\mpieff \approx 150\,\mathrm{MeV}$, which can be thought of as partway between $\mpieff \sim \mpi$ and the Epelbaum results.
Contrariwise, EMN 500 MeV favors distinctly lower values for $\mpieff$ around 80~MeV at \NNNNLO, which is not in accordance with previous investigations of the value of the soft scale in \chiEFT.
The corresponding value of $\Lambdab$ is in keeping with that found with other potentials, but is 20\% lower than that found for the same family of potentials with $\mpieff=138$ MeV.

\section{Summary and outlook}
\label{sec:outlook}

In this paper, we have extended an assessment of the BUQEYE model from one \chiEFT\ NN potential in Ref.~\cite{Millican:2024yuz} to potentials with different choices of regulator scheme and scale.
With an analytic framework that proceeds from that previous work, we first evaluated the impact of scheme and scale choices on the \chiEFT\ convergence pattern and found that soft potentials (i.e., those whose regulator scale is too small in momentum space or too large in coordinate space) have irregular convergence patterns, which is consistent with the shifting of pion-exchange physics from odd to even orders in the Weinberg counting scheme.
If it is to be suitable for estimating theory uncertainties for these potentials the BUQEYE model will have to be modified to accommodate this alteration in the \chiEFT\ convergence pattern.

Then, we looked
for mismatches between the actual behavior of the coefficients and the stationarity we assumed in our GP.
We found that the length scale in the scattering angle of the 2D observables diminished with rising relative momentum, which can be explained semi-classically.
Two options for remediating the length-scale nonstationarity inherent in the BUQEYE model's GP approach to EFT truncation error were considered---a nonstationary kernel (NSK) and a warping/downsampling approach.
After verifying that the two methods produced the same results (as gauged by the statistical diagnostics) if subject to the same training and testing split, we ultimately decided to employ the warping/downsampling method.
Once this method was incorporated into our analysis framework, our model's consistency with the data showed marked improvement as demonstrated by a suite of three statistical diagnostics.

Finally, with the warping/downsampling method for remediating nonstationarity in effect, we sought to optimize the breakdown scale $\Lambdab$ and the soft scale $\mpieff$  as tuning parameters for a statistically consistent GP description of EFT observable coefficients.
We found that fixing the value of $\mpieff$ and using the MAP value of the $\Lambdab$ pdf 
tended to produce statistical diagnostics that reflected consistency between our model and the underlying data.
However, order-by-order posteriors for $\Lambdab$ are not statistically consistent in most cases.
The exceptions are the SMS 450\,MeV and (excepting \NNLO) 500\,MeV potentials.
The results when both $\mpieff$ and $\Lambdab$ were treated as random variables showed evidence of overfitting: We are simply not able to extract them both.

The methods employed here enable assessments of the BUQEYE model in other domains,
including $\Delta$-ful $np$ \chiEFT\ potentials as compared to their $\Delta$-less alternatives, $pp$ scattering with its attendant Coulomb interactions, 
and many-body (especially 3N) interactions.
Success in quantifying 
the truncation error of EFT models across these domains, which might entail modifications of the BUQEYE model, in turn enables cross-comparisons to validate the EFT power counting or allows Bayesian model-mixing to rigorously combine predictions with propagated uncertainties.

\begin{acknowledgments}
We thank Matt Pratola, Matthias Heinz, and Evgeny Epelbaum for useful discussions.
The work of PJM and RJF was supported in part by the National Science Foundation Award No.~PHY-2209442 and the NUCLEI SciDAC Collaboration under U.S. Department of Energy under contract DE-FG02-96ER40963.
The work of DRP was supported by the US Department of Energy under
contract DE-FG02-93ER-40756 and by the Swedish Research Council via a Tage Erlander Professorship (Grant No 2022-00215).
The work of RJF, DRP, and MTP was supported in part by the National Science Foundation CSSI program under Award
No.~OAC-2004601 (BAND Collaboration~\cite{BAND_Framework}).
RJF also acknowledges support from the ExtreMe Matter Institute EMMI
at the GSI Helmholtzzentrum für Schwerionenforschung GmbH, Darmstadt, Germany.
\end{acknowledgments}

\bibliography{EMN_Correlations_Refs,bayesian_refs}

\clearpage
\appendix

\section{Supplemental Material}
\label{app:additional_potentials}

\begin{figure*}
    \centering
    \includegraphics{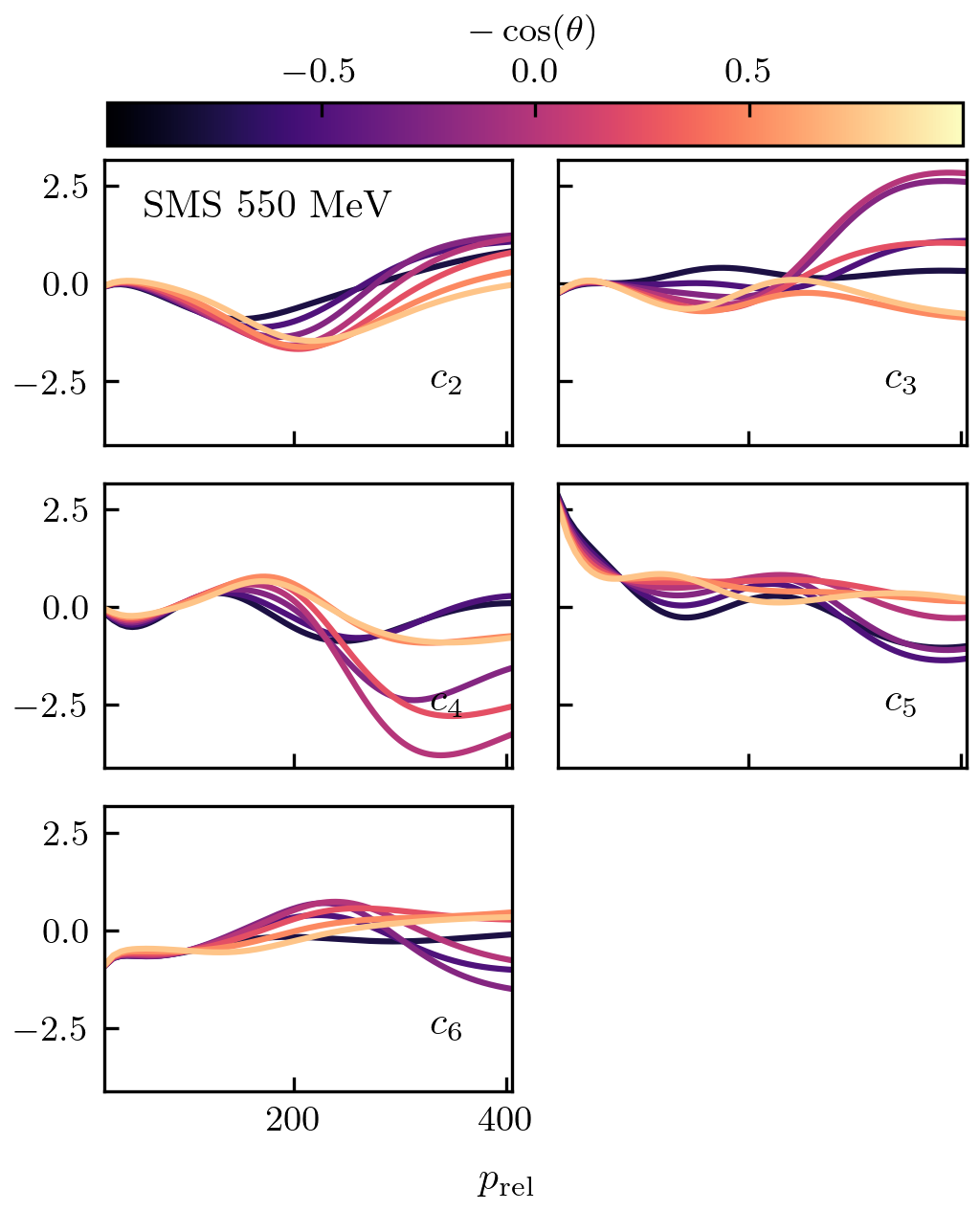}
    \includegraphics{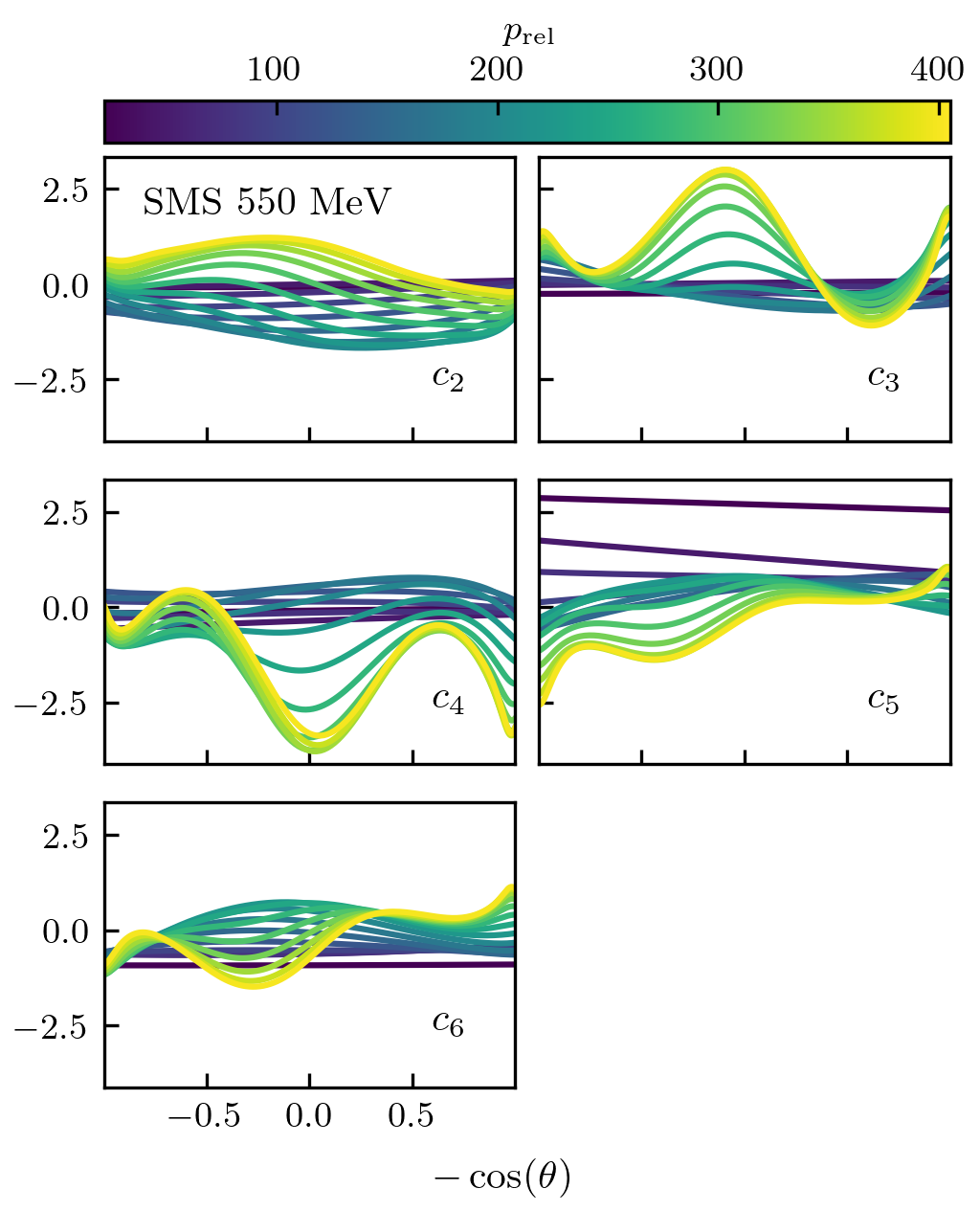}
    \includegraphics{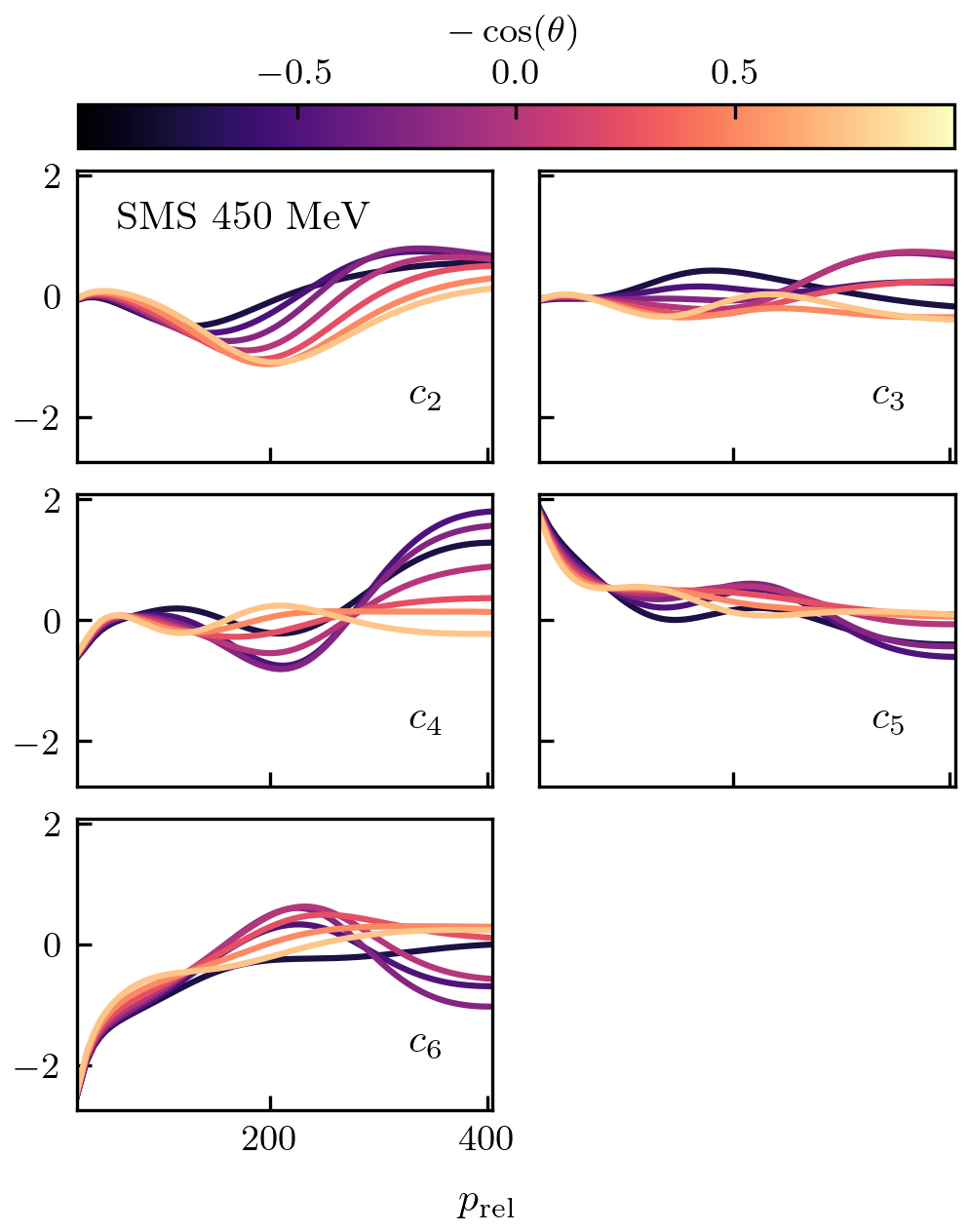}
    \includegraphics{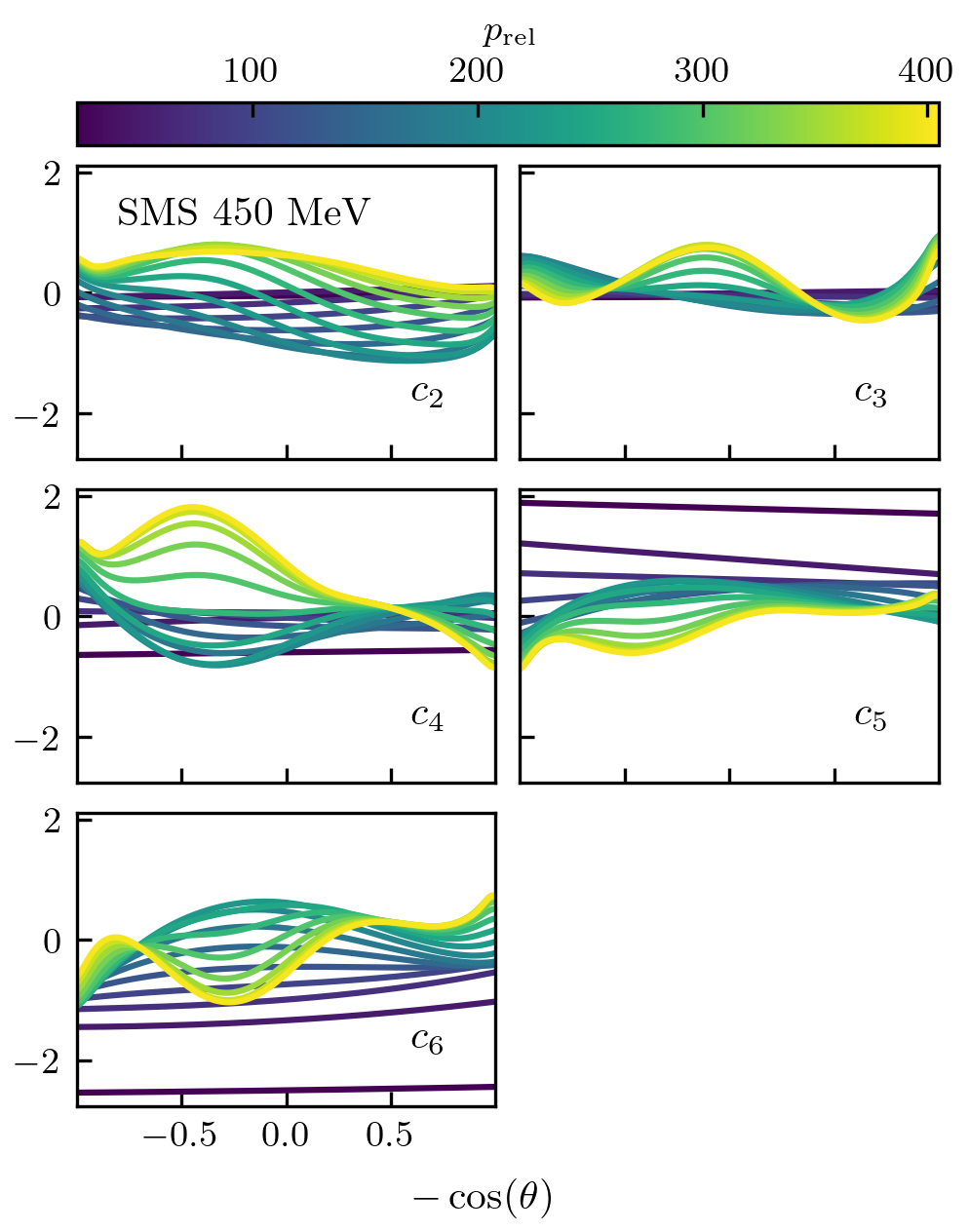}
    \caption{Order-by-order coefficients for $d\sigma / d\Omega$ for SMS 550 MeV and 450 MeV at fixed $\negcos$ and at fixed $\prel$ extracted with the choice $\left(\Lambdab = 570\,\mathrm{MeV}, \mpieff = 138\,\mathrm{MeV}\right)$ from Ref.~\cite{Millican:2024yuz}.}
    \label{fig:coeffs_eachorder_sms}
\end{figure*}

\begin{figure*}
    \centering
    \includegraphics{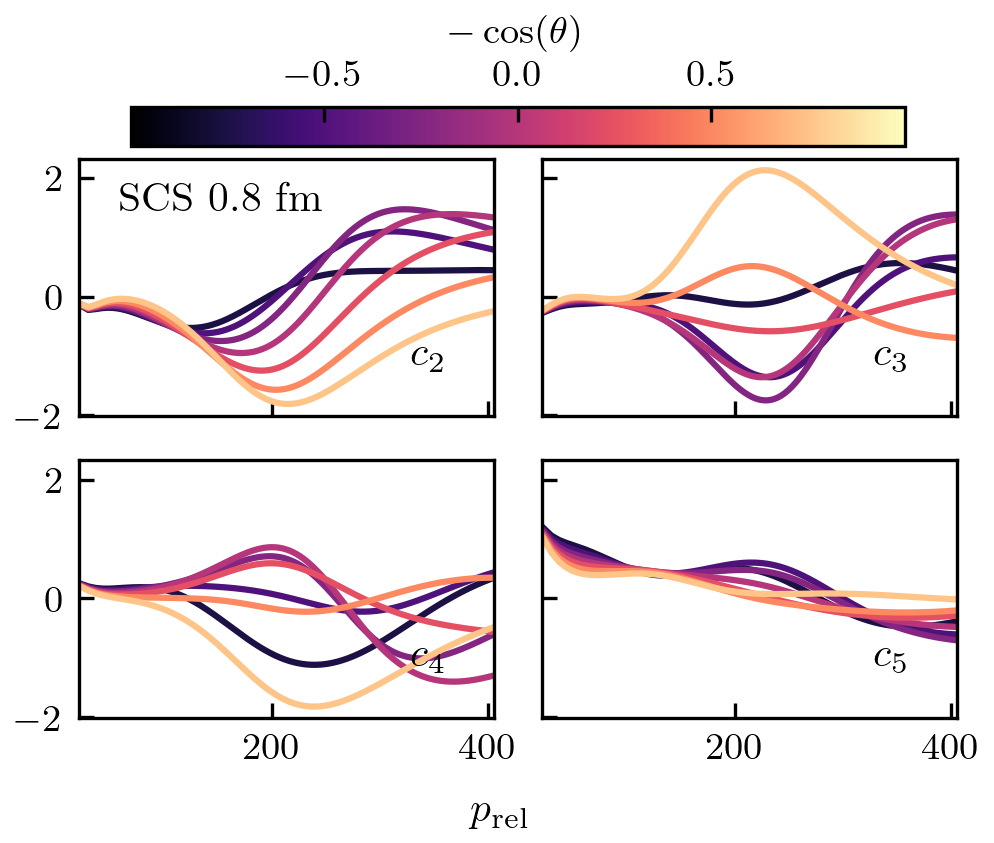}
    \includegraphics{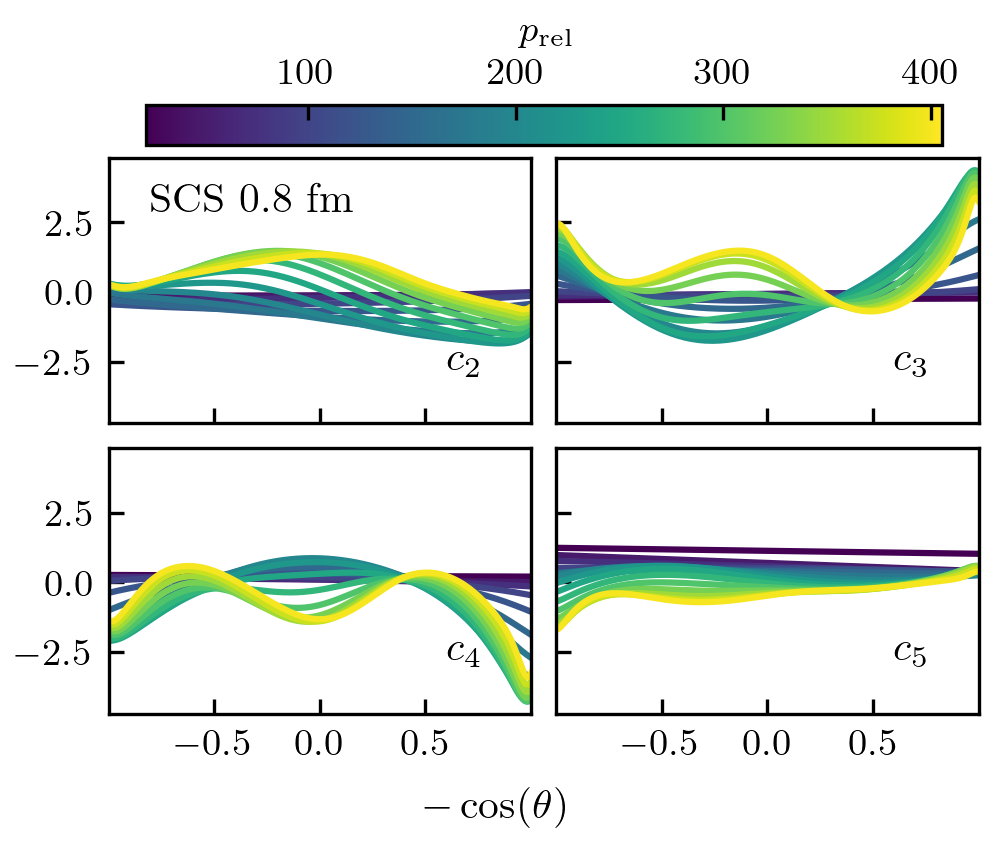}
    \includegraphics{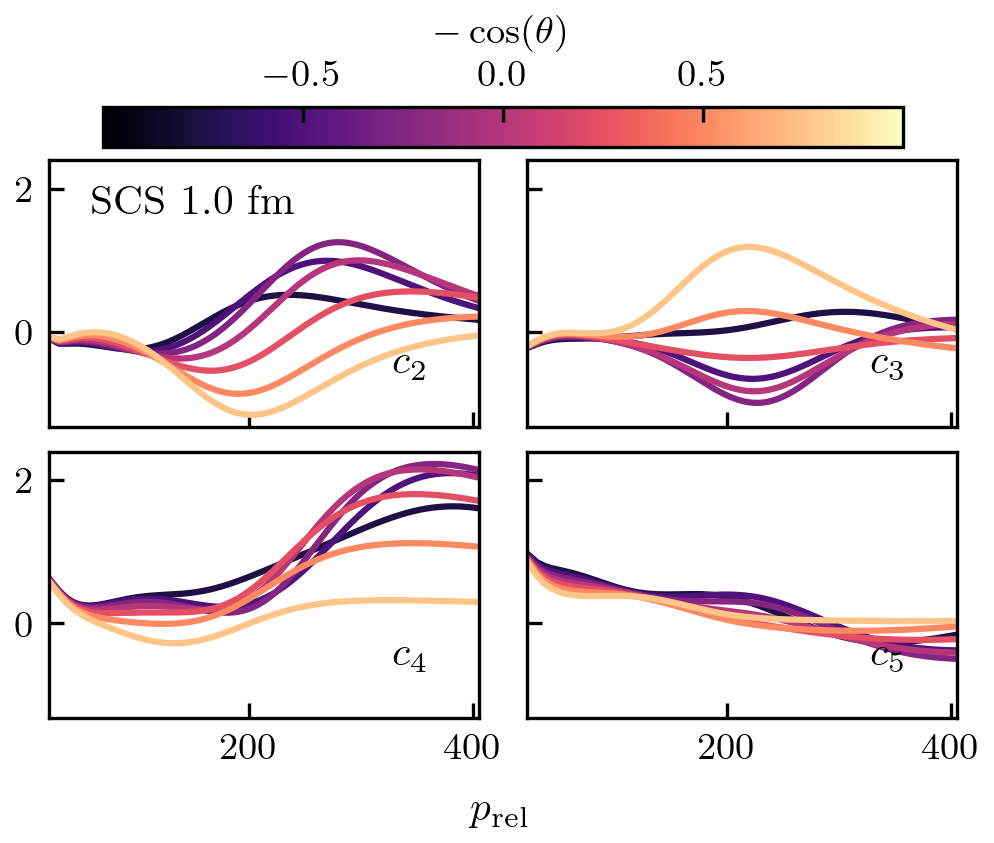}
    \includegraphics{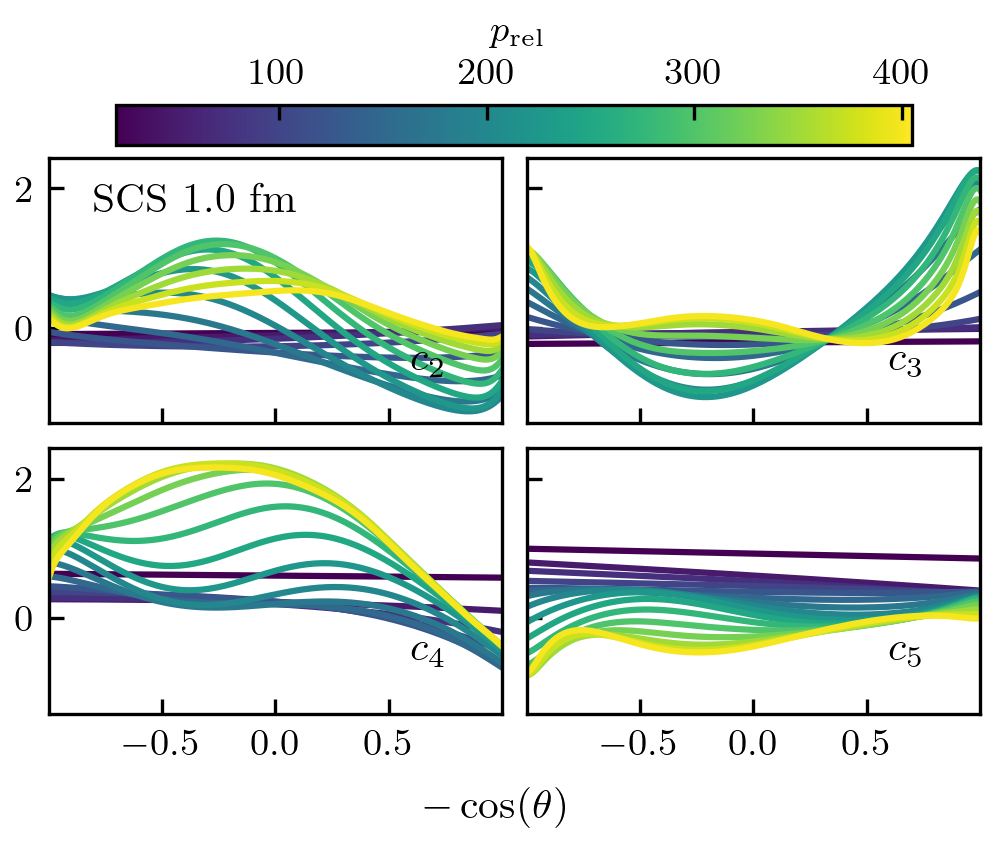}
    \includegraphics{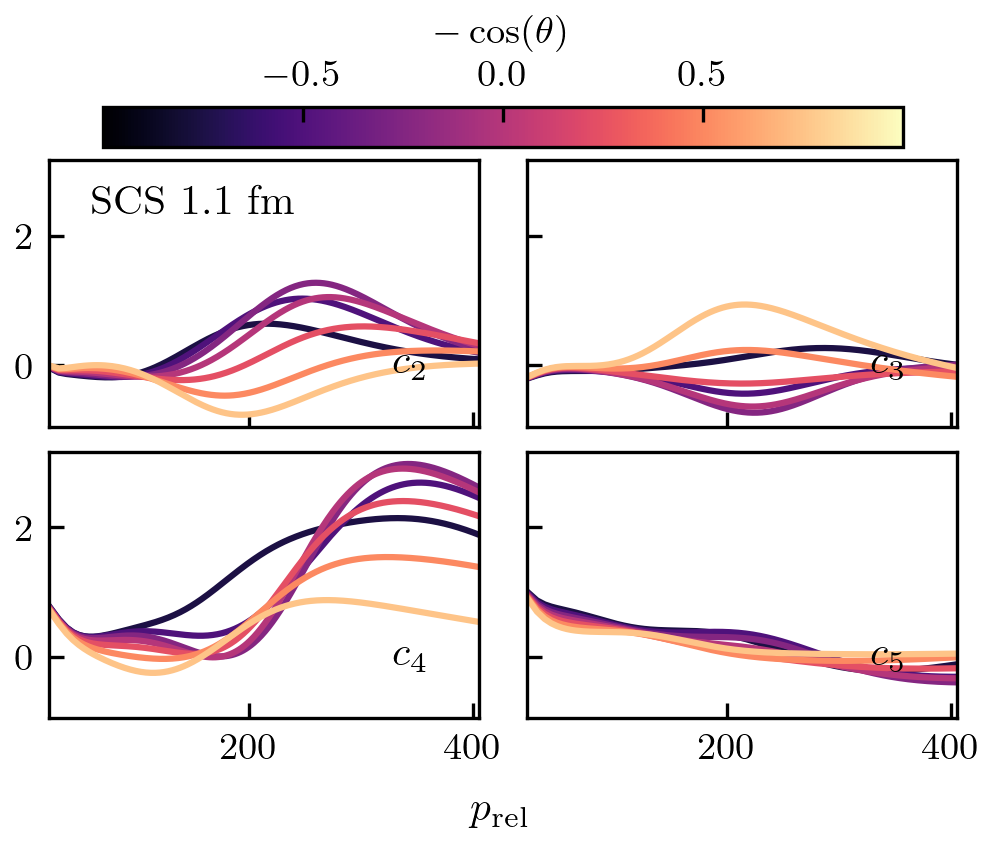}
    \includegraphics{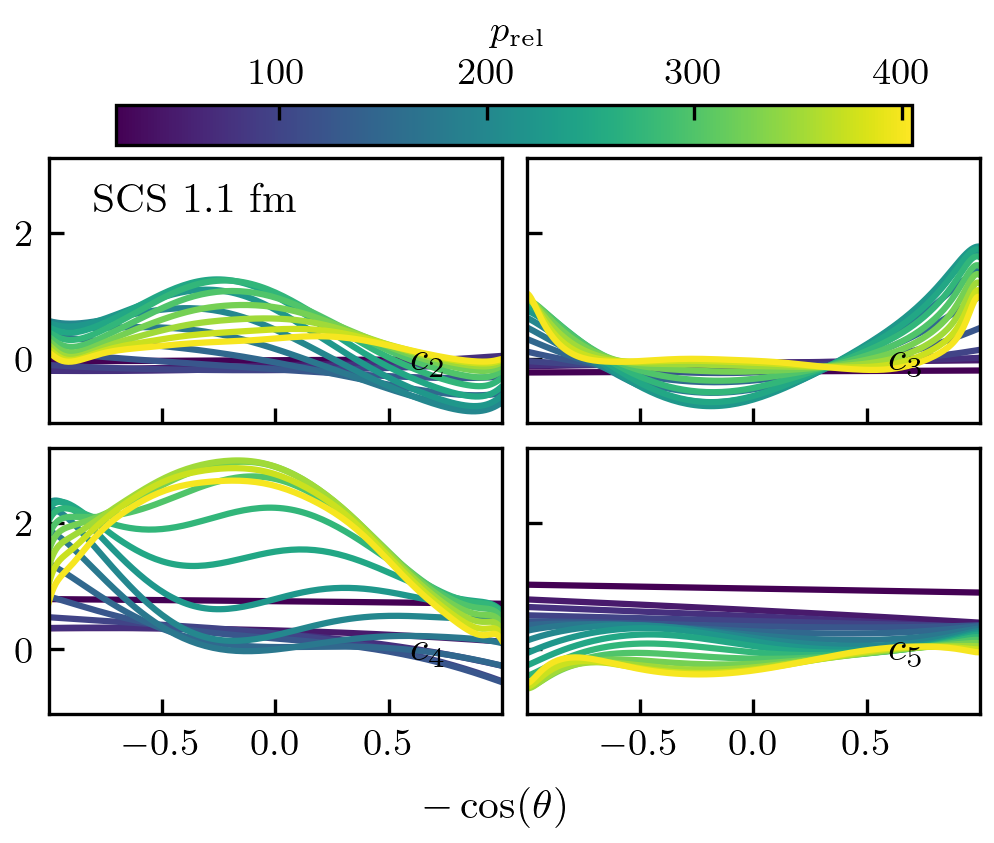}
    \caption{Order-by-order coefficients for $d\sigma / d\Omega$ for SCS 0.8 fm, 1.0 fm, and 1.1 fm at fixed $\negcos$ and at fixed $\prel$ extracted with the choice $\left(\Lambdab = 570\,\mathrm{MeV}, \mpieff = 138\,\mathrm{MeV}\right)$ from Ref.~\cite{Millican:2024yuz}.}
    \label{fig:coeffs_eachorder_scs}
\end{figure*}

\begin{figure*}
    \centering
    \includegraphics{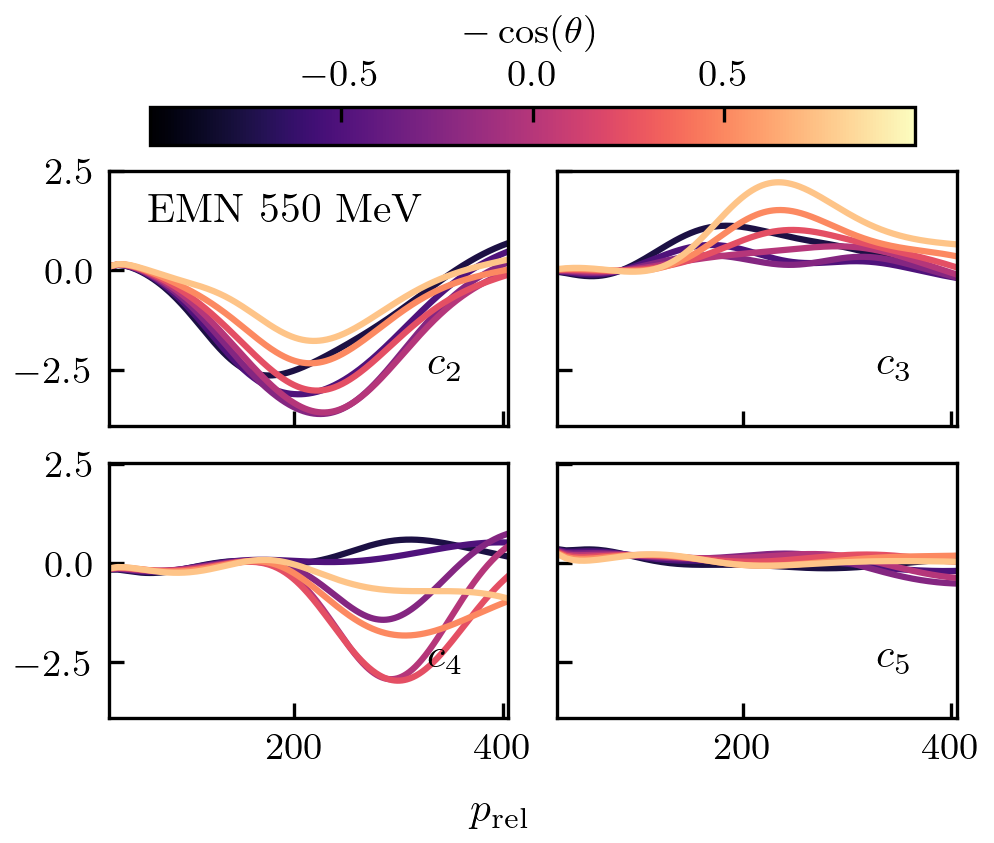}
    \includegraphics{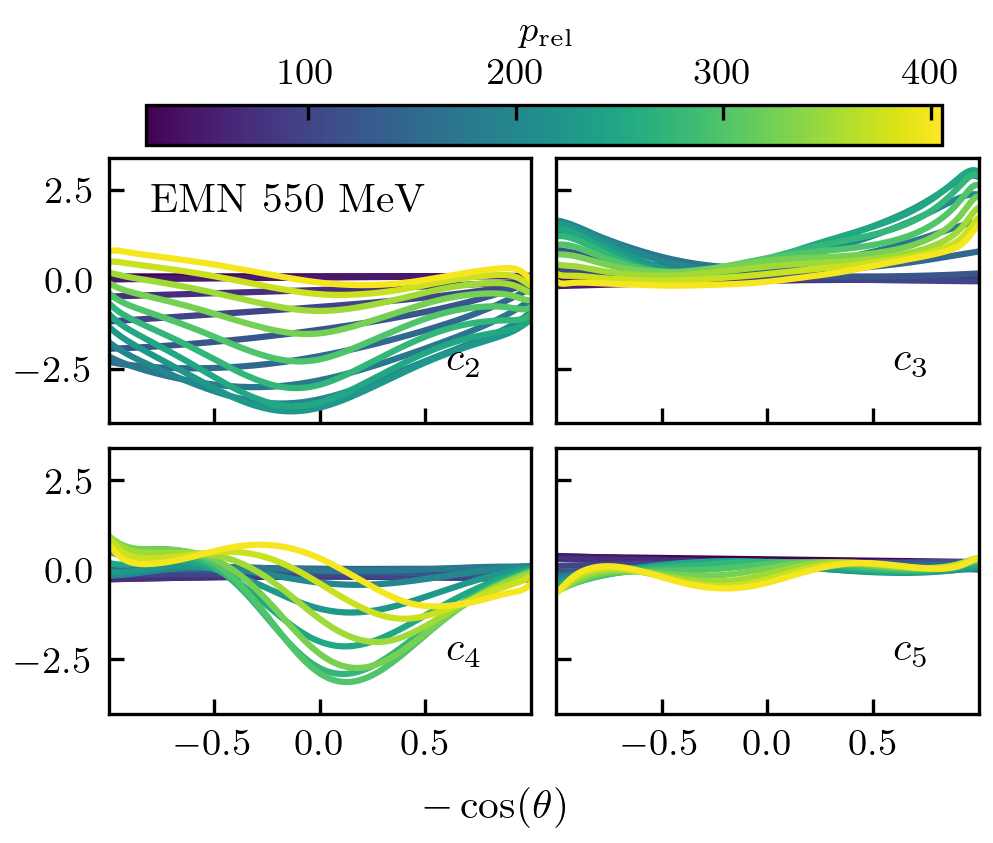}
    \includegraphics{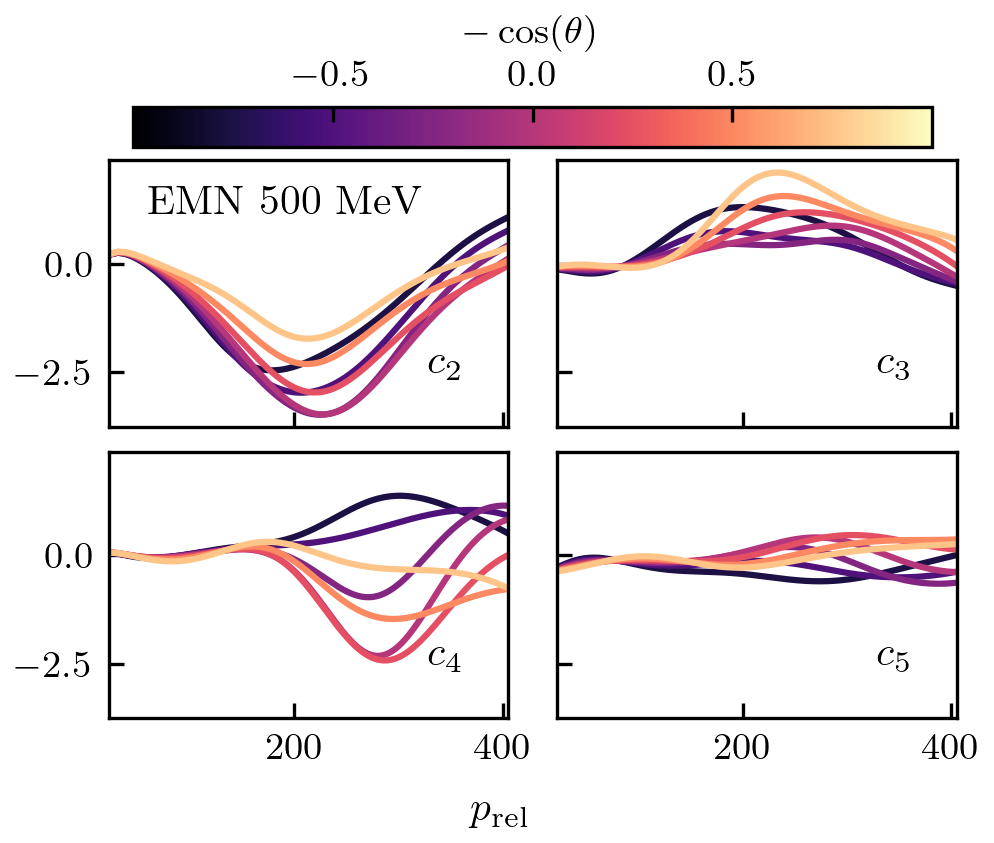}
    \includegraphics{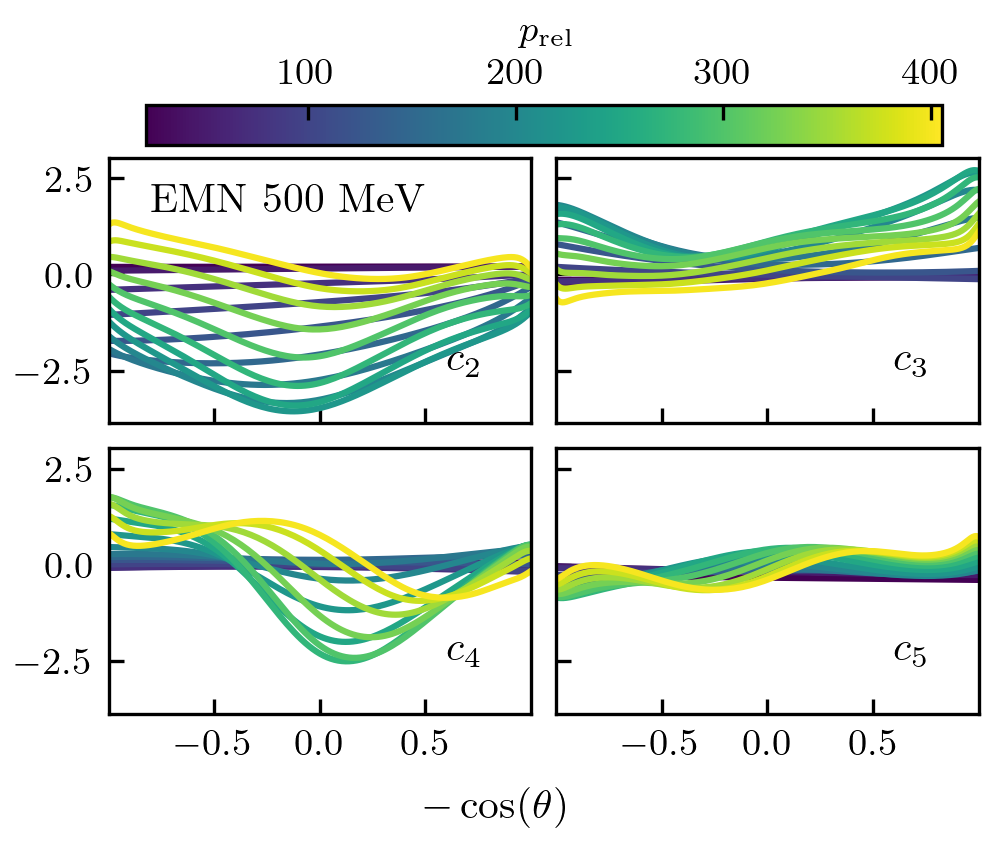}
    \includegraphics{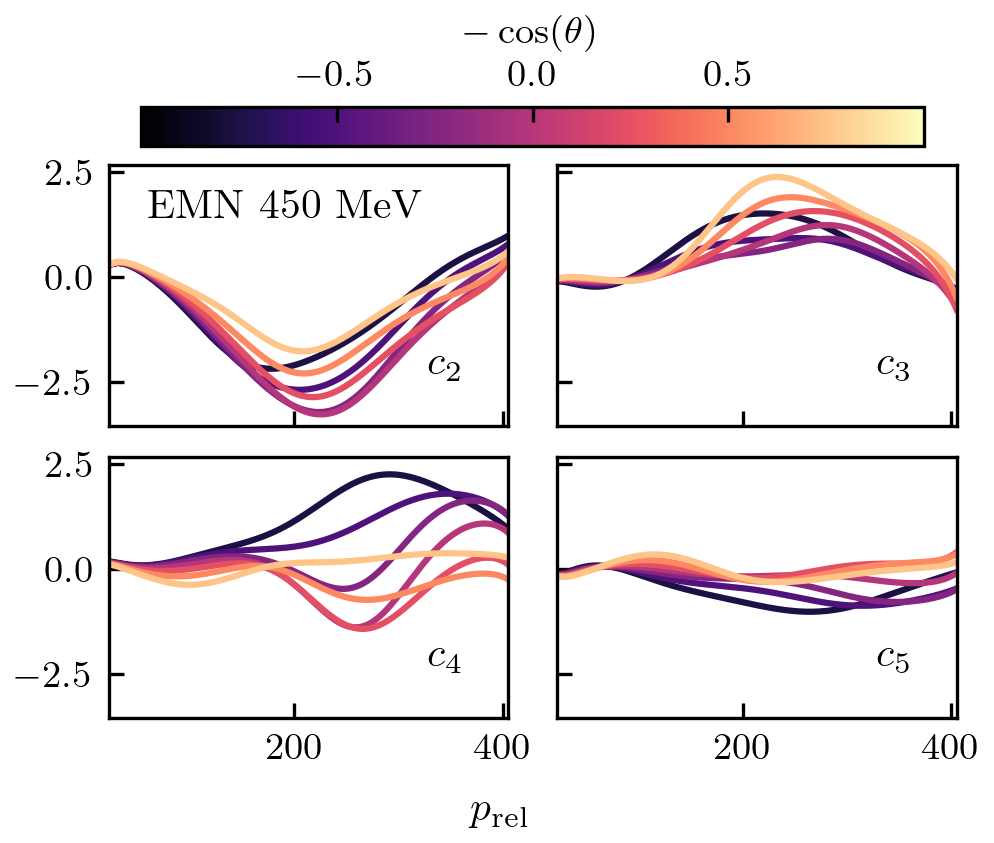}
    \includegraphics{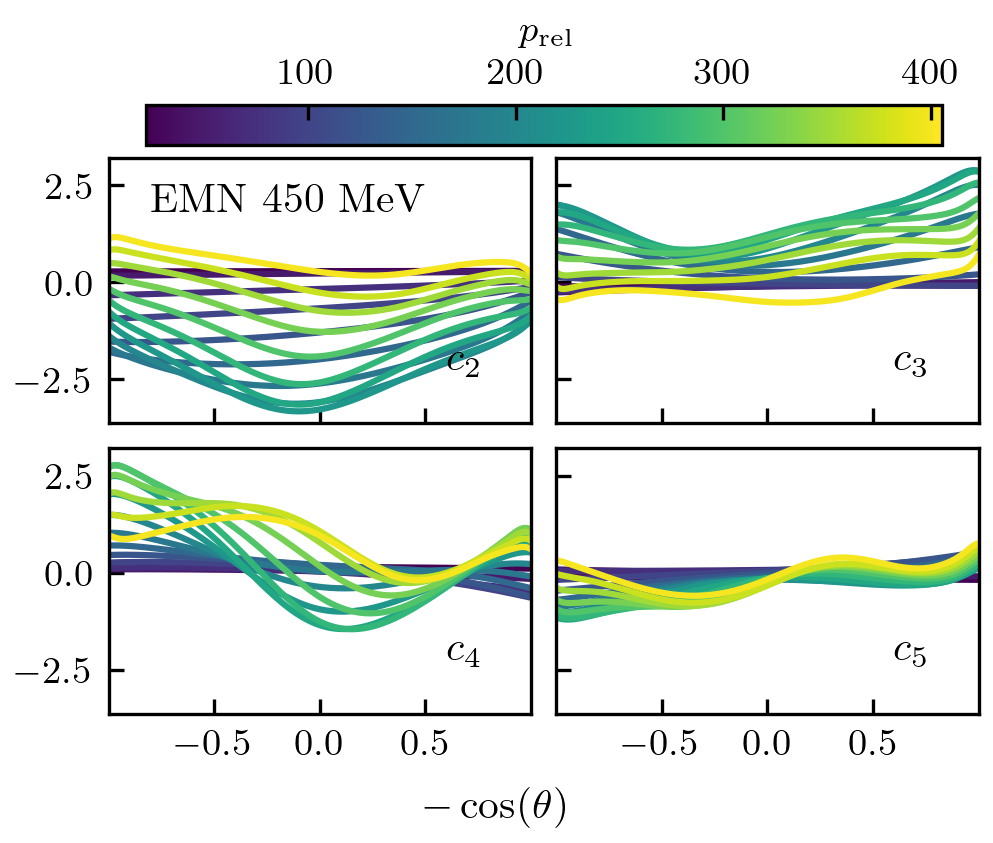}
    \caption{Order-by-order coefficients for $d\sigma / d\Omega$ for EMN 550 MeV, 500 MeV, and 450 MeV at fixed $\negcos$ and at fixed $\prel$ extracted with the choice $\left(\Lambdab = 570\,\mathrm{MeV}, \mpieff = 138\,\mathrm{MeV}\right)$ from Ref.~\cite{Millican:2024yuz}.}
    \label{fig:coeffs_eachorder_emn}
\end{figure*}

\begin{figure*}
    \centering
    \includegraphics{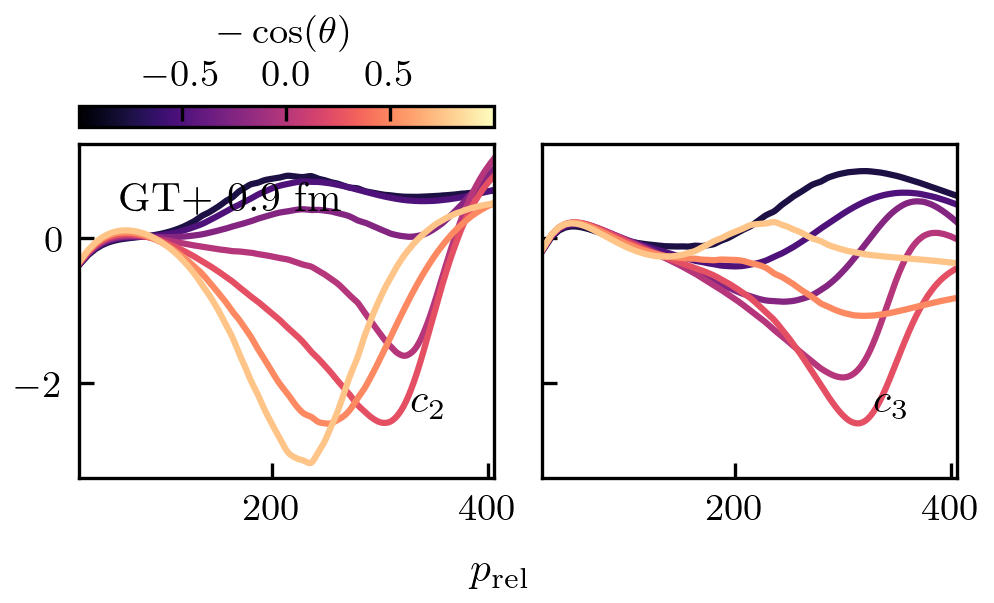}
    \includegraphics{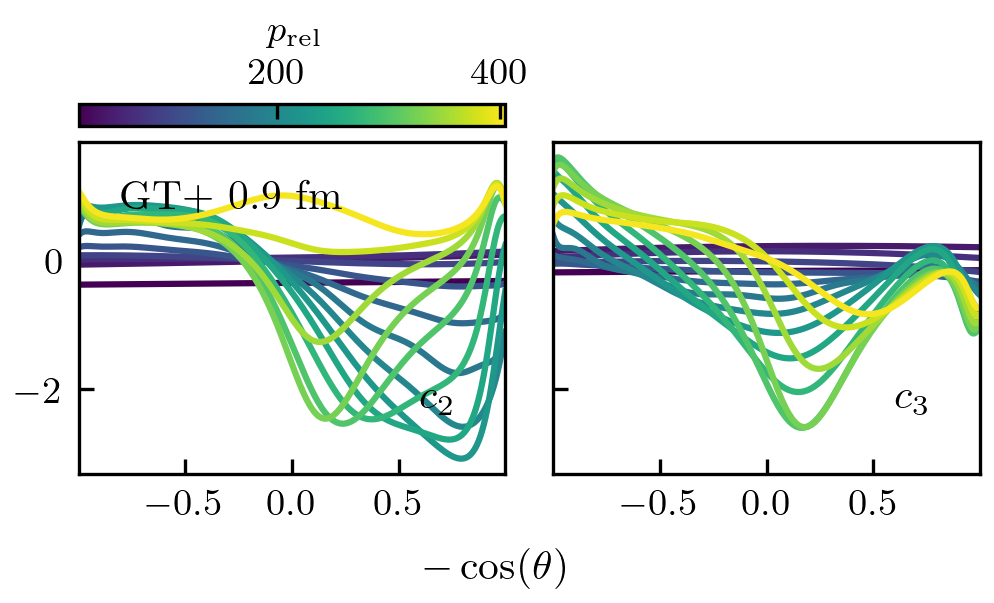}
    \includegraphics{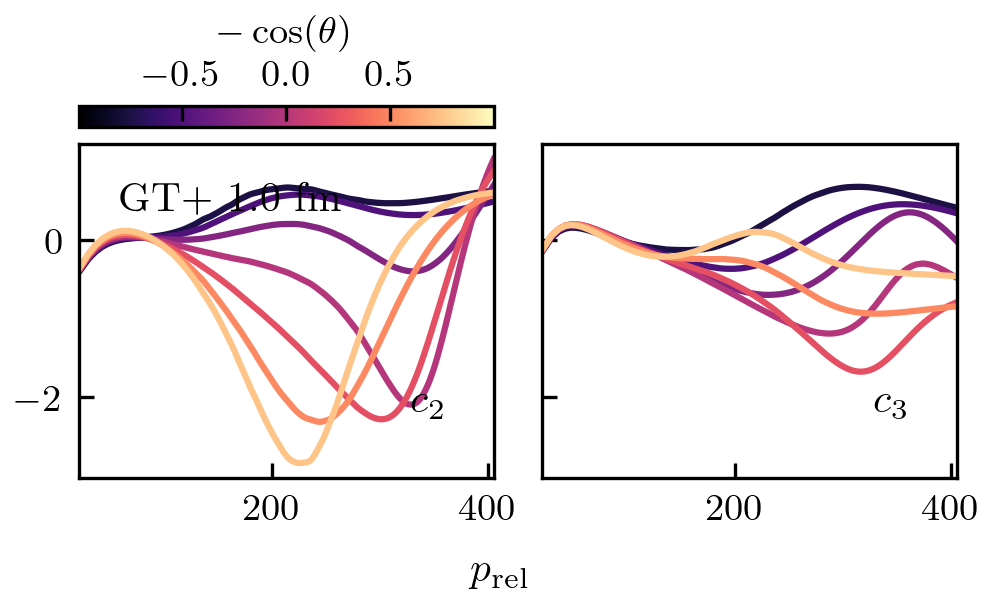}
    \includegraphics{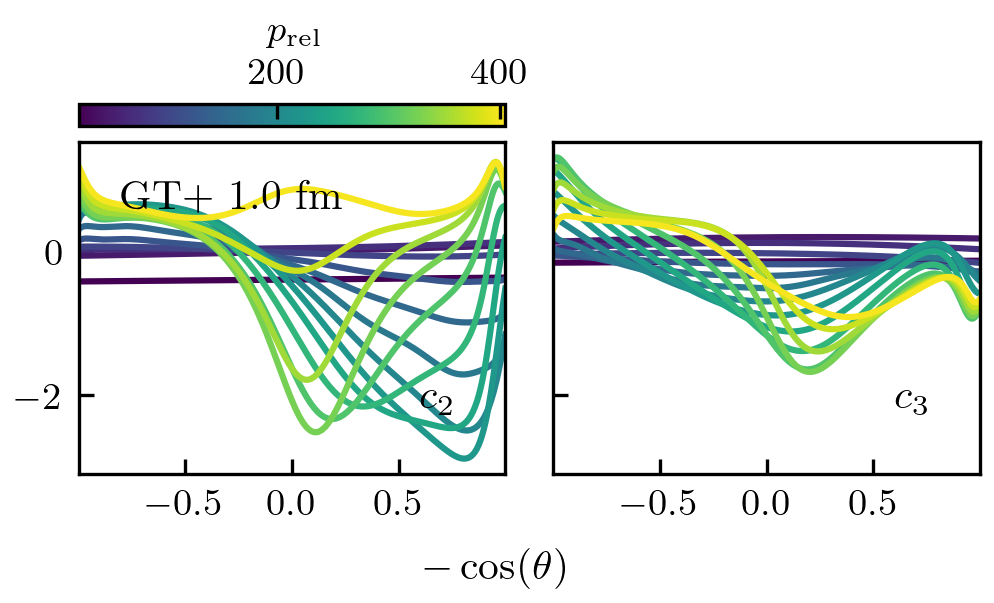}
    \includegraphics{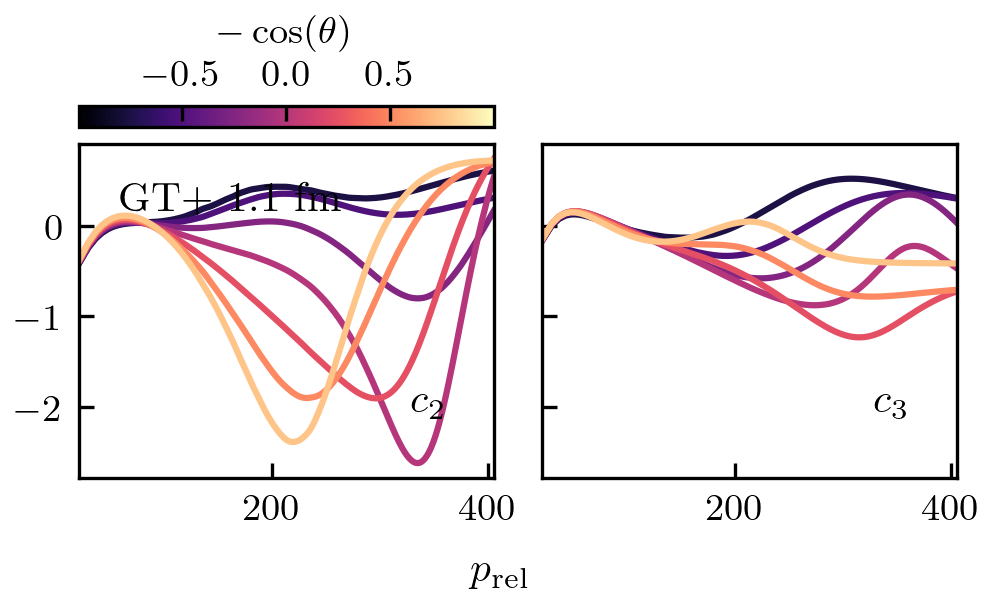}
    \includegraphics{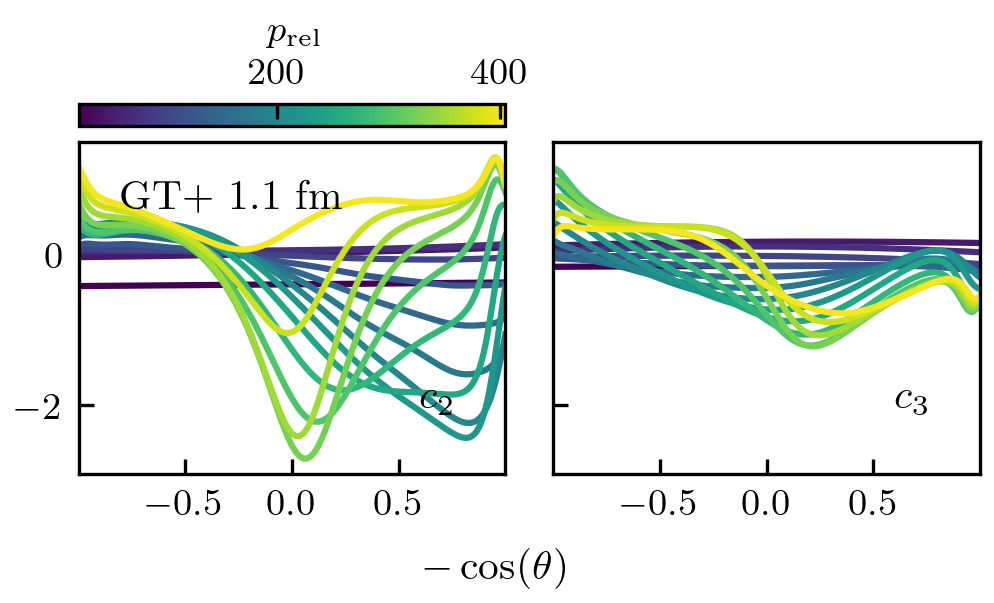}
    \includegraphics{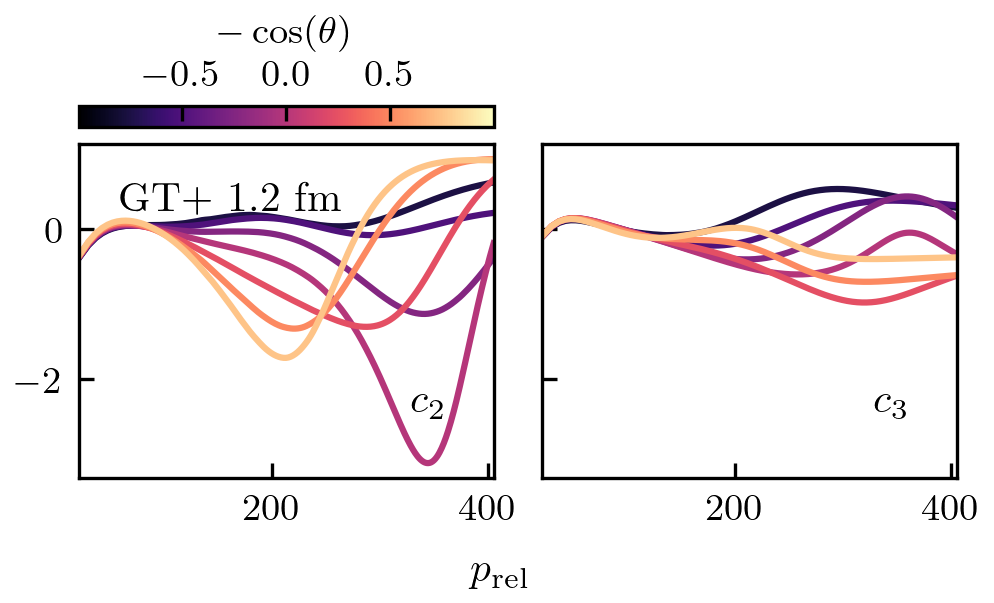}
    \includegraphics{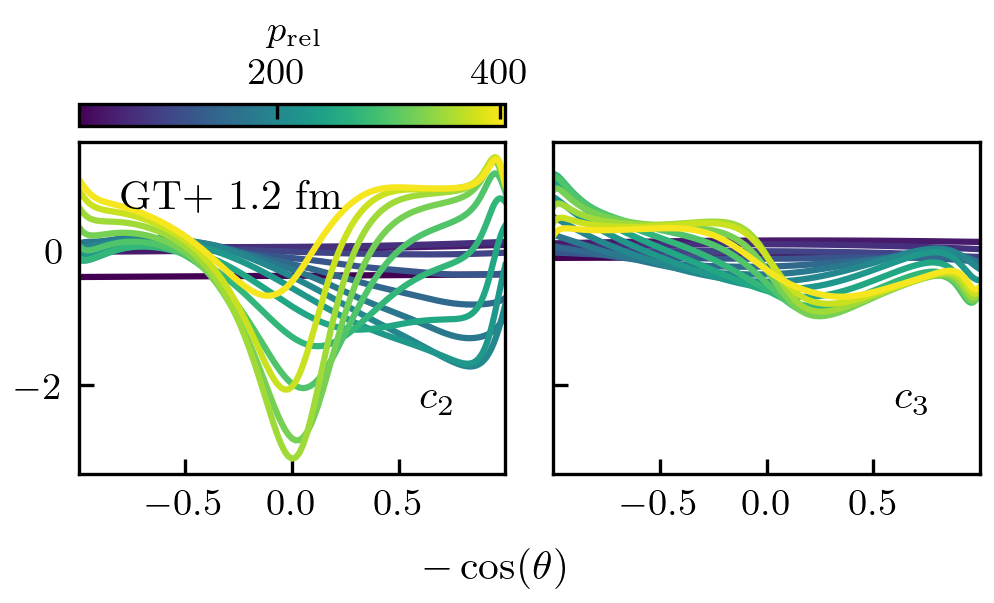}
    \caption{Order-by-order coefficients for $d\sigma / d\Omega$ for GT+ 0.9 fm, 1.0 fm, 1.1 fm, and 1.2 fm at fixed $\negcos$ and at fixed $\prel$ extracted with the choice $\left(\Lambdab = 570\,\mathrm{MeV}, \mpieff = 138\,\mathrm{MeV}\right)$ from Ref.~\cite{Millican:2024yuz}.}
    \label{fig:coeffs_eachorder_gt}
\end{figure*}

The content of this Supplemental Material sheds additional light on the coefficients not displayed in Figs.~\ref{fig:coeffs_eachorder_smsscs_vsprel} and~\ref{fig:coeffs_eachorder_smsscs_vscos} (see Figs.~\ref{fig:coeffs_eachorder_sms}--\ref{fig:coeffs_eachorder_gt}) and on the order-by-order sizes of all potentials' coefficients as quantified by rms values (see Table~\ref{tab:rms_deg_slices_full}).

\begin{figure*}[h]
    \centering
    \includegraphics{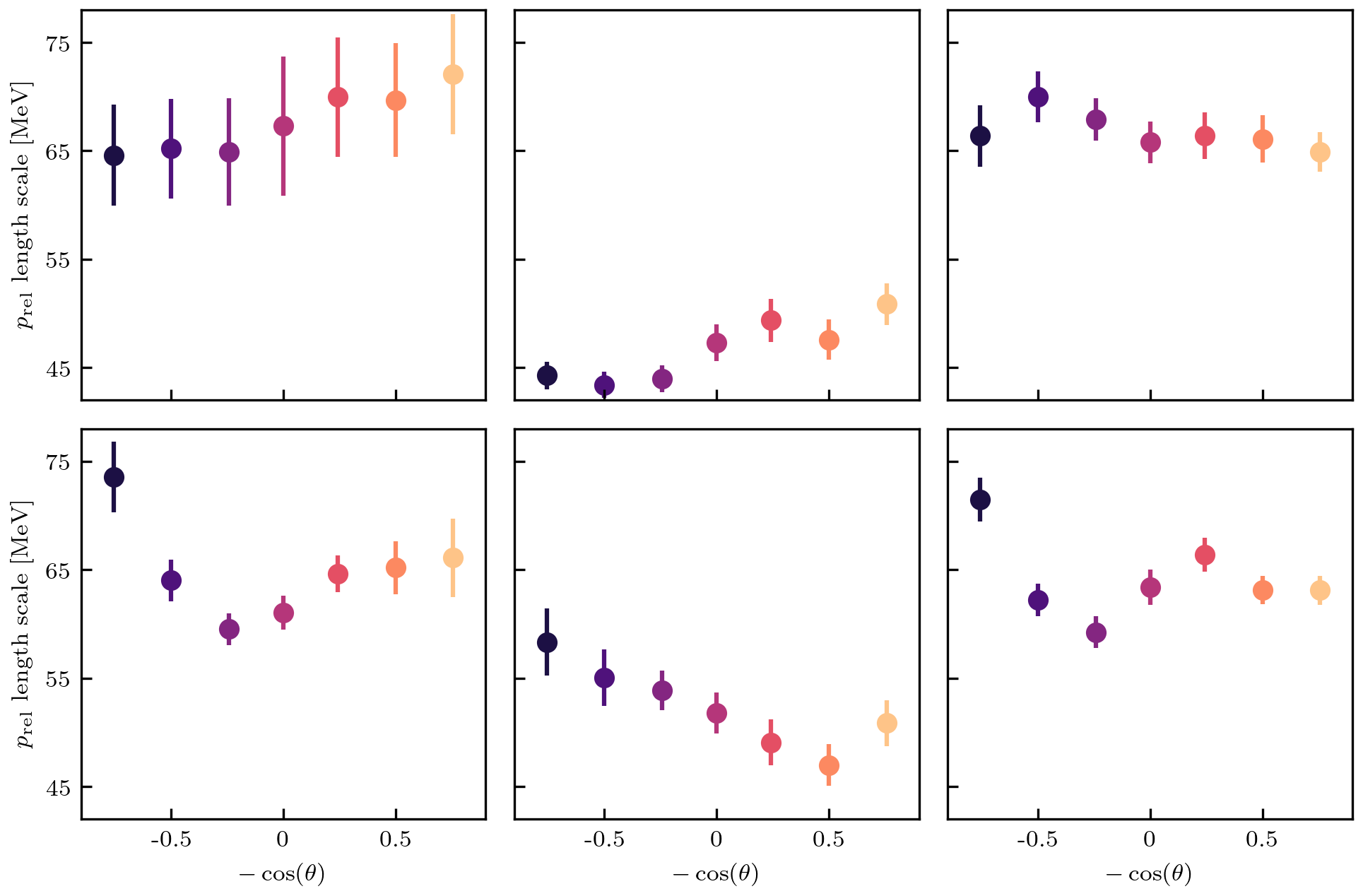}
    \phantomsublabel{-4.45}{3.10}{fig:ls_deg_slices_eachobs_dsg}
    \phantomsublabel{-2.45}{3.10}{fig:ls_deg_slices_eachobs_d}
    \phantomsublabel{-0.45}{3.10}{fig:ls_deg_slices_eachobs_axx}
    \phantomsublabel{-4.45}{1.10}{fig:ls_deg_slices_eachobs_ayy}
    \phantomsublabel{-2.45}{1.10}{fig:ls_deg_slices_eachobs_a}
    \phantomsublabel{-0.45}{1.10}{fig:ls_deg_slices_eachobs_ay}
    \caption{Plots of $\ell_{E}$ in $\prel$ at fixed $\negcos$ for \protect\subref{fig:ls_deg_slices_eachobs_dsg} $d\sigma / d\Omega$, 
    \protect\subref{fig:ls_deg_slices_eachobs_d} $D$, 
    \protect\subref{fig:ls_deg_slices_eachobs_axx} $A_{xx}$, 
    \protect\subref{fig:ls_deg_slices_eachobs_ayy} $A_{yy}$, 
    \protect\subref{fig:ls_deg_slices_eachobs_a} $A$, and 
    \protect\subref{fig:ls_deg_slices_eachobs_ay} $A_{y}$ for SMS 500 MeV extracted with the choice $\left(\Lambdab = 570\,\mathrm{MeV}, \mpieff = 138\,\mathrm{MeV}\right)$ from Ref.~\cite{Millican:2024yuz}.
    Note the stationarity of $\ell_{E}$ in $\xtheta = \negcos$.}
    \label{fig:ls_deg_slices_eachobs}
\end{figure*}

\begin{figure*}[h]
    \centering
    \includegraphics{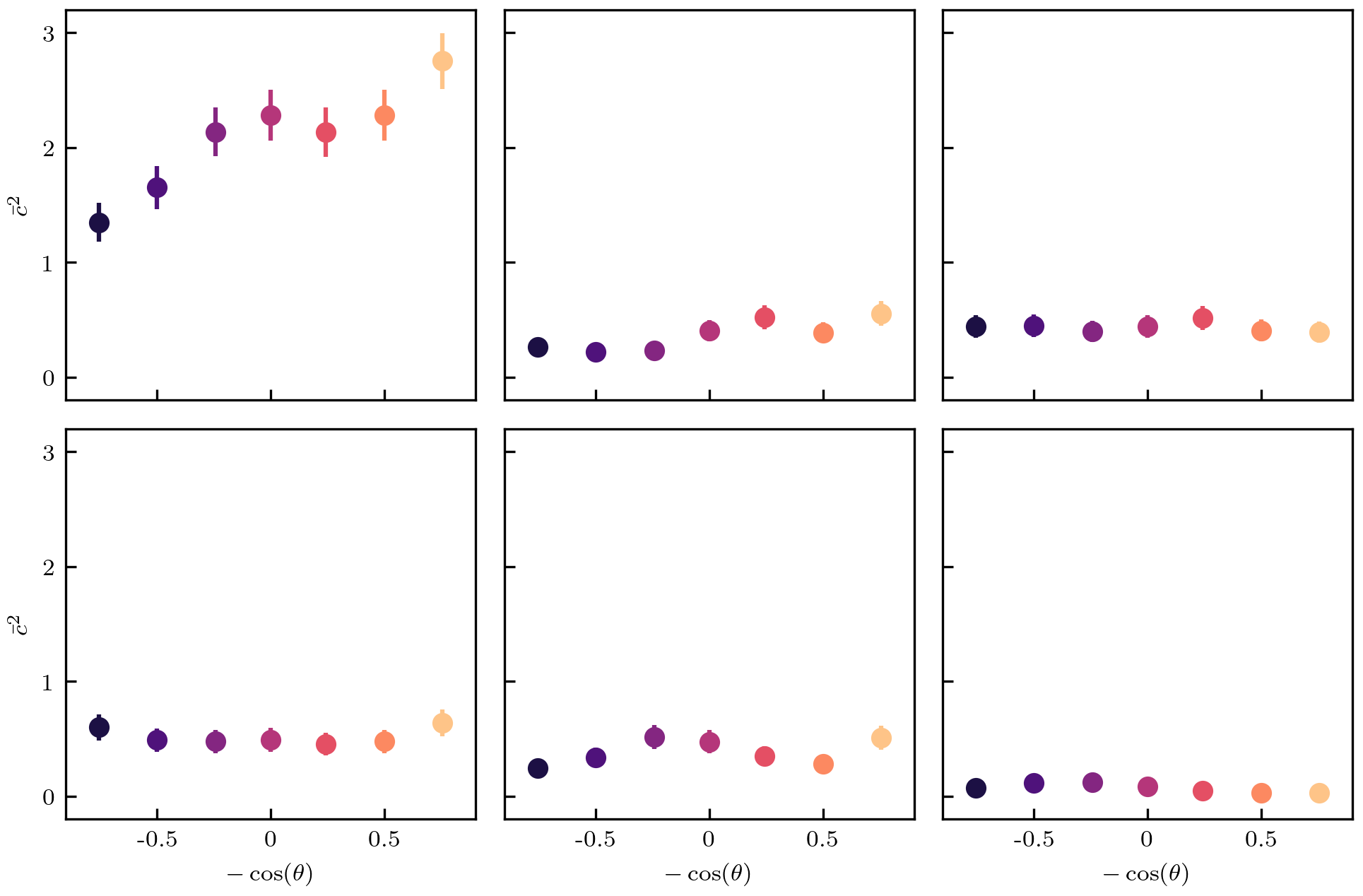}
    \phantomsublabel{-4.55}{3.10}{fig:var_deg_slices_eachobs_dsg}
    \phantomsublabel{-2.49}{3.10}{fig:var_deg_slices_eachobs_d}
    \phantomsublabel{-0.42}{3.10}{fig:var_deg_slices_eachobs_axx}
    \phantomsublabel{-4.55}{1.10}{fig:var_deg_slices_eachobs_ayy}
    \phantomsublabel{-2.49}{1.10}{fig:var_deg_slices_eachobs_a}
    \phantomsublabel{-0.42}{1.10}{fig:var_deg_slices_eachobs_ay}
    \caption{Plots of $\cbar^{2}$ at fixed $\negcos$ for \protect\subref{fig:var_deg_slices_eachobs_dsg} $d\sigma / d\Omega$, 
    \protect\subref{fig:var_deg_slices_eachobs_d} $D$, 
    \protect\subref{fig:var_deg_slices_eachobs_axx} $A_{xx}$, 
    \protect\subref{fig:var_deg_slices_eachobs_ayy} $A_{yy}$, 
    \protect\subref{fig:var_deg_slices_eachobs_a} $A$, and 
    \protect\subref{fig:var_deg_slices_eachobs_ay} $A_{y}$ for SMS 500 MeV extracted with the choice $\left(\Lambdab = 570\,\mathrm{MeV}, \mpieff = 138\,\mathrm{MeV}\right)$ from Ref.~\cite{Millican:2024yuz}.
    Note the stationarity of $\cbar^{2}$ in $\xtheta = \negcos$.}
    \label{fig:var_deg_slices_eachobs}
\end{figure*}

Additionally, it contains plots demonstrating the stationarity of $\ell_{E}$ and $\cbar^{2}$ in $\xtheta = \negcos$ (see Figs.~\ref{fig:ls_deg_slices_eachobs} and~\ref{fig:var_deg_slices_eachobs}, respectively).

\clearpage

\begin{table*}[h!]
    \centering
    \renewcommand{\arraystretch}{1.2}
    \setlength{\tabcolsep}{1pt}
    \begin{ruledtabular}
    \begin{tabular}{ c c c c c c c c c }
         & \makecell{\textbf{SMS} \\ \textbf{550 MeV}} & \makecell{\textbf{SMS} \\ \textbf{500 MeV}} & \makecell{\textbf{SMS} \\ \textbf{450 MeV}} & \makecell{SMS \\ 400 MeV} & \makecell{GT+ \\ 0.9 fm} & \makecell{GT+ \\ 1.0 fm} & \makecell{GT+ \\ 1.1 fm} & \makecell{GT+ \\ 1.2 fm} \\
        \hline
        $c_{2}$ & $0.8 \pm 0.1$ & $0.7 \pm 0.1$ & $0.57 \pm 0.07$ & $0.42 \pm 0.05$ & $1.3 \pm 0.6$ & $2.0 \pm 0.9$ & $3 \pm 1$ & $3 \pm 1$ \\
        $c_{3}$ & $0.7 \pm 0.4$ & $0.4 \pm 0.2$ & $0.26 \pm 0.09$ & $0.15 \pm 0.04$ & $1.3 \pm 0.4$ & $2.1 \pm 0.5$ & $3.2 \pm 0.7$ & $3.8 \pm 0.8$ \\
        $c_{4}$ & $1.0 \pm 0.6$ & $0.5 \pm 0.2$ & $0.5 \pm 0.3$ & $1.0 \pm 0.4$ & --- & --- & --- & --- \\
        $c_{5}$ & $0.88 \pm 0.08$ & $0.6 \pm 0.1$ & $0.44 \pm 0.04$ & $0.31 \pm 0.05$ & --- & --- & --- & --- \\
        $c_{6}$ & $0.5 \pm 0.1$ & $0.5 \pm 0.1$ & $0.56 \pm 0.08$ & $0.72 \pm 0.07$ & --- & --- & --- & --- \\
        \hline
         & \makecell{SCS \\ 0.8 fm} & \makecell{\textbf{SCS} \\ \textbf{0.9 fm}} & \makecell{\textbf{SCS} \\ \textbf{1.0 fm}} & \makecell{SCS \\ 1.1 fm} & \makecell{SCS \\ 1.2 fm} & \makecell{EMN \\ 550 MeV} & \makecell{\textbf{EMN} \\ \textbf{500 MeV}} & \makecell{EMN \\ 450 MeV} \\
        \hline
        $c_{2}$ & $0.9 \pm 0.2$ & $1.0 \pm 0.2$ & $0.7 \pm 0.2$ & $0.7 \pm 0.2$ & $0.5 \pm 0.2$ & $1.8 \pm 0.4$ & $1.7 \pm 0.4$ & $1.5 \pm 0.3$ \\
        $c_{3}$ & $0.8 \pm 0.4$ & $0.9 \pm 0.5$ & $0.6 \pm 0.3$ & $0.4 \pm 0.2$ & $0.3 \pm 0.1$ & $0.7 \pm 0.3$ & $0.7 \pm 0.3$ & $0.8 \pm 0.2$ \\
        $c_{4}$ & $0.7 \pm 0.4$ & $1.3 \pm 0.2$ & $1.8 \pm 0.6$ & $2.0 \pm 0.6$ & $1.9 \pm 0.5$ & $0.8 \pm 0.5$ & $0.7 \pm 0.3$ & $0.6 \pm 0.3$ \\
        $c_{5}$ & $0.9 \pm 0.1$ & $1.4 \pm 0.2$ & $1.0 \pm 0.1$ & $0.7 \pm 0.1$ & $0.59 \pm 0.07$ & $0.38 \pm 0.08$ & $0.56 \pm 0.04$ & $0.3 \pm 0.1$ \\
    \end{tabular}
    \end{ruledtabular}
    \caption{
    Root-mean-square (rms) values for the coefficients derived from $d\sigma / d\Omega$ for all potentials under test.
    In each case, the rms value is calculated for each coefficient, and then all the coefficients for a given potential and order are averaged to obtain these values.
    As elsewhere, the bolded potentials are those we focus on in this work (see the discussion in Sec.~\ref{subsec:convergence_diagnosis}).
    }
    \label{tab:rms_deg_slices_full}
\end{table*}

\end{document}